\documentclass{aa} 
\usepackage[varg]{txfonts}
\usepackage{graphicx}
\usepackage{url}
\usepackage{hyperref}
\usepackage[usenames,dvipsnames,svgnames,table]{xcolor}
\usepackage{upquote} 

\usepackage{soul,xcolor}
\setstcolor{red}

\titlerunning{\gaia~DR2: All-sky classification of high-amplitude pulsating stars} 

\def\gaia{\textit{Gaia}\xspace}
\def\gmag{$G$\xspace}
\def\gbp{$G_{\rm BP}$\xspace}
\def\grp{$G_{\rm RP}$\xspace}
\def\bpminrp{$G_{\rm BP}-G_{\rm RP}$\xspace}

\begin{document} 

\title{\gaia Data Release 2}
\subtitle{All-sky classification of high-amplitude pulsating stars}

\author{L.~Rimoldini\inst{\ref{inst4}}\fnmsep\thanks{\email{Lorenzo.Rimoldini@unige.ch}}
\and
B.~Holl\inst{\ref{inst4},\ref{inst1}}
\and
M.~Audard\inst{\ref{inst4},\ref{inst1}}
\and
N.~Mowlavi\inst{\ref{inst4},\ref{inst1}}
\and
K.~Nienartowicz\inst{\ref{inst3}}
\and
D.~W.~Evans\inst{\ref{inst2}}
\and
L.~P.~Guy\inst{\ref{inst4},\ref{inst-lsst}}
\and
I.~Lecoeur-Ta\"ibi\inst{\ref{inst4}}
\and
G.~Jevardat~de~Fombelle\inst{\ref{inst3}}
\and
O.~Marchal\inst{\ref{inst4},\ref{inst25}}
\and
M.~Roelens\inst{\ref{inst4},\ref{inst1},\ref{inst-MR}}
\and
J.~De~Ridder\inst{\ref{inst6}}
\and
L.~M.~Sarro\inst{\ref{inst9}}
\and
S.~Regibo\inst{\ref{inst6}}
\and
M.~Lopez\inst{\ref{inst10}}
\and
G.~Clementini\inst{\ref{inst22}}
\and
V.~Ripepi\inst{\ref{inst12}}
\and
R.~Molinaro\inst{\ref{inst12}}
\and
A.~Garofalo\inst{\ref{inst22}}
\and
L.~Moln\'{a}r\inst{\ref{inst11},\ref{inst11b}}
\and
E.~Plachy\inst{\ref{inst11},\ref{inst11b}}
\and
\'A.~Juh\'asz\inst{\ref{inst11},\ref{inst27}}
\and
L.~Szabados\inst{\ref{inst11}}
\and
T.~Lebzelter\inst{\ref{inst21}}
\and
D.~Teyssier\inst{\ref{instESA}}  
\and
L.~Eyer\inst{\ref{inst1}}
}
\authorrunning{Rimoldini et al.}

\institute{Department of Astronomy, University of Geneva, Chemin d'Ecogia 16, CH-1290 Versoix, Switzerland\label{inst4}
\and
Department of Astronomy, University of Geneva, Chemin des Maillettes 51, CH-1290 Versoix, Switzerland\label{inst1}
\and 
SixSq, Rue du Bois-du-Lan 8, CH-1217 Meyrin, Switzerland\label{inst3}
\and
Institute of Astronomy, University of Cambridge, Madingley Road, Cambridge CB3 0HA, United Kingdom\label{inst2}
\and
AURA/LSST, 950 N Cherry Avenue, Tucson, Arizona 85719, USA\label{inst-lsst}
\and
G\'EPI, Observatoire de Paris, Universit\'{e} PSL, CNRS, Place Jules Janssen 5, F-92195 Meudon, France\label{inst25}
\and 
Institute of Global Health, 9 Chemin des Mines, CH-1202 Geneva, Switzerland\label{inst-MR}
\and
Institute of Astronomy, KU Leuven, Celestijnenlaan 200D, B-3001 Leuven, Belgium\label{inst6}
\and
Departamento Inteligencia Artificial, UNED, Calle Juan del Rosal 16, E-28040 Madrid, Spain\label{inst9} 
\and
Departamento de Astrof\'isica, Centro de Astrobiolog\'ia (INTA-CSIC), PO Box 78, E-28691 Villanueva de la Ca\~nada, Spain\label{inst10}
\and
INAF - Osservatorio di Astrofisica e Scienza dello Spazio di Bologna, Via Gobetti 93/3, I-40129 Bologna, Italy\label{inst22}
\and
INAF - Osservatorio Astronomico di Capodimonte, Via Moiariello 16, I-80131 Napoli, Italy\label{inst12}
\and
Konkoly Observatory, Research Centre for Astronomy \& Earth Sciences, Hungarian Academy of Sciences, Konkoly Thege Mikl\'os \'ut 15-17, H-1121 Budapest, Hungary\label{inst11}
\and
MTA CSFK Lend\"ulet Near-Field Cosmology Research Group, Konkoly Thege Mikl\'os \'ut 15-17, H-1121 Budapest, Hungary\label{inst11b}
\and
Department of Astronomy, E\"otv\"os Lor\'and University, P\'azm\'any P\'eter s\'et\'any 1/a, H-1117 Budapest, Hungary\label{inst27}
\and
Department of Astrophysics, University of Vienna, Tuerkenschanz\-strasse 17, A-1180 Vienna, Austria\label{inst21}
\and
Telespazio Vega UK Ltd for ESA/ESAC, Camino bajo del Castillo, s/n, Urbanizacion Villafranca del Castillo, Villanueva de la Ca\~{n}ada, E-28692 Madrid, Spain \label{instESA}
}

\date{Received 9 November 2018 / Accepted 17 March 2019}

\abstract 
   {More than half a million of the 1.69 billion sources in \gaia Data Release 2 (DR2) are published with photometric time series that exhibit light variations during the 22 months of observation.} 
%
   {An all-sky classification of common high-amplitude pulsators (Cepheids, long-period variables, $\delta$\,Scuti/SX\,Phoenicis, and RR\,Lyrae stars) is provided for stars with brightness variations greater than 0.1 mag in \gmag band.}
%
   {A semi-supervised classification approach was employed, firstly training multi-stage random forest classifiers with sources of known types in the literature, followed by a preliminary classification of the \gaia data and a second training phase that included a selection of the first classification results to improve the representation of some classes, before the improved classifiers were applied to the \gaia data.  Dedicated validation classifiers were used to reduce the level of contamination in the published results. 
 A relevant fraction of objects were not yet sufficiently sampled for reliable Fourier series decomposition, consequently classifiers were based on features derived from statistics of photometric time series in the \gmag, \gbp, and \grp bands, as well as from some astrometric parameters.}
%
   {The published classification results include 195\,780 RR\,Lyrae stars, 150\,757 long-period variables, 8550 Cepheids, and 8882 $\delta$\,Scuti/SX\,Phoenicis stars. All of these results represent candidates whose completeness and contamination are described 
   as a function of variability type and classification reliability. 
   Results are expressed in terms of class labels and classification scores, which are available in the \texttt{vari\_classifier\_result} table of the \gaia archive.}
%
   {}


\maketitle

\section{Introduction \label{sec:introduction}} 
 
The light curves of variable stars exhibit features that can reveal valuable information on the physical causes of brightness variations.
They help us improve our understanding of stellar properties, some of which can be used to develop and refine methods that turn variable stars into astrophysical tools applicable to Galactic or even extra-galactic scales \citep{2006ASPC..349.....S}.

The brightness variations of pulsating stars are caused by periodic expansion and contraction, which may alternate throughout the whole star (causing a uniform swelling, shrinking, and temperature changes across the entire stellar surface) or in the case of non-radial oscillations occur simultaneously but in different regions of the star (typically associated with lower brightness variations).  
The amplitude, periodicity, regularity, lifetime, and other features of stellar pulsations depend on stellar evolution stages, which can be mapped on the Hertzsprung-Russell diagram and are found to correspond to different observed types of variability \citep[e.g.][]{2018arXiv180409382G}, which are commonly identified by the light-curve shapes, pulsation period(s), amplitude(s), and intrinsic colours, among other factors.

Pulsating variables became particularly important for the information that can be inferred on stellar interiors with asteroseismology \citep[e.g.][]{2010aste.book.....A, 2004ESASP.559....1C} and the relationships between the pulsation periods and stellar luminosities \citep[e.g.][]{2006ARA&A..44...93S, 2017A&A...605A..79G, 2018arXiv180502079C}, which can be used to determine the distance, and the three-dimensional distributions of these stars can thus outline the structures they belong to within our Galaxy and beyond \citep[e.g.][]{2017ApJ...850...96H, 2013ApJ...765..154D}.

Examples of pulsating variables in recent surveys include the Panoramic Survey Telescope and Rapid Response System \citep[Pan-STARRS;][]{2017AJ....153..204S, 2016ApJ...817...73H}, Catalina \citep{2015MNRAS.446.2251T, 2014ApJS..213....9D, 2013ApJ...763...32D, 2013ApJ...765..154D}, the Optical Gravitational Lensing Experiment \citep[OGLE;][]{2017AcA....67..297S, 2016AcA....66..131S, 2015AcA....65..233S, 2015AcA....65..297S, 2014AcA....64..177S}, \textit{Kepler} \citep{2015AJ....149...68B, 2011A&A...529A..89D}, the Asteroid Terrestrial-impact Last Alert System \citep[ATLAS;][]{2018arXiv180402132H}, \gaia~DR1 \citep{2017arXiv170203295E, 2016A&A...595A.133C}, the Lincoln Near-Earth Asteroid Research \citep[LINEAR;][]{2013AJ....146..101P}, the Northern Sky Variability Survey \citep[NSVS;][]{2006AJ....132.1202K, 2004AJ....128.2965W}, the All Sky Automated Survey \citep[ASAS;][]{2002AcA....52..397P}, and ASAS for Supernovae \citep[ASAS-SN;][]{2018MNRAS.477.3145J, 2018arXiv180907329J}.  
Surveys that cover large regions of the sky tend to collect remarkable volumes of data, which help improve our understanding of known objects, increase the knowledge of rare objects, enable discoveries, and raise new questions, while providing all the data in a consistent context.
\gaia pursues all this across the whole sky with the benefits of a rich set of instruments \citep{2016A&A...595A...1G} that repeatedly provide the astrometry, photometry, and spectroscopy of the observed objects. This increases the measurement accuracy and follows variations in time.

The second Data Release (DR2) of \gaia \citep{2018arXiv180409365G} includes an earlier-than-planned publication of variable sources and related photometric time series in the \gmag, \gbp, and \grp bands. 
In particular, 550\,737 variable stars are published \citep[as summarised in][]{2018arXiv180409373H}, among which 363\,969 objects are identified by the all-sky classification pipeline \citep[see section~7.3 of][]{2018gdr2.reptE...7E} as candidate 
RR\,Lyrae stars, Cepheids, $\delta$\,Scuti/SX\,Phoenicis stars, and long-period variables, 
which constitute the first published classification results from the \gaia variability pipeline, covering the whole sky and 
without a priori selections of sources based on their sampling (unless only a single measurement was available). 
Other variable stars published in \gaia~DR2 originated from special variability detection algorithms \citep{2018A&A...616A..16L, 2018arXiv180500747R} and from another classification of sources dedicated to well-sampled sources \citep[see section~7.2.3.6 in][]{2018gdr2.reptE...7E}. 
The list of \gaia~DR2 articles related to the data processing and analysis or validation of the published variable objects is presented in \citet{2018arXiv180409373H}.

The \gaia variability processing and analysis includes several modules dedicated to different tasks, as described in \citet{2017arXiv170203295E}, such as the computation of statistical parameters, the detection of variability, the characterisation of light curves, the classification of variable objects, and specific object studies (SOS) that confirm the identifications from previous stages (and in some cases reclassify) and refine the description of variability of specific types. 
The large number of sources and the sparse sampling of \gaia time series do not allow for `perfect' classifications of all sources.
Classification results of types that are verified by SOS are published  
for transparency of the pipeline processing, but also to make it possible to publish results of sources that SOS could not confirm or that were improperly rejected for various reasons (e.g.\ because of spurious information, insufficient observations or photometric precision, inaccurate astrometric parameters, or other factors). 
Consequently, classification results are generally more complete but also more contaminated and associated with a less accurate light-curve characterisation than the one derived from SOS.

For \gaia~DR2, two independent classifications of variable objects were performed: one with at least 20~field-of-view (FoV) transits in the \gmag~band \citep[described in section~7.2.3.6 of][]{2018gdr2.reptE...7E}, which included period search and Fourier modelling in the computation of classification attributes, and one that required at least 2~FoV transits in the \gmag~band, which covered the whole sky more homogeneously but was limited to statistical parameters to characterise source features. The latter was published in the \gaia~DR2 archive and is described in detail in this article; a general overview  and some of its technical details are presented also in section~7.3 of \citet[][]{2018gdr2.reptE...7E}. 
The (unpublished) classification that made use of modelling attributes added a total of 37\,016 variable stars that were confirmed by different SOS packages.
Thus, for \gaia~DR2, the published SOS results were not strictly a subset of the published classifier results \citep[see fig.~3 in][for an illustration per variability type]{2018arXiv180409373H}.

The quality of the classification results was limited by time constraints and by the dependence on data produced by \gaia pipelines that were executed in parallel to ours, which in some cases forced us to use preliminary (not published) \gaia data products and reduced the options available for variability processing.  
Without restricting the publication of classifications to objects associated with well-sampled and fully understood time series (according to current knowledge), the effect of selection biases is reduced and the community is granted early access to additional time series,  
adding discovery potential and facilitating progress in currently debated topics 
\citep[e.g.][]{2017MNRAS.466.4711B,2018arXiv180401755J}.
In addition to the classification scores and other indications given in this article, subsets of sources with enhanced reliability and sufficient sampling are selected by subsequent pipeline modules (SOS) for some of the variability types, whose detailed analyses and validations 
are presented in separate articles (\citealt{2018arXiv180502035M}, for long-period variables, and \citealt{2018arXiv180502079C} and \citealt{2018arXiv180511395M}, for Cepheids and RR\,Lyrae stars). 

This article emphasises the method employed for the classification of stars covering the whole sky and presents its results with indications relevant to their usage. Following the internal agreement of the \gaia data processing and analysis consortium, no scientific exploitation of these results is made herein.
The \gaia data set is summarised in Section~\ref{sec:data}, the method employed to identify four main classes of pulsating variables is described in Section~\ref{sec:method}, results and comparisons with the literature are outlined in Section~\ref{sec:results}, and conclusions are drawn in Section~\ref{sec:conclusions}. 
Appendix~\ref{app:classes} recalls the classes of the objects employed in the training set, and 
Appendix~\ref{app:queries} provides a few sample queries applicable to classification-related searches in the \gaia archive\footnote{\url{http://gea.esac.esa.int/archive}\label{foot:archive}}.

\section{Data \label{sec:data}}

The data of \gaia~DR2 are accessible from the ESA \gaia archive, from four partner data centres\footnote{\url{https://www.cosmos.esa.int/web/gaia/data-access}}, and from a number of affiliated data centres around the world.
Due to time constraints and to multiple \gaia modules preparing for DR2 in parallel, only some of the published data products were available at the time of variability processing. 
The processing of variable sources employed photometric data of FoV transits \citep{2018arXiv180409368E, 2018arXiv180409367R} in the \gmag, \gbp, and \grp bands, with a small fraction of per-CCD \gmag-band photometry \citep{2018arXiv180500747R}, and a preliminary version of the astrometric solution \citep[the final version is described in][]{2018arXiv180409366L}, 
while it could not make use of (preliminary nor published) spectroscopic data \citep{2018arXiv180409371S}, radial velocities \citep{2018arXiv180409372K}, and astrophysical parameters \citep{2018arXiv180409374A}. 
Moreover, the initial stages of the all-sky classification activities (the cross match of \gaia data with catalogues of variable objects from the literature, described in Section~\ref{sec:xm}, and the determination of parameters for quasar attributes, mentioned in Section~\ref{sec:attributes}) relied on preliminary per-FoV photometry in the \gmag, \gbp, and \grp bands in order to gain time for subsequent (more critical) classification phases.
Despite these short-comings, we believe that the advantages of an early publication of time series outweighs the benefits of a fully consistent data product,  
as the published classifications are not critically affected
and the community can start working on these time series (together with the other DR2 data products) without waiting for the next data release. Eventually, these inconsistencies and their effects on results are expected to be reduced (or vanish) in the future.

The time series were reconstructed from FoV transit data with times referred to the solar system barycentre (Barycentric Coordinate Time), which means that they do not encode the motion of \gaia around the Sun (nor the time dilation from the gravitational potential of the latter).
All sources with at least two FoV transits in the \gmag~band were processed, and suspicious time-series measurements were filtered out as outlined in \citet{2018arXiv180409373H} and detailed in section~7.2.3.2 of \citet{2018gdr2.reptE...7E}.
The final \gaia validation stage, subsequent to the variability pipeline processing, 
removed duplicated sources and objects affected by imprecise astrometric solutions \citep{2018arXiv180409375A}, which increased the minimum number of \gmag-band FoV transits of the published \gaia variables to five.
The global properties of the published variable sources, an overview of their statistical parameters, magnitude and sky distributions per variability type, and the sky coverage as a function of the number of FoV transits per source are shown in \citet{2018arXiv180409373H}.

Section~\ref{sec:method} describes the methods and their implementation for the classification of the \gaia variable objects across the whole sky, while Section~\ref{sec:results} describes these classification results as they appear in the \gaia archive (i.e.\ after the validation cuts that followed the variability processing).

\section{Methods \label{sec:method}}

In the current era of big data, it is not possible to inspect every single light curve (at least not within the human resources and time allocated for the \gaia data releases).  Machine-learning gives the possibility of automating decisions and thus processing a large number of sources based on patterns that are automatically recognised from much smaller controlled samples of training objects.

The initial goal for an advance classification across the whole sky with the \gaia~DR2 data (thus including poorly sampled sources) was represented by RR\,Lyrae stars because they are important as standard candles and stellar population tracers \citep[e.g.][and references therein]{2014IAUS..301..129C}, and because they are easily identified through their large light variation amplitudes and short periods (typically shorter than a day), as other works classifying RR\,Lyrae stars with a low number of observations have shown \citep[e.g.][]{2000AJ....120..963I, 2007AJ....134.2236S}.
The classifier developed for RR\,Lyrae stars included several variability types by design (to reduce the contamination of the targeted class). After an assessment of its results, it was decided that it was worth to extend the publication of the classification results to additional types of pulsating stars (Cepheids, $\delta$\,Scuti/SX\,Phoenicis stars, and long-period variables) so that the community could further benefit from the publication of their time series as well.  As this decision was reached after the construction of the classifier model, 
the training sources of these additional variability types were not as representative in the sky and in the magnitude distributions as the RR\,Lyrae training sample, and thus the associated classification results were more likely to be affected by training biases.

\begin{figure}
 \centering
 \includegraphics[width=0.75\hsize]{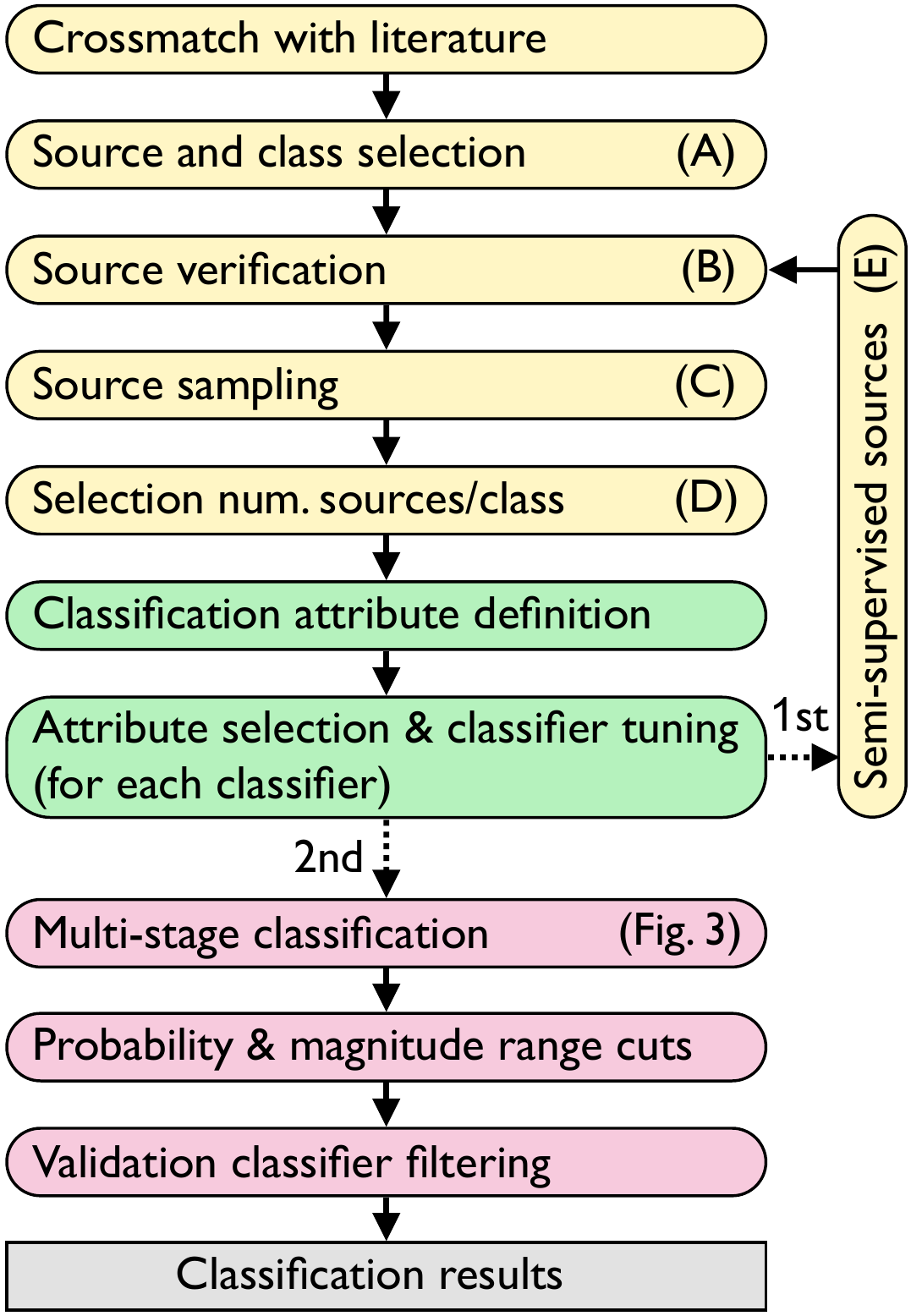}
  \caption{General flow of the procedures employed to achieve the classification results. Labels (A)--(E) refer to the items described in Section~\ref{sec:source-select}. Exceptions (as in the case of the long-period variables, which were not filtered by a validation classifier) are described in the text. The box colours group the procedures by topic: training source/class selection (yellow), classification attributes (green), and classifiers and result selection (red).}
  \label{fig:flow}
\end{figure}

As described in \citet{2017arXiv170203295E}, the classification of variable stars in the \gaia pipeline was based on attributes that characterised the objects (to train and then classify), such as the statistics of photometric time series, the associated modelling parameters, the astrometric properties, and other features if available.
The classification results published in \gaia~DR2 were obtained by means of supervised classification, which depended crucially on the selection of training-set objects (whose rationale is explained in Section~\ref{sec:training}), on the combination of attributes that best described the distinctive features of the related classes, although without period information (Section~\ref{sec:attributes}), and on the organisation of the classifier (Section~\ref{sec:classifier}).
The implementation details of these procedures applied to various classifiers are presented in Section~\ref{sec:impl}, followed by an outline of the automated validation of the first classification results (Section~\ref{sec:validation}) and the definition of the classification score (Section~\ref{sec:class_score}).
A summary of the steps described in this section is shown in Fig.~\ref{fig:flow}, which presents the general flow of various procedures, while the details and exceptions are explained in the text. 
The classification quality estimators often mentioned in this work include the per-class completeness (i.e.\ true positive or recall) rates and contamination (i.e.\ $1-$\,precision or $1-$\,purity).

\subsection{Training set \label{sec:training}}

In supervised classification, the importance of training-set objects of known classes is paramount because results will be as good, poor, or biased as the training set.
The data to train a classifier should be as similar as possible to the unlabelled data that are to be classified.
In order to embed the \gaia properties (time sampling, photometric bands, and some of the data imperfections) in the training set, the latter was built using \gaia sources of independently known classes that resulted from the cross match of known objects in the literature with \gaia sources.
The cross match with a large number of surveys, catalogues, sources, and variability types from the literature was necessary to gather sufficient objects from which training sources could be selected, and included a multitude of classes, covered all relevant sky regions per class (to probe location-dependent effects), and have magnitudes, colours, brightness variation amplitudes, periods (if periodic), number of FoV transits, and other parameters that are distributed according to the full range admitted for each class, convolved with the detectability of \gaia.
It is usually not possible to achieve this level of detail for every parameter and class, but it remains important to pursue it within the available resources in order to limit the effect of training-set biases.

The representation of classes as a function of the number of FoV transits or sampling could be improved for those classes that suffered from insufficient sky coverage (which also reduced the diversity of time-series sampling), for example, by adding new training sources derived from downsampling better sampled time series. This approach was not employed in this work in order to preserve the attribute distributions and relations within the data, and also to prevent new biases and artefacts from the introduction of artificial sources. However, future classifications of \gaia variables might include such training extensions, depending on the outcome of analyses evaluating the benefits and costs of similar procedures for each class \citep[e.g.][]{2012PASP..124..280L}.

\subsubsection{Cross match \label{sec:xm}}

The classifier method that was employed to cross match variable objects with \gaia~DR1 sources \citep{2017arXiv170204165R} required significant preparatory work for each data set from the literature.  Considering the large number of catalogues of variable sources targeted for cross match in the variability processing of \gaia~DR2, a simplified method was developed to keep using multiple dimensions in order to reduce the number of incorrect and missed matches, but without the aid of machine-learning.
A single multi-dimensional distance was built from different metrics, depending on their availability: angular separation (for all cases), colour (represented by the difference of time-series medians in the \gbp and \grp bands for the \gaia data), magnitude (computed as the median time-series magnitude in the \gmag band for the \gaia data), and amplitude estimators such as the range or standard deviation of \gmag-band time-series magnitudes. 
These photometric parameters were computed after a simple filtering of \gaia FoV transits to reduce the possible influence of outliers: magnitudes above or below 15 times the median absolute deviation (MAD) from the median were removed (for all bands) and the top and bottom 10\%\ of magnitudes in the \gmag-band time series were filtered out for the amplitude estimation. 

In order to compute a multi-dimensional distance, the metrics derived from \gaia photometric time series were compared with those from data sets of other surveys (with different bands) by means of empirical relations, limited to linear models and established iteratively: in the first iteration, matches were derived simply from the nearest neighbours (only astrometry) within one arcsecond, making the first cross-survey empirical relations available for use and refinement in subsequent iterations (that were not limited to astrometric comparisons, nor to the same cross match radius).
For the second and subsequent iterations, all neighbours within a preset radius (typically a few arcseconds, depending on the survey) were considered as potential matches.
A single match for each source was selected according to the smallest multi-dimensional distance\footnote{The cross-match distance was defined as the sum in quadrature of the differences of metric values of two surveys; each component was normalised by the MAD of the distribution of the differences of the related metric that was derived from the matches of the previous iteration.}, after the removal of likely duplicate \gaia sources \citep{2018arXiv180409375A}.
If the same match was associated with multiple sources (as may happen in crowded fields), it was assigned to the source with the smallest cross match distance, 
and new potential matches (if available) were (re-)considered for the other sources (and the same procedure was applied recursively in case of multiple sources for each match).
The distributions of the differences of all pairs of metrics of the final iteration were visually inspected for the possibility of applying further constraints and excluding suspicious outliers. 
The time series of matched \gaia sources were plotted for verification, adjustments of the thresholds of the metrics, and their possible reassessment.

This method enabled us to cross match efficiently about 70 different data sets from the literature (for a total of about two million sources of known classes). The subset of catalogues considered sufficiently reliable for training purposes included over 750 thousand objects from the data sets listed in table~7.1 of \citet{2018gdr2.reptE...7E}, together with literature references and the cross-match metrics employed. Although this large number of cross-matched sources included 29 classes, the constant stars, eclipsing binaries, and RR\,Lyrae variables accounted for 92\%\ (about 54, 19, and 19\%, respectively) of all sources; the largest contributions came from the Sloan Digital Sky Survey standard star catalogue for Stripe 82 \citep{2007AJ....134..973I}, from the OGLE-IV eclipsing binaries and RR\,Lyrae stars identified in the Galactic bulge \citep{2016AcA....66..405S, 2014AcA....64..177S} and in the Magellanic Clouds \citep{2016AcA....66..421P, 2016AcA....66..131S}, and from the Pan-STARRS1 classification of RR\,Lyrae stars \citep{2017AJ....153..204S}. These catalogues accounted for about 84\%\ of all cross-matched objects.
In the special case of OGLE-IV GSEP constant objects \citep{2012AcA....62..219S}, no specific catalogue for constant stars is available. About 10~thousand constant star candidates were derived from sources with the smallest variations in both $V$ and $I$ bands as a function of magnitude, after objects were removed because they were considered less reliable (with $V-I$ bluer than $-1$~mag or redder than 3~mag, with fewer than 20 or 200 observations in the $V$ and $I$ band, respectively, with measurements flagged as `bad', with standard deviation greater than 0.1~mag in either band, and fainter than 22 or 21~mag in the $V$ and $I$ band, respectively) or they were identified as galaxies by OGLE.

\subsubsection{Classes and sources \label{sec:source-select}}

The selections of classes and of training sources that represent them are pivotal to the purpose of the training set.
In supervised classification, it is important to train with a variety of classes to reduce the chance of contamination from sources that belong to classes that are missing in the training set. 
In addition to the classes targeted for publication, those that are commonly confused with them as well as others that are well separated are expected to be part of a training set.
Objects of untrained classes will be classified as one of the trained classes, possibly associated with low classification scores and/or significant anomaly or outlier indicators (such as the proximity\footnote{\url{https://www.stat.berkeley.edu/~breiman/RandomForests/cc_home.htm}} in random forest or the Mahalanobis distance, \citealt{Mahalanobis}, in Gaussian mixture classifiers), which can help in the rejection of false positives.
As a consequence of the literature data sets that were selected to classify variable stars for \gaia~DR2, different combinations of 29 classes (see Section~\ref{sec:multistage}) were included in the training set with at least 10 representatives per class for a minimum class definition (although poorly sampled classes remained poorly defined and likely biased).
These classes comprised constant and variable objects, whose labels and definitions were listed in section~7.3.3.1 of \citet{2018gdr2.reptE...7E}; they are replicated herein for convenience in Appendix~\ref{app:classes}. They include 11 subclasses that belonged to four families of pulsating stars (Cepheids, long-period variables, RR\,Lyrae, and $\delta$\,Scuti/SX\,Phoenicis stars), which were published in the classification results of variables in the ESA \gaia archive$^{\ref{foot:archive}}$.

The large number of cross-matched sources allowed us to be very selective and better control the distribution of objects in the sky, in magnitude, and in the number of \gmag-band FoV transits for several classes and consequently reduce the effect of training-set biases on classifications.
The final training set included about 33~thousand \gaia sources.
Their selection involved the following semi-automated procedures interleaved with visual inspections to assess the quality of each stage and the subsequent steps.

\paragraph{(A) Literature data.} After uniforming the class labels of different cross-matched data sets, the latter were first selected and then prioritised according to evaluations of their classification reliability, in order to resolve contradictions of different literature classes associated with the same \gaia source (as found in 0.2\%\ of all matches, which included types that described different manifestations of the same phenomenon, such as flaring and rotating spotted stars).
The reliability of literature classifications was based on survey features, on the location of sources in the sky, on the appearance of their light curves in the \gaia photometry (limited to samples of sources with contradictory classes in the literature), and on qualitative rankings from the experience of a subset of co-authors in using these data sets for some classes. This assessment did not correct for every classification disagreement in the literature, but it was deemed to reduce the effect of the remaining incorrect class assignments well below the effect of other sources of classification confusion.

\paragraph{(B) Source verification.} Simple statistics (such as the median colour and magnitude, the range of magnitude variations, the skewness, and the Abbe value computed on magnitudes sorted in time and in phase with the literature period when available) were computed from the \gaia time series, and sources that did not satisfy constraints typical of their own class were rejected.  
These constraints were typically applied to two-dimensional projections of the statistics mentioned above for each trained class, but they remained rather permissive (i) to allow for a wide range of possible distance, extinction, and reddening (with consequent effects on magnitudes and colours), (ii) to consider the limited sampling of a good fraction of \gaia sources (with few FoV transits), and (iii) to include the occasional effect of spurious measurements that remained after the initial time-series cleaning (so that the classifier could expect such artefacts as well).
In the case of eclipsing binaries, the reduced and sparse sampling might miss the signal, therefore a positive skewness was required in addition to a minimum \gmag-band range of 0.1~mag.

\paragraph{(C) Sampling.} Cross-matched sources incorporate the footprints of the original data sets, which can cause artificial (survey-specific) peaks in the distribution of several parameters.  In order to alleviate predictable training biases, subsets of cross-matched sources were selected for more representative distributions in the sky, in magnitude, and in the number of FoV transits (contemporaneously), pursued separately for each class, and conditioned on the availability of at least about 10\,000 sources per class (which was possible for classes denoted by the following labels, sorted by increasing occurrence: CEP, ELL, ROT, RRC, RRAB, ECL, and CONSTANT).  

\paragraph{(D) Classifier priors.}
The classification method described in Section~\ref{sec:classifier} was based on a set of random forest classifiers \citep{Breiman.Random.Forest} that were generally sensitive to the number of sources for each class relative to each other. 
For better control of the relative importances of classes, the sources of each class were further selected up to preset quotas for each class. 
This selection took into account the literature origins of sources and applied different caps to the random draw of sources from specific cross-matched data sets in order to favour the representation of a variety of information in the training set and reduce the influence of large surveys whose properties or targeted locations only partially matched those of the \gaia data.

\paragraph{(E) Semi-supervised sources.} Despite the large number of cross-matched sources, their number density (per class) decreased quickly in multi-dimensional volumes, leaving under-represented intervals in several cases.  For the classes labelled CONSTANT, MIRA, RRAB, RRC, and SR, gaps in the magnitude and/or sky distributions could be filled with sources classified as such classes from a first execution of the classification module. The latter made no use of the attributes whose distributions were meant to be corrected (such as the \gmag magnitude or sky location), nor of the attributes that were obviously correlated with them (such as the \gbp and \grp magnitudes, inter-band correlations, and all astrometric attributes) in order to reduce classifier biases, although other attributes might have suffered from biases as well (some of which may be related to residual correlations with the targeted attributes as well). The remaining attributes were selected according to the techniques described in Section~\ref{sec:attributes}.

In this semi-supervised approach, classified sources within magnitude intervals and sky regions that lacked representation were added to the training set if their classification confidence and the associated class completeness rate were estimated sufficiently high (to limit the chance of introducing contaminants). Such a dependence on the classification reliability might bias the sampling of sources. To mitigate this (less severe) effect and make the contribution of new training information possible, the classification confidence of new sources was also constrained by an upper limit. The same values of uncalibrated classification `probabilities' had different meaning for different classes, therefore these thresholds were assessed as a function of class. The new candidate training sources were verified by the same per-class statistics that were applied to the other training sources (item~B), sampled to smooth distributions (item~C, when applicable), or to fit the classifier priors (item~D), and finally amounted to 11\%\ of the training set globally (in particular, 14, 50, 13, 9, and 71\%\ of CONSTANT, MIRA, RRAB, RRC, and SR classes, respectively). Reliable \gaia classifications are expected to increase their footprint (in terms of number of classes and contribution to each of them) in the training sets of variable objects in future \gaia data releases.

\begin{figure}
\centering
  \includegraphics[width=\hsize]{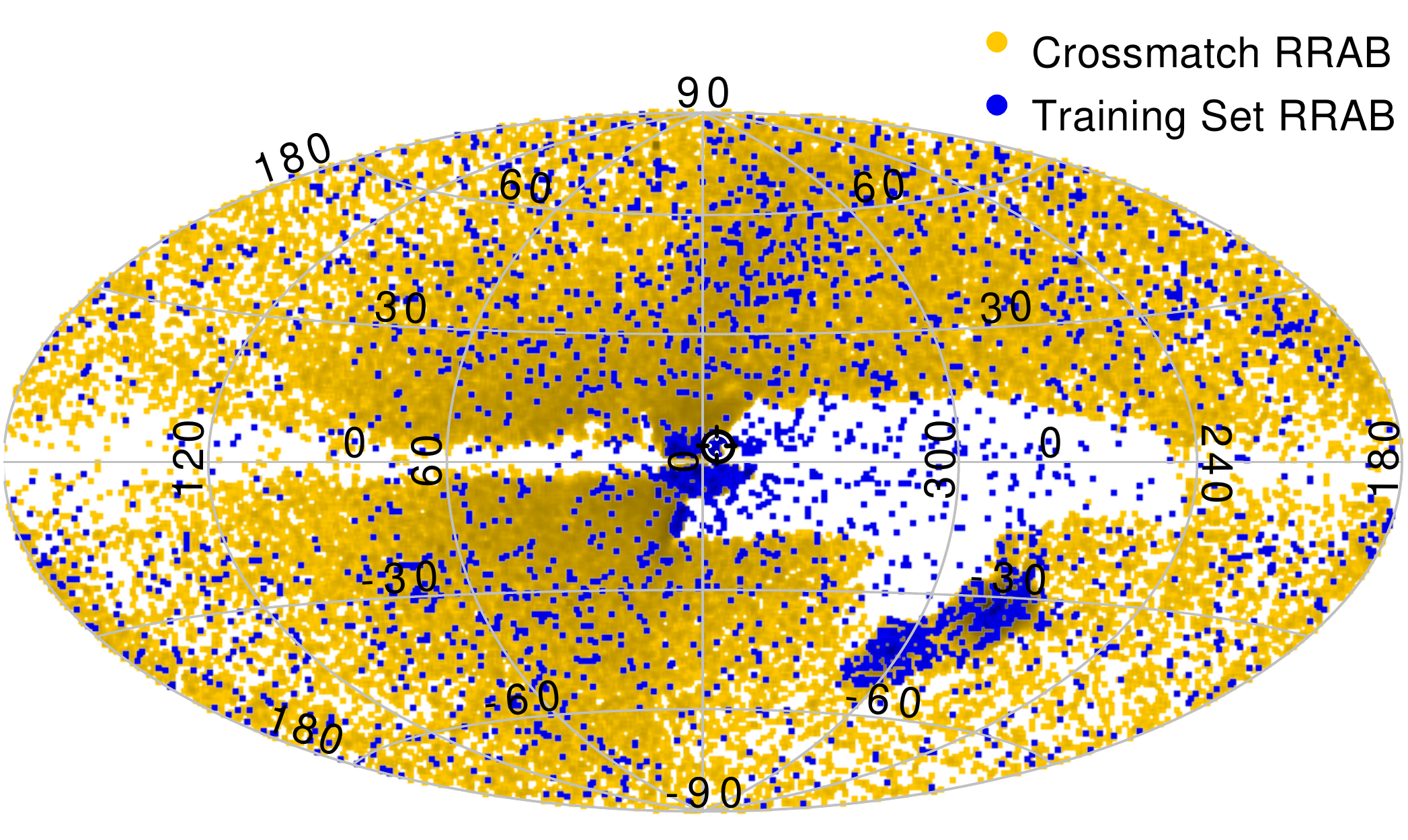}\\(a)\\ \vspace*{0.2cm}
  \includegraphics[width=\hsize]{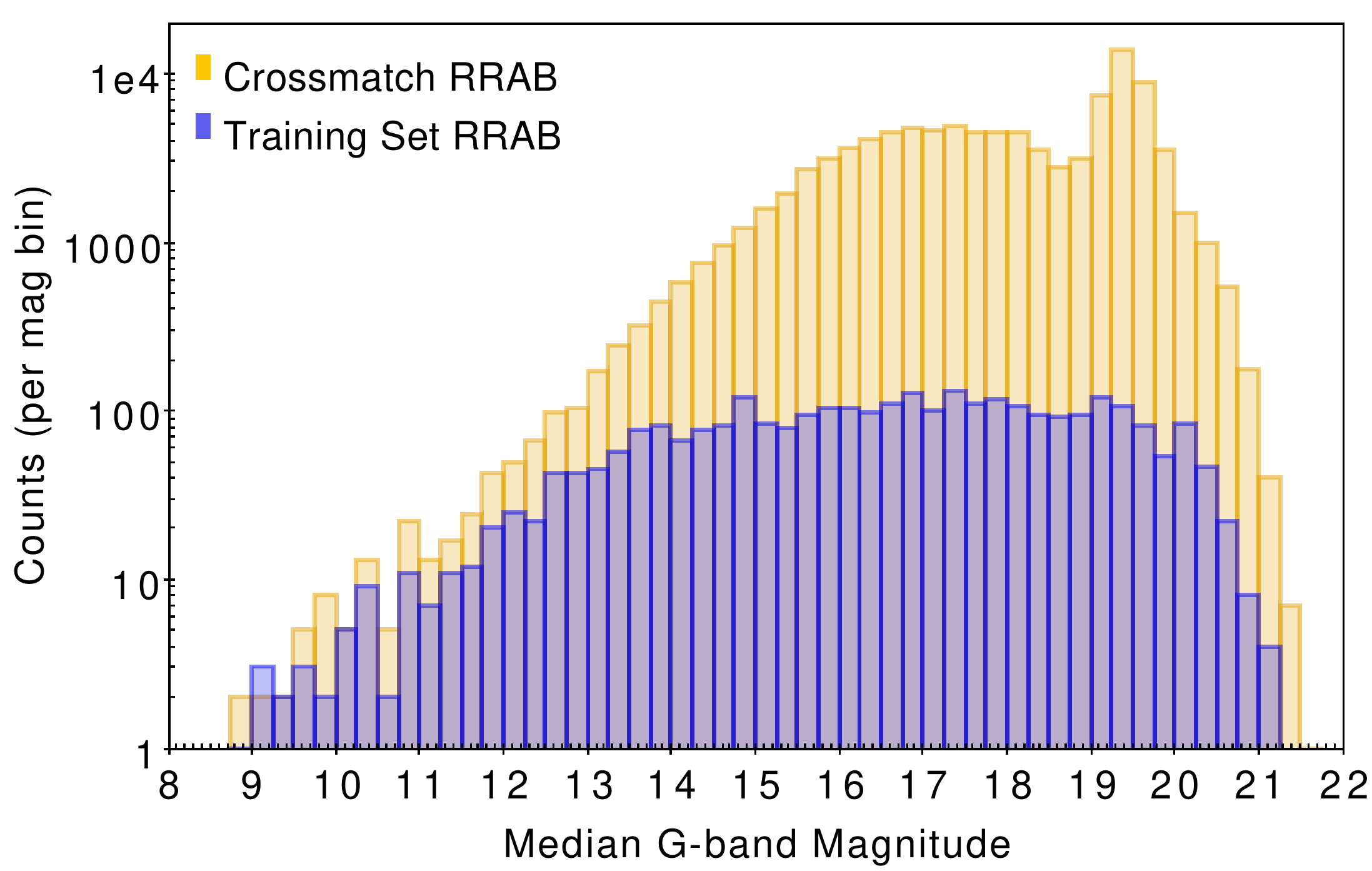}\\(b)
  \caption{Distributions of cross-match (orange) vs.\ training-set (blue) sources of RRAB type in the sky~(a), in Galactic coordinates (degrees), and in the \gmag-band magnitude~(b). The source sampling helped reduce the number of sources for training and smoothed both magnitude and sky distributions  (with intended over-densities in the regions of the Galactic bulge and Magellanic Clouds), while semi-supervised sources filled the under-represented region in the cross match centred around the Galactic longitude of 300~deg.}
  \label{fig:illustration_XM_to_training}
\end{figure}

Figure~\ref{fig:illustration_XM_to_training} illustrates an example of the distributions of stars in the sky and in the \gmag-band magnitude as they appeared in the cross match and in the training set in the particular case of the fundamental-mode RR\,Lyrae (RRAB) stars. 
The reduction of the number of sources and the smoothing of density peaks in both sky and magnitude distributions, based on the multi-dimensional source sampling described in item~(C), are clearly visible in Fig.~\ref{fig:illustration_XM_to_training}.
Moreover, the addition of semi-supervised sources, following the procedure detailed in item~(E), is shown to fill a region of the sky that previously lacked known representatives.

The final composition of the training-set classes included super- and sub-classes according to the hierarchical classifier organisation presented in Fig.~\ref{fig:multistage}, and their representation was provided together with the assessments of classifier models in Figs.~\ref{fig:CM_12}--\ref{fig:CM_45} (because the number of sources for each class and the competing classes are meaningful in the context of each classifier and the classes targeted for publication).

\subsubsection{Attributes \label{sec:attributes}}

Classification attributes describe the source features that help us distinguish the classes to which the sources belong.
The extraction of such attributes may involve simple or complicated methods, statistics, models, and data types of different nature (e.g.\ photometric, astrometric, spectroscopic, global, or epoch specific).
The translation of all this information into numerical values in a homogeneous way and in a common context makes it possible to compare sources, identify attributes with typical values for specific classes and other attributes (or combinations thereof) with values that differ the most for sources of different classes. 
Class models are defined by algorithms that are based on the classification attributes of training sources and are subsequently used to classify unlabelled objects using the same set of attribute definitions.

The effectiveness of attributes is determined by the relevance of the features to describe, convolved with data properties (such as sampling, accuracy, and rate of spurious measurements), which may also depend on source characteristics such as sky location, brightness, and others, making it difficult to guess a priori the most efficient attribute definitions.
In the attempt to capture all possible features of each class and at the same time include different expressions for the same features, the number of attributes can grow quickly.
Attributes are necessary elements of classification models, but too many of them can have an adverse effect.
Although data sample sizes have increased enormously in recent years,  
the volume of the attribute space grows rapidly as the dimensionality increases, and the number density of sources represented in such volumes (and their statistical significance) decreases just as rapidly \citep[this effect is also known as the curse of dimensionality, see][]{Bellman1961,Hastie2009}.
Too many attributes can also lead to overfitting the training-set features, so that the model loses generality when applied to unlabelled data.
Even though the susceptibility to overfitting depends on the classification technique too (and random forest proved to be one of the most robust methods in this context), a smaller number of attributes can improve the model in terms of learning accuracy, general applicability, and interpretability, in addition to lower computational requirements (i.e.\ time and storage resources).

Common techniques to reduce dimensionality include the combination of multiple attributes and/or the selection of a subset of highly discriminant attributes that optimally split different classes. 
We focussed on the latter and tested different methods to efficiently optimise the training sets of five classifiers (described in Section~\ref{sec:classifier}) that initially contained 150~attributes for a total of 33~thousand training sources.  
Some of these techniques were implemented in the literature \citep{AUCRF,GENUER20102225,2005q.bio.....3025D}, while others were coded according to the principles of forward selection and backward elimination of attributes \citep{Guyon.Elisseeff.Variable.Selection}, that is, by progressively adding the most useful attributes or removing the least useful ones (where the usefulness of each tested attribute was evaluated by the change in the classifier accuracy rate). 
The backward elimination can capture more synergies among attributes than the forward-selection technique (which evaluates only subsets of possible attribute combinations), but the former can be very demanding in terms of processing time, and computationally less intensive solutions are often pursued.
Attributes are then ranked by their selection order, which rewards the truly useful ones and penalises the inefficient, noisy, and redundant ones. 
Every attribute selection method employed random forest as classification technique, and the trained model assessment was based on `out-of-bag' sources (unused training objects from the random draw with replacement to build each tree).

Given the time requirements of the iterative attribute testing on a relatively large volume of training data, the forward selection of the most useful attributes was limited to the identification of the top 12 attributes, while the backward elimination evaluated the classifier accuracy after each removal of the least important attributes and reassessed the importance of the remaining attributes after each iteration. 
The attribute importance is available for all attributes of a given random forest classifier (computationally less intensive than assessing the effect of each attribute by retraining classifiers without each of them) and is evaluated by the mean decrease in accuracy after shuffling the values of each attribute (one per time) among the out-of-bag objects. 

The highest ranked attributes from each method as a function of classifier accuracy were compared, combined, and their individual impact on random forest classifiers was tested manually by forward selections of attributes (from the pre-identified subset of the most useful ones) until the classifier accuracy (i.e.\ completeness) reached a maximum or did not increase significantly. The tuning of random forest parameters (the numbers of trees and of attributes tested at each node) was automatically included in all of the optimisation procedures (because attribute changes imply new classifiers).

The attributes were selected in both initial (supervised) and final (semi-supervised) classification runs with the restrictions mentioned in the first paragraph of item (E) in Section~\ref{sec:source-select}.
For the final classification, 40~attributes were selected from the union of the attributes employed in the five random forest classifiers detailed in Section~\ref{sec:classifier}.
Except for a few astrometric parameters, most of the attributes described features in the \gmag, \gbp, and \grp photometric time series (and from their combinations) in terms of statistical values. For a homogeneous treatment of all sources, considering that about half of the sources (or more, depending on the variability type) had fewer than 20~FoV transits in the \gmag band, time series were not modelled by Fourier series decompositions, thus attributes did not include the characteristics that are typically employed to identify periodic variable objects, such as periods and comparisons of amplitudes and phases of different harmonics. 
All of the attributes related to the photometry were computed on cleaned photometric time series \citep{2018arXiv180409373H}. 
The definitions of each attribute and of the subsets of attributes for each classifier are presented in section~7.3.3.3 of \citet{2018gdr2.reptE...7E}.

As mentioned in Section~\ref{sec:data}, the calculation of astrometric attributes (PARALLAX, PROPER\_MOTION, PROPER\_MO\-TION\_ERROR\_TO\_VALUE\_RATIO) could only make use of a preliminary (not published) astrometric solution. The quasar-specific attributes (LOG\_QSO\_VAR, LOG\_NONQSO\_VAR, NONQSO\_PROB) were computed from a parameterised quasar variance model \citep{2011AJ....141...93B} with parameter values that were determined from a preliminary per-FoV \gmag-band photometry.

\subsection{Classifier \label{sec:classifier}}

The machine-learning algorithm employed for this classification was random forest \citep{Breiman.Random.Forest}, which averages the results of multiple decision trees with randomness in the selection of the data for each tree and in the selection of attributes at each node, typically leading to high accuracy (low bias and variance), robustness to noisy and correlated attributes, reduced susceptibility to overfitting, and other advantages \citep[e.g.\ see Chapter 15 of][]{Hastie2009}.

The ESA \gaia archive$^{\ref{foot:archive}}$ includes two tables that describe the global characteristics of a classifier:
\begin{itemize}
\item \texttt{gaiadr2.vari\_classifier\_definition}: this includes the classifier name and a brief description of the classifier (when multiple classification results are published for the same sources in the future);
\item \texttt{gaiadr2.vari\_classifier\_class\_definition}: this includes labels and brief descriptions of the variability classes of the classified objects.
\end{itemize}

\subsubsection{Multi-stage classification \label{sec:multistage}}
As a consequence of the numerous classes in the training set (Section~\ref{sec:source-select}), a single classifier is not an optimal choice in general because the multiple features necessary to recognise all of the details of all classes and subclasses at the same time unavoidably dilute the training data in attribute space and propagate the reduced statistical significance to the classification results. Moreover, failures in main classifications or subtle subclassifications are equally penalised, so that sources that belong to challenging subtypes can  easily be assigned to completely different parent classes. 
On the other hand, a hierarchical organisation of classifiers dedicated to solve simpler problems in sequence can return more trustworthy results because
(i)~fewer attributes are needed to split a smaller number of (super)classes, further protecting from overfitting, 
(ii)~each (super)class representation is statistically more significant, and
(iii)~each classifier model is easier to understand, which is an important advantage in the interpretation of results.
However, the cumulated accuracy of a sequence of unavoidably imperfect classifiers decreases after each stage, so that multi-stage classifiers should only split class groups with very little confusion among them, and the high accuracy of classifiers is especially important at the nodes preceding the targeted classes in order to limit false positives  and negatives as much as possible, which irreversibly propagate in the wrong branches of the tree.

Dedicated classifiers 
were configured at the nodes of a multi-stage decision tree 
and controlled the levels at which (subsets of) classes could be compared and then separated.
Classes were grouped because of similar physical origins or common observational challenges. For example, low-amplitude variables are likely associated with noisy attributes and it is worth separating their classification from the one of other objects because a classifier trained with sources of similar quality is better suited to recognise their features \citep{2012PASP..124..280L}.
Moreover, it is often easier to identify the common characteristics of a superclass (e.g.\ RR\,Lyrae) than its exact subclass (such as a double-mode RR\,Lyrae subtype, which typically requires more measurements than are available for most sources in \gaia~DR2).
The multi-stage top-down approach enables distinguishing groups of classes that share similar global features and to progressively split them by employing more detailed information, until the targeted (sub)class levels are reached. 
The classification probabilities of sources belonging to specific (groups of) classes are distributed from the top to the bottom levels according to the probabilities of such (groups of) classes to be assigned by the classifiers in each node \citep[classifier probabilities express the fractions of the previous node probabilities to be passed to the next stage for the same groups of classes; see also section~4.5.3 in][]{2017arXiv170203295E}.

In the multi-stage classification employed here, classification results followed the paths of the nodes with the highest probabilities along the \textit{same} branches (i.e., lower-level nodes did not override the classifications of higher-level nodes): results were associated with the highest classifier probabilities at the end nodes of each branch, although they did not necessarily correspond to the highest probabilities of the nodes at the end of all branches. This decision followed from the following considerations:
\begin{enumerate}
\item Classifier probabilities do not express real (calibrated) probabilities (see Section~\ref{sec:class_score}) and the meaning of their values depends strongly on each class because of the specific sources in the training set, the relative number of trained sources per class, the setup of the multi-stage tree of classifiers, and the different number of nodes in different branches (lower optimal probabilities are expected from longer sequences of classifiers), among other factors (including the classification algorithm). Thus, the products of such probabilities (after each node) should not be interpreted as real probabilities either. 
\item Classifiers that are higher in the hierarchy are deemed more trustworthy than those at lower levels because the required information for detailed classification might not be available for some sources (e.g.\ if poorly sampled), because key attributes for identifying specific classes might not be used for every node preceding these classes, and because high accuracy is typically required for higher-level classifiers.
\end{enumerate}
This multi-stage classification guaranteed that insufficient representation or information for subtype identification did not jeopardise the parent-class membership, although the final (sub)type identification could be mistaken. 

\subsubsection{Implementation \label{sec:impl}}

\begin{figure}
 \centering
 \includegraphics[width=0.9\hsize]{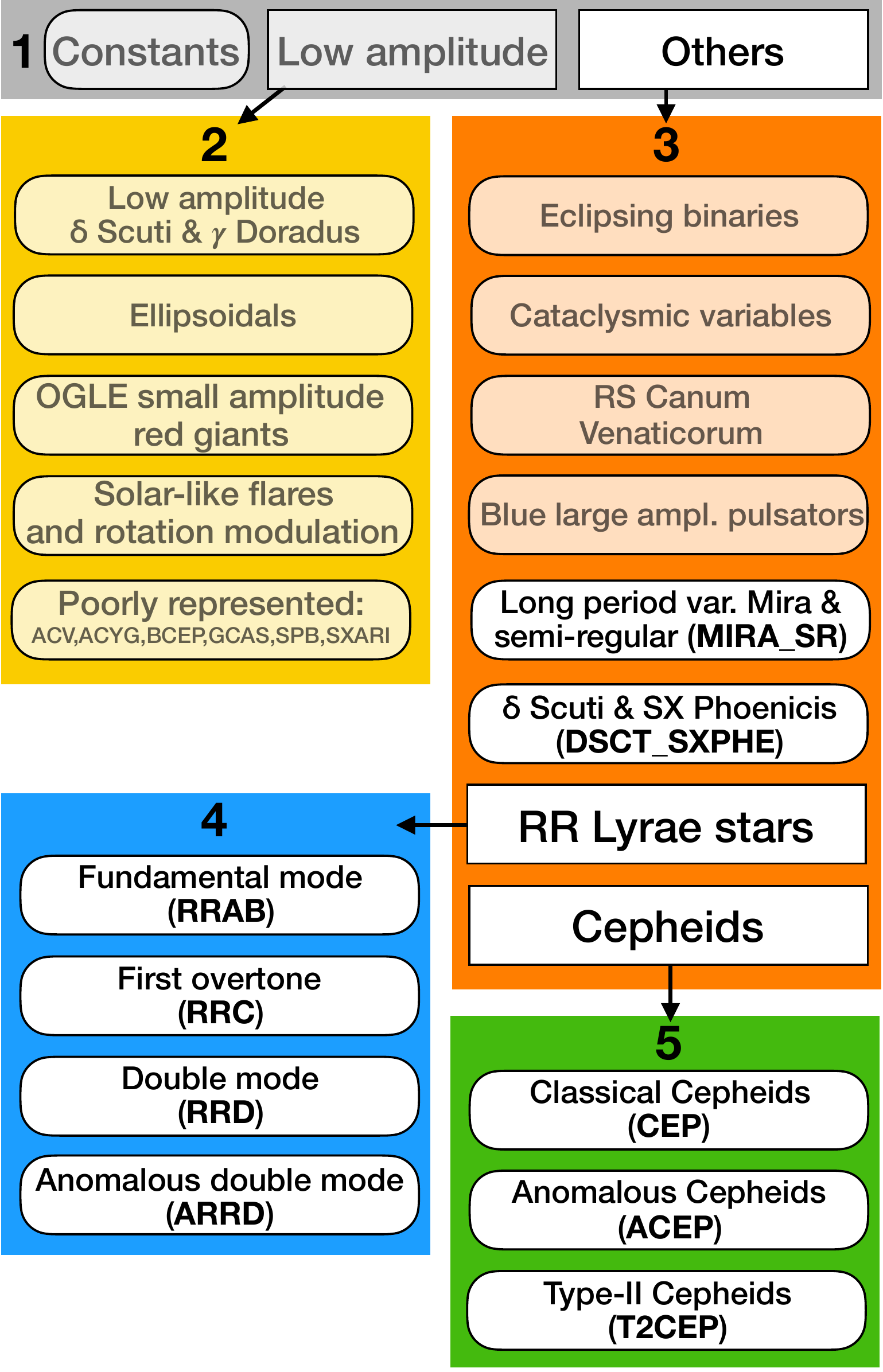}
  \caption{Multi-stage classification tree of five random forest classifiers. Each classifier is identified by a number within the shaded region of the same colour that encompasses the (sub)types of objects to classify.  The square boxes denote superclasses that are further split in subsequent stages. The names of the final classification types (which include subtypes and class combinations) appear in round-corner boxes, and only those with a white background are published in \gaia~DR2.}
  \label{fig:multistage}
\end{figure}

The multi-stage tree with the respective classifiers after the semi-supervised phase described in item~(E) of Section~\ref{sec:source-select} are depicted in Fig.~\ref{fig:multistage}, where numbered shaded regions (from~1 to~5) and their contents denote classifiers and the classes they are meant to classify, respectively.
These five classifiers were assessed through confusion matrices (which verify the classifications of training objects, accounting for the known and classified sources in rows and columns of a matrix, respectively), which were estimated from the out-of-bag sources in random forest.
The random forest method was applied to the classifiers employing the implementation of Weka \citep{Weka-citation}. The configuration parameters and the subsets of classification attributes employed for each classifier were listed in section~7.3.3.4 of \citet{2018gdr2.reptE...7E}, while the most useful features for each classifier are mentioned in the following items \citep[for the definitions of attributes, see section~7.3.3.3 of][]{2018gdr2.reptE...7E}.

\begin{figure}
\centering
  \hspace{-2.5cm}\includegraphics[width=0.53\hsize]{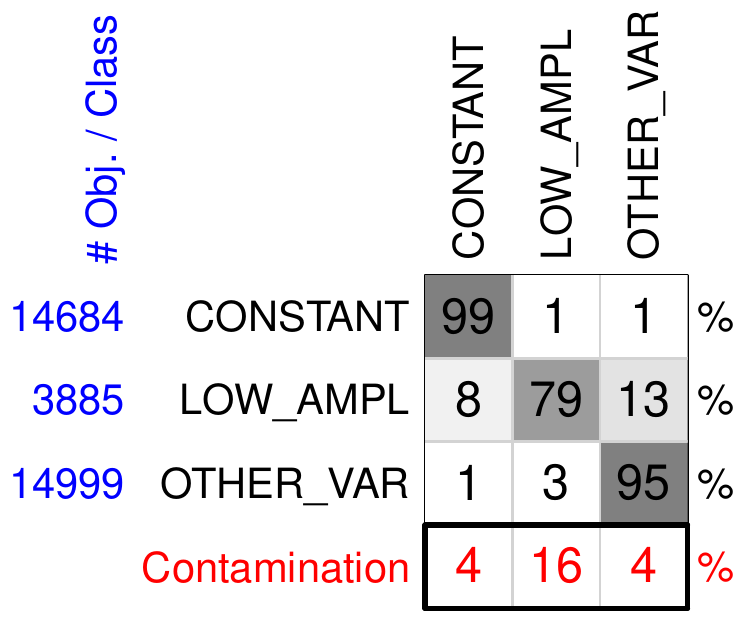}\\ \hspace*{0cm}(a)
  \vspace{-4.6cm}\\
  \includegraphics[width=\hsize]{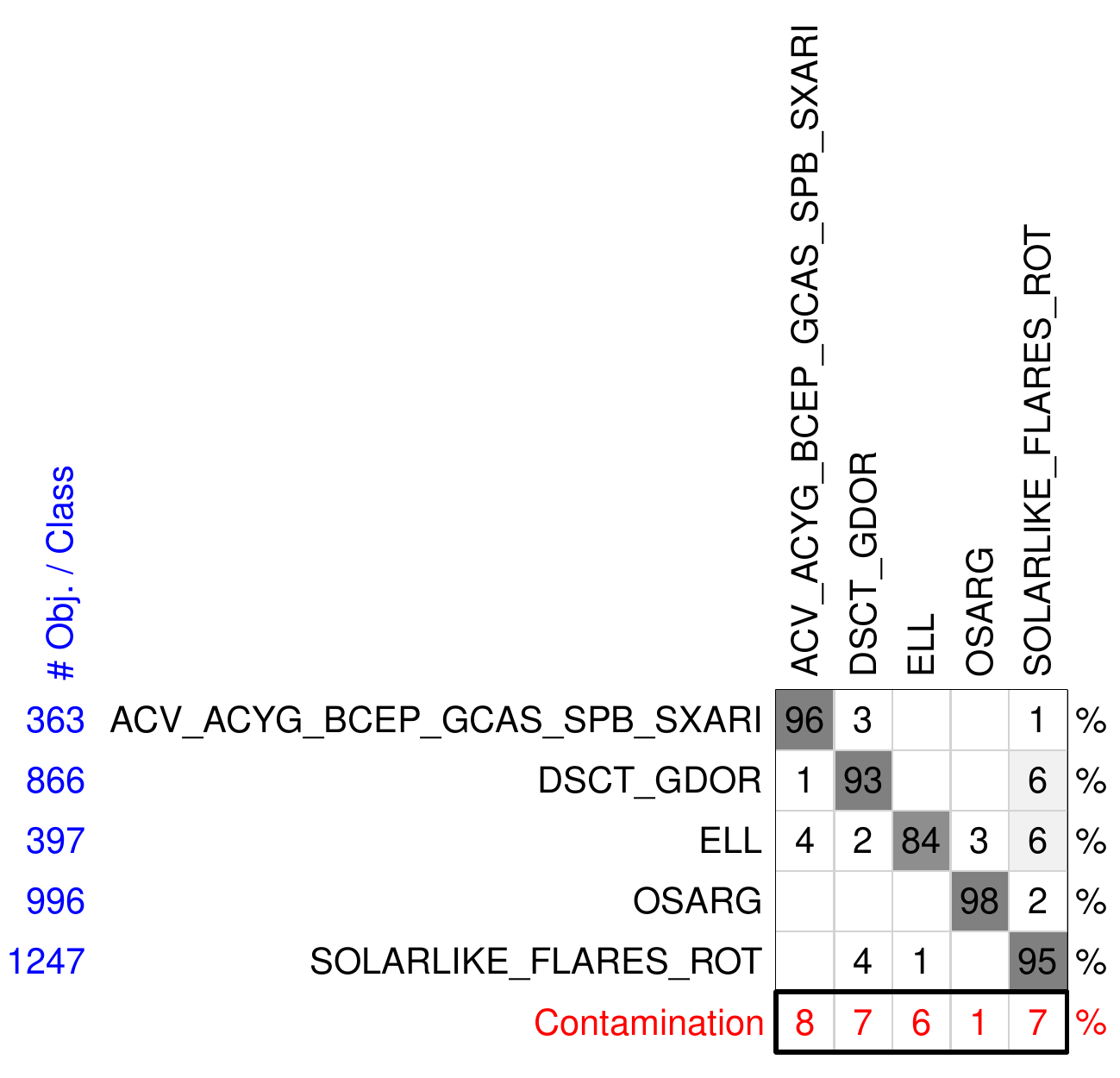}\\ \hspace*{6cm}(b)
  \caption{Confusion matrices of classifiers 1 (a) and 2 (b), as denoted in Fig.~\ref{fig:multistage}. The classifications of training objects (in rows) are compared with classifier results (in columns), which are estimated from the out-of-bag sources in random forest. 
  Given the amount of true positives (TP), false positives (FP), and false negatives (FN), the completeness [TP/(TP+FN)] and contamination [FP/(TP+FP)] rates, expressed as rounded per-cent values, appear in the diagonal (in black) and the bottom row (in red), respectively, while the numbers of training objects per class are listed in blue on the left-hand side of each matrix. Rounded rates imply that not all rows sum to 100\%. Rates below 0.5\%\ are not shown to facilitate the reading of the most relevant parts. Darker shaded squares are used to highlight higher occurrence rates. }
  \label{fig:CM_12}
\end{figure}

\begin{figure}
\centering
  \includegraphics[width=\hsize]{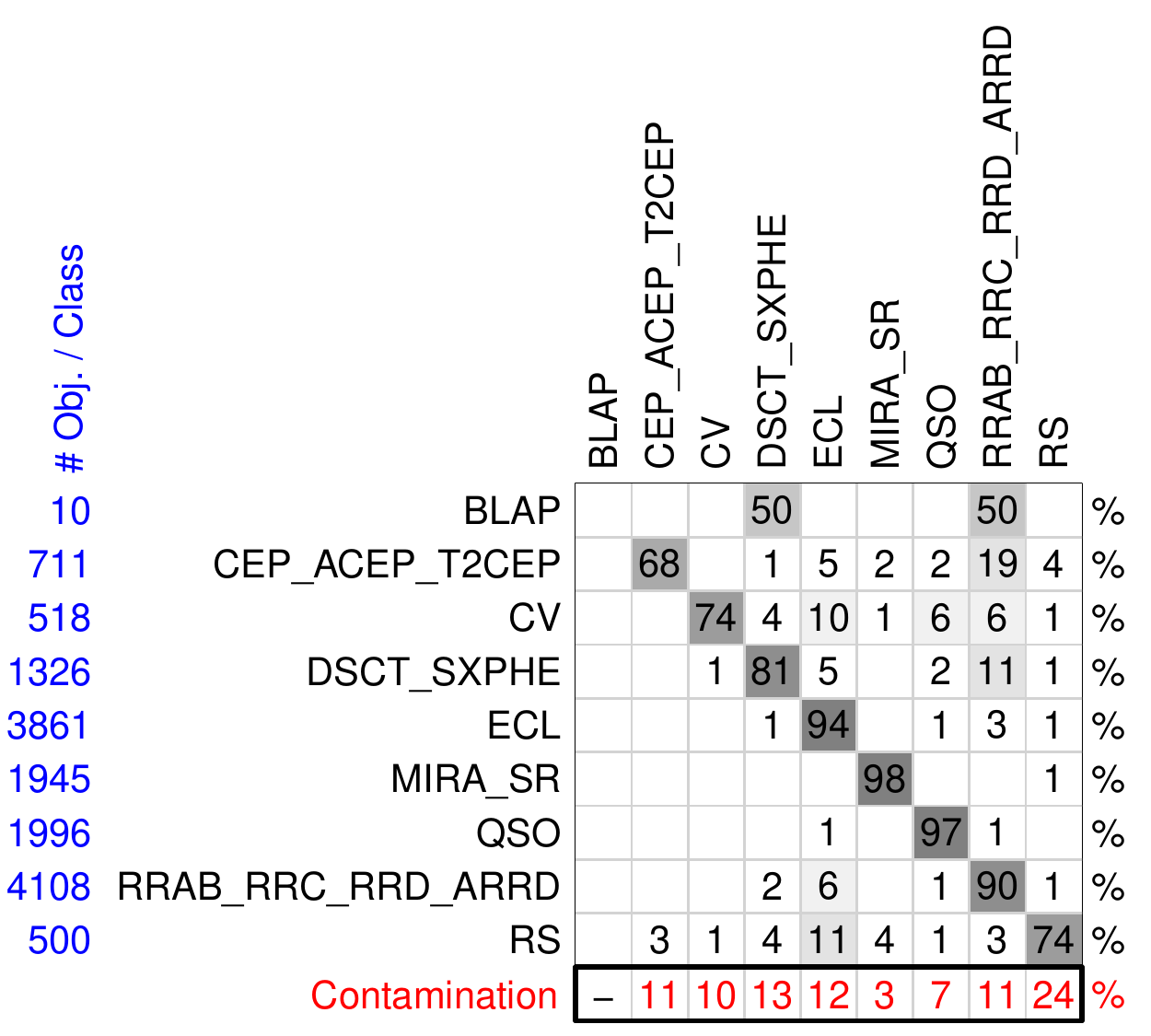}
  \caption{Same as Fig.~\ref{fig:CM_12}, but for classifier 3 (as denoted in Fig.~\ref{fig:multistage}).}
  \label{fig:CM_3}
\end{figure}

\begin{figure}
\centering
  \includegraphics[width=0.51\hsize]{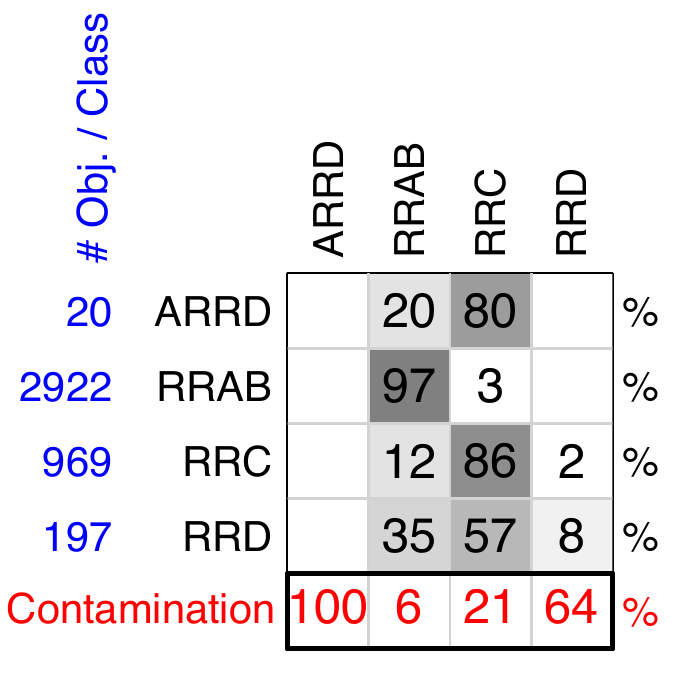}
  \includegraphics[width=0.47\hsize]{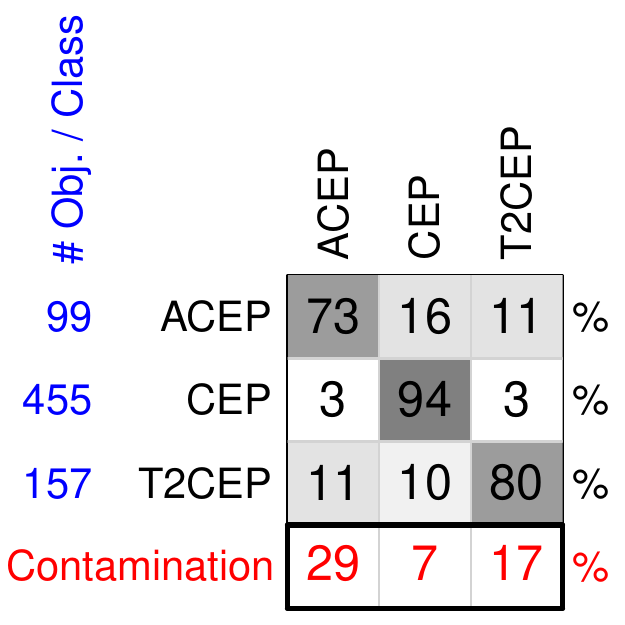}
  \\ \hspace*{1.7cm}(a)\hspace*{4.1cm}(b)
  \caption{Same as Fig.~\ref{fig:CM_12}, but for classifiers 4 (a) and 5 (b), as denoted in Fig.~\ref{fig:multistage}.}
  \label{fig:CM_45}
\end{figure}

\begin{itemize}
\item Classifier~1. The variable objects of interest were separated from constant and typically low-amplitude (percent level) variables. Because the calibration and the uncertainties of epoch photometry \citep{2018arXiv180409368E, 2018arXiv180409367R} were not yet optimal \citep{2018gdr2.reptE...5B} and the presence of spurious measurements rendered variability detection with standard metrics and theoretical expectations impractical \citep[see section~7.2.3.4 in][]{2018gdr2.reptE...7E}, a data-driven approach with a classifier provided a viable alternative to infer source variability levels. As the perceived constancy of objects depended on the precision of measurements,  sources close to the variability detection limit could be classified as either constant or variable. In order to further distinguish this group of sources (and reduce the contamination of other classes), a low-amplitude class was introduced in addition to those for constant and other variable objects. The confusion matrix of this classifier is shown in Fig.~\ref{fig:CM_12}(a). Constant objects were recovered with very high completeness (99\%) and with contamination rates of 8 and 1\%\ from the low-amplitude and other variables, respectively. The lower weight of low-amplitude variables, implicitly assigned by the smaller number of representatives with respect to the other two classes, was designed to reduce contamination in the class of the other variables without seriously competing against the class containing the targeted variables. The loss of 3\%\ of the other variables to low-amplitude objects was expected because a small fraction of low-amplitude objects was not removed from the other variables (where 3\%\ of the objects had a \gmag range of less than 0.03~mag, 97\%\ of which were $\delta$\,Scuti/SX\,Phoenicis stars, in addition to a handful of RS~Canum Venaticorum-type stars, because the sub-classifications of these stars based on amplitude was often not available in the literature). 
The most useful attributes of this classifier were REDUCED\_CHI2\_G, DENOISED\_UNBIASED\_UNWEIGHTED\_VARIANCE, NORMALIZED\_CHI\_SQUARE\_EXCESS, DURATION,\\ RANGE\_G, MAD\_G, STETSON\_G, G\_VS\_TIME\_MEDI-AN\_ABS\_SLOPE, the colours from the three \gaia bands, and the median magnitudes.
\item Classifier~2. Although the classifications of time series that belonged to classes with typically low magnitude ranges in the \gmag band (dominated by percent level variations, over 80\%\ of which were below 0.1~mag) were not published, the confusion matrix of the classifier is still shown in Fig.~\ref{fig:CM_12}(b) to highlight the composition of low-amplitude classes, which included a set of $\delta$\,Scuti stars (merged with $\gamma$\,Doradus stars and $\delta$\,Scuti-$\gamma$\,Doradus hybrids) that exhibited low-amplitude variations. 
Some classes (such as $\gamma$\,Cassiopeiae) are listed among low-amplitude variables because no high-amplitude sample was available in the training set. 
Low-amplitude $\gamma$\,Cassiopeiae stars were merged with pulsating and other multi-periodic types because eruptions and apparent irregularities due to multi-periodicity (in addition to similar colours and amplitudes) can cause their time series to look rather similar, with the sampling of \gaia~DR2. 
The most useful attributes included MEAN\_G, PARALLAX, BP\_MINUS\_RP\_COLOUR, and MEAN\_BP.
\item Classifier~3. The key identifications of the published high-amplitude pulsating variables were returned by this classifier. Focusing on the classes relevant to \gaia~DR2,  the confusion matrix in Fig.~\ref{fig:CM_3} shows that the RR\,Lyrae stars and long-period variables (Miras and semiregulars) were identified with an accuracy of at least 90\%, while the less well represented $\delta$\,Scuti/SX\,Phoenicis and Cepheid classes reached a completeness of 81 and 68\%, respectively. The overall contamination rates of these four groups of classes were limited to 3 to 13\%.  The main sources of contamination of RR\,Lyrae classifications (for which the RR\,Lyrae class was also the most common misclassification) were represented by Cepheids, $\delta$\,Scuti/SX\,Phoenicis stars, and eclipsing binaries, as expected from a classifier that did not use Fourier parameters as attributes. The blue large-amplitude pulsators \citep[BLAP,][]{2017NatAs...1E.166P} were not sufficiently representative in the training set, and the confusion matrix confirmed that they had no chance of detection, hence it was decided to merge the BLAP classifications with those of $\delta$\,Scuti/SX\,Phoenicis stars, where potential BLAP candidates (if any) were most likely to be found. 
Some of the attributes that were particularly useful in identifying the published classes in those trained for this classifier were as follows: G\_VS\_TIME\_MAX\_SLOPE and NORMALIZED\_CHI\_SQUARE\_EXCESS~for~Cepheids; RANGE\_G and DENOISED\_UNBIASED\_UNWEIGHTED\_VARIANCE for $\delta$\,Scuti/SX\,Phoenicis stars; BP\_MINUS\_RP\_COLOUR, MEDIAN\_RANGE\_HALFDAY\_TO\_ALL, and G\_MINUS\\\_RP\_COLOUR for Miras/Semiregulars; SKEWNESS\_G and MEDIAN\_ABS\_SLOPE\_ONEDAY for RR\,Lyrae stars. 
\item Classifier~4. A sub-classification of RR\,Lyrae stars into fundamental mode (RRAB), first-overtone (RRC), and (anomalous) double-mode types (A/RRD) was attempted, and the confusion matrix in Fig.~\ref{fig:CM_45}(a) clearly shows that (anomalous) double-mode RR\,Lyrae types were almost always confused with the much more numerous fundamental mode and first-overtone ones, of which the former was identified with high completeness and limited contamination rates (97 and 6\%, respectively). 
The most useful attributes were DENOISED\_UNBIASED\_UNWEIGHTED\_KURTOSIS\_MOMENT, TRIMMED\_RANGE\_G, and G\_VS\_TIME\_MAX\_SLOPE.
\item Classifier~5. A sub-classification of Cepheids was also attempted, and the confusion matrix shown in Fig.~\ref{fig:CM_45}(b) indicates decreasing levels of completeness,  accompanied by increasing contamination rates, for classical, type-II, and anomalous Cepheids (in this order). The best performance was achieved by the classical Cepheids with 94 and 7\%\ completeness and contamination, respectively.
The classification attributes that proved to be the most useful included MEDIAN\_RP, LOG\_NONQSO\_VAR, and BP\_MINUS\_RP\_COLOUR.
\end{itemize}
Classifier validations from the confusion matrices presented in Figs.~\ref{fig:CM_12}--\ref{fig:CM_45} included all classification probabilities and were limited to training objects, these assessments therefore depended on the choice of sources, classes, and their relative representation, and they were naturally biased (by definition) against untrained sources of the same classes that looked different for some reason (e.g.\ in case the effects of interstellar extinction were not fully accounted for in the training set). These validations gave valuable indications for building classifier models, but classification results should be assessed independently of these preliminary estimates.

This multi-stage classifier was applied to sources with at least two FoV transits in the \gmag band and with the trimmed range (from the 5th to the 95th percentile) of epoch photometry in the \gmag band greater than 0.1~mag.
The results were assessed as a function of classification probability, and excessive numbers of weak candidates were excluded by setting minimum probability thresholds, if needed. In particular, class labels related to candidates that were not affected by probability thresholds were ACEP, ARRD, and RRD, while minimum classification probabilities of 0.3, 0.4, 0.4, 0.58, 0.6, and 0.6 were applied to candidates of classes labelled as MIRA\_SR, CEP, T2CEP, DSCT\_SXPHE (after summing the probabilities of BLAP candidates), RRAB, and RRC, respectively.
This subset of classification results was then filtered by validation classifiers (unpublished), as described in Section~\ref{sec:validation}.

\subsection{Validation \label{sec:validation}}

The first selection of classification results, described in Section~\ref{sec:impl}, still included significant contamination (just by looking at the number of candidates for each class). 
In order to alleviate the presence of contaminants in an automated way, new dedicated (validation) classifiers were built to help 
separate false positives from true positives.
This binary classification was pursued with random forest classifiers, one for each of the three groups of classes labelled as CEP\_ACEP\_T2CEP, DSCT\_SXPHE, and RRAB\_RRC\_RRD\_ARRD.
The training sets employed a similar number of true positives and false positives, which were sampled from the classified \gaia sources in common with the 750~thousand cross-matched objects of known types in the literature (defined in Section~\ref{sec:xm}). Except for the DSCT\_SXPHE candidates (which were classified after training with all of the $\delta$\,Scuti/SX\,Phoenicis stars cross matched with the literature), the validation classifiers were trained with cross-matched objects that were not used by the preceding classification stage (Section~\ref{sec:classifier}).
Similar to the other classifiers, the validation training sources were sampled for a representative distribution in the sky, and their attributes were selected as described in Section~\ref{sec:attributes}, with the consequent optimisation of random forest parameters.
Validation classifiers were applied to the preliminary selection of classification results (Section~\ref{sec:impl}) and further classified the candidates of the three superclasses mentioned above as true positive or false positive (with a validation classification probability threshold of 0.5). 
Only the true-positive identifications were published in the variability classification table of \gaia~DR2 and are described in Section~\ref{sec:results}.

The completeness rates of true positives were 94, 98, and 91\%, with corresponding contamination rates of 3, 3, and 9\%, for the CEP\_ACEP\_T2CEP, DSCT\_SXPHE, and the RRAB\_RRC\_RRD\_ARRD superclasses, respectively.  Thus, contamination was reduced, at the cost of 2 to 9\%\ of further reduction in completeness, with the greatest loss of true positives for the RR\,Lyrae candidates.

In the special case of MIRA\_SR candidates, the training sources of long-period variables were not considered sufficiently representative to further validate the preliminary  results. The published classifications therefore reflect the selection of the most likely candidates performed by the dedicated SOS module \citep{2018arXiv180502035M}.

Known misclassified objects from the literature were not removed from the classification results in order to preserve the consistency of the results and prevent the appearance of cross-match footprints, or statistical studies would face additional challenges to distinguish real from artificial features. Moreover, not all of the classifications available in the literature are necessarily correct, and in some cases, the \gaia data provide the additional information that can lead to an improved judgement. 

\subsection{Classification score \label{sec:class_score}}

The classification scores and the corresponding class labels assigned to the classified variable sources are stored in the \texttt{vari\_classifier\_result} table of the ESA \gaia archive$^{\ref{foot:archive}}$ under the field names \texttt{best\_class\_score} and \texttt{best\_class\_name}, respectively.
The classifier score is a numerical quantity between 0 and 1 that expresses a linear transformation of the classification probabilities as follows:
\[
\texttt{best\_class\_score}  = (P_{\rm class} - P_{\min,\, \rm class}) / (1 - P_{\min,\, \rm class}),
\]
where $P_{\rm class}$ denotes the classification probability (in the range from 0 to 1) of the best class from the multi-stage classifier (Section~\ref{sec:classifier}) and $P_{\min, \rm class}$ refers to the minimum probability thresholds as a function of class, listed in the last paragraph of Section~\ref{sec:impl}.

The motivation for this transformation followed from the attempt of assigning similar scores to classifications with similar reliability, considering that the probabilities of the weakest candidates depended on class. This simplistic approach should not be interpreted as returning true (calibrated) probabilities \citep[e.g.][]{2012ApJS..203...32R}, which were not pursued for this data release.
The classifier `probabilities' do not represent the real probabilities of classifications to be true positives because of their dependence on the classifier method and the training set (such as the number of classes, the number of sources for each class, the extent of representation of the selected sources, and the setup of the multi-stage tree, as  mentioned in Section~\ref{sec:multistage}).

\section{Results \label{sec:results}}

The classifications of pulsating variable stars of Cepheid, Mira/semiregular, $\delta$\,Scuti/SX\,Phoenicis, and RR\,Lyrae types, with light-variation ranges greater than 0.1~mag in the \gmag band (increased to 0.2~mag for long-period variables), are published in the table \texttt{vari\_classifier\_result} of the ESA \gaia archive$^{\ref{foot:archive}}$, which includes the source identifier, the classifier name (in case of multiple independent classifiers in the future), the class label, and the associated classification score.

\begin{figure}
\centering
  \includegraphics[width=\hsize]{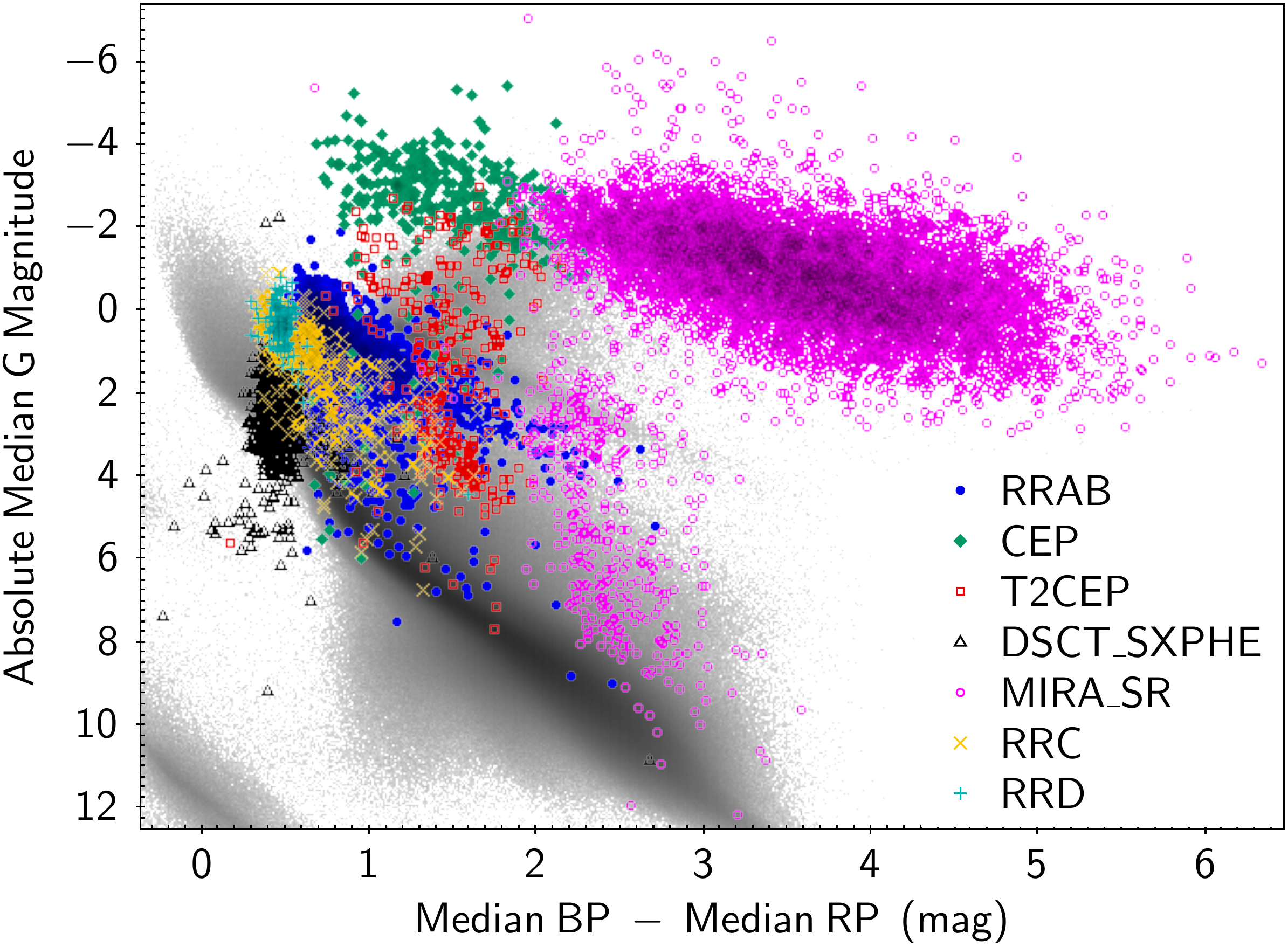}
  \caption{Colour-absolute magnitude diagram (as absolute median \gmag-band magnitude vs.\ median~\gbp$-$~median~\grp) of a selection of classification candidates: fundamental mode (RRAB, blue dots), first-overtone (RRC, orange `$\times$'~marks), and double-mode (RRD, cyan `$+$'~marks) RR\,Lyrae types, classical (CEP, green rhombi) and type-II (T2CEP, red squares) Cepheids, $\delta$\,Scuti/SX\,Phoenicis types (DSCT\_SXPHE, black triangles), and Mira/semiregular types (MIRA\_SR, magenta circles). 
 Faint outliers typically denote contaminating objects, which in the case of long-period variables, are represented mostly by young stellar objects.
 All of these sources satisfy the conditions listed in appendix~B of \citet{2018arXiv180409382G}, among which a relative parallax precision better than 20\%, with no correction for interstellar or circumstellar extinction or reddening. The background points in grey act as a reference of objects within a radius of 1~kpc from the Sun.}
  \label{fig:HR}
\end{figure}

An overview of the classification results that satisfy the astrometric and photometric requirements for an observational Hertzsprung-Russell diagram \citep[see appendix~B of][]{2018arXiv180409382G} is presented in Fig.~\ref{fig:HR}.
The effect of extinction is visible for several classes (in particular, long-period variables, classical Cepheids, and fundamental mode and first-overtone RR\,Lyrae stars).
In addition, faint outliers, with respect to the loci of other candidates of the same class, identify contaminants for most of the classes (assuming accurate parallax values), which are discussed in more detail for each variability type in this section.

Samples of light curves for each class are shown in the summary article of variables in \gaia~DR2 \citep{2018arXiv180409373H} and in the related SOS articles \citep{2018arXiv180502079C,2018arXiv180502035M}; they are therefore not reproduced here. 
The goal of this section is to present the global properties of the classified candidates for each class group (colour-magnitude diagrams, distributions in the sky versus magnitude, and classification score), outline specific features of the candidates unconfirmed or reclassified by SOS, and finally compare the classifications of a subset of sources with those that are known in the literature.

The comparison of \gaia classifications with the literature can be useful in the assessment of some aspects related to completeness and contamination.
However, results in the literature are not exempt from misclassifications, and they depend relevantly on the observational properties (e.g.\ photometric bands, sky location, magnitude limits, signal-to-noise ratio, astrometric resolution, availability of spectroscopic and astrometric information, and number of epochs and their sampling), on the classification processing (e.g.\ reliability level, targeted completeness, and purity levels), and on the variability types (some of which may elude detection or proper classification depending on observation times or stellar evolution) and their occurrence relative to each other (i.e.\ in the literature versus in reality).
Comparisons with cross-matched sources of known types from the literature are therefore influenced by the different survey features and criteria.
On the other hand, no survey provides a perfect reference in all circumstances, and the diversity of a multitude of literature catalogues can help overcome some of the limitations of single surveys.

We here compare the \gaia classifications with a subset of 494~thousand sources that have a \gmag-band range greater than 0.1~mag of
the 750~thousand cross-matched objects defined in Section~\ref{sec:xm}, in addition to new \gaia source cross matches with the active galactic nuclei identified in the mid-infrared using the final catalogue release of the Wide-field Infrared Survey Explorer \citep[AllWISEAGN;][]{2015ApJS..221...12S} and the quasars listed in the third release of the Large Quasar Astrometric Catalog \citep[LQAC;][]{2015A&A...583A..75S}.
Despite the variety of classes considered, 94\%\ of the known objects were represented by constant, eclipsing binary, and RR\,Lyrae stars. 
From this set of cross-matched sources, we expect biases in the apparent rates of completeness and contamination (especially for the latter as it relies on the inclusion of all contaminating classes, on the adoption of realistic relative class proportions, and on a negligible contribution from unexpected sources), in addition to the optimistic selection bias from using classifications from the literatures (part of which might be easier to identify). 
Nevertheless, we present completeness and contamination rates for each class (or subclass, when applicable) and as a function of minimum classification score in order to show trends (such as the most or least pure subclass) and give insights into likely sources of contamination.

As mentioned in Section~\ref{sec:impl}, four (super)classes constitute the primary results of relevance. Subclassifications are provided, but their use should be limited to the identification of stars that belong to the most common subtypes (for reduced contamination levels) or simply to the membership of their parent classes.

\begin{figure}[ht!]
\centering
  \includegraphics[width=\hsize]{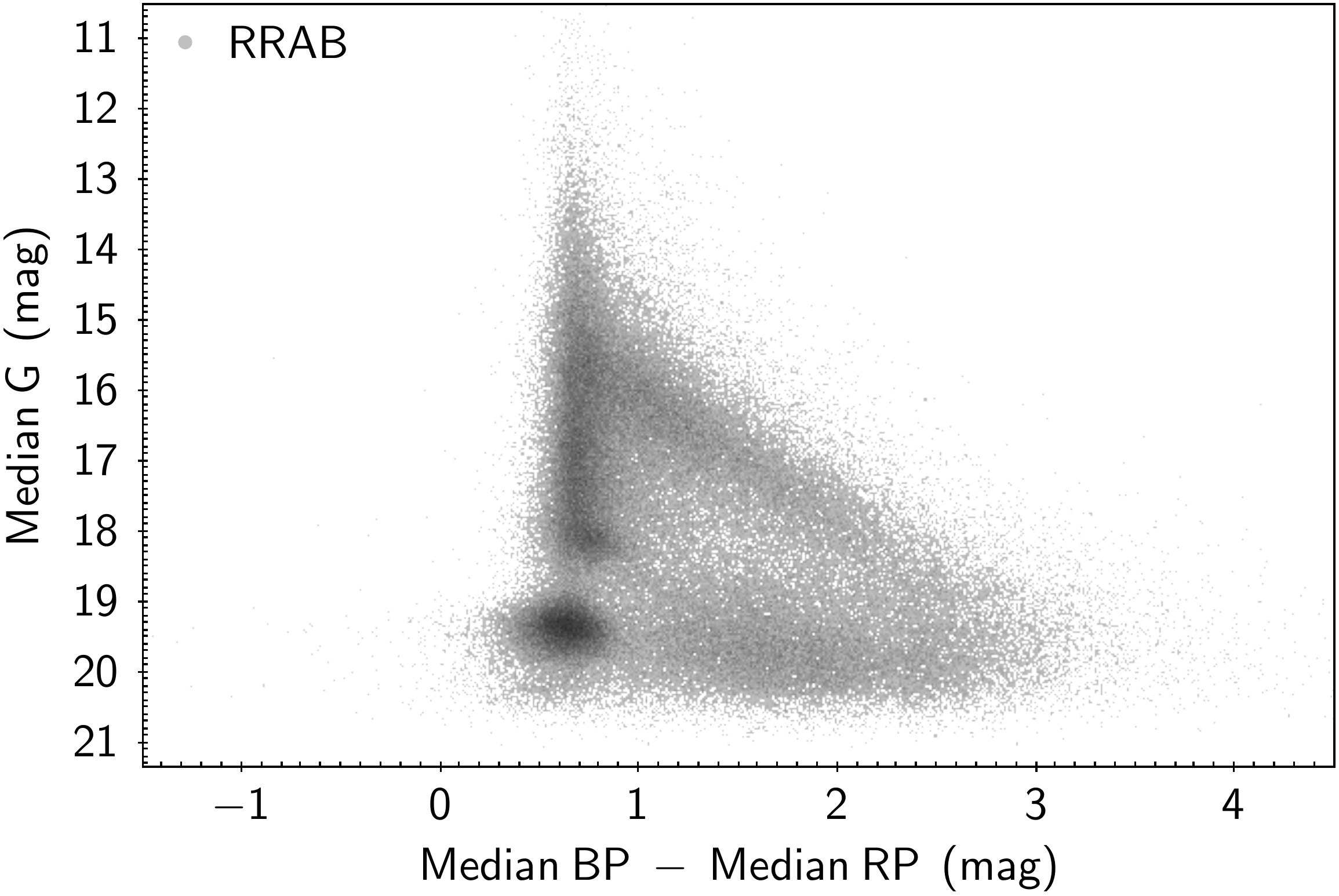}\\(a)\\ \vspace*{0.2cm}
  \includegraphics[width=\hsize]{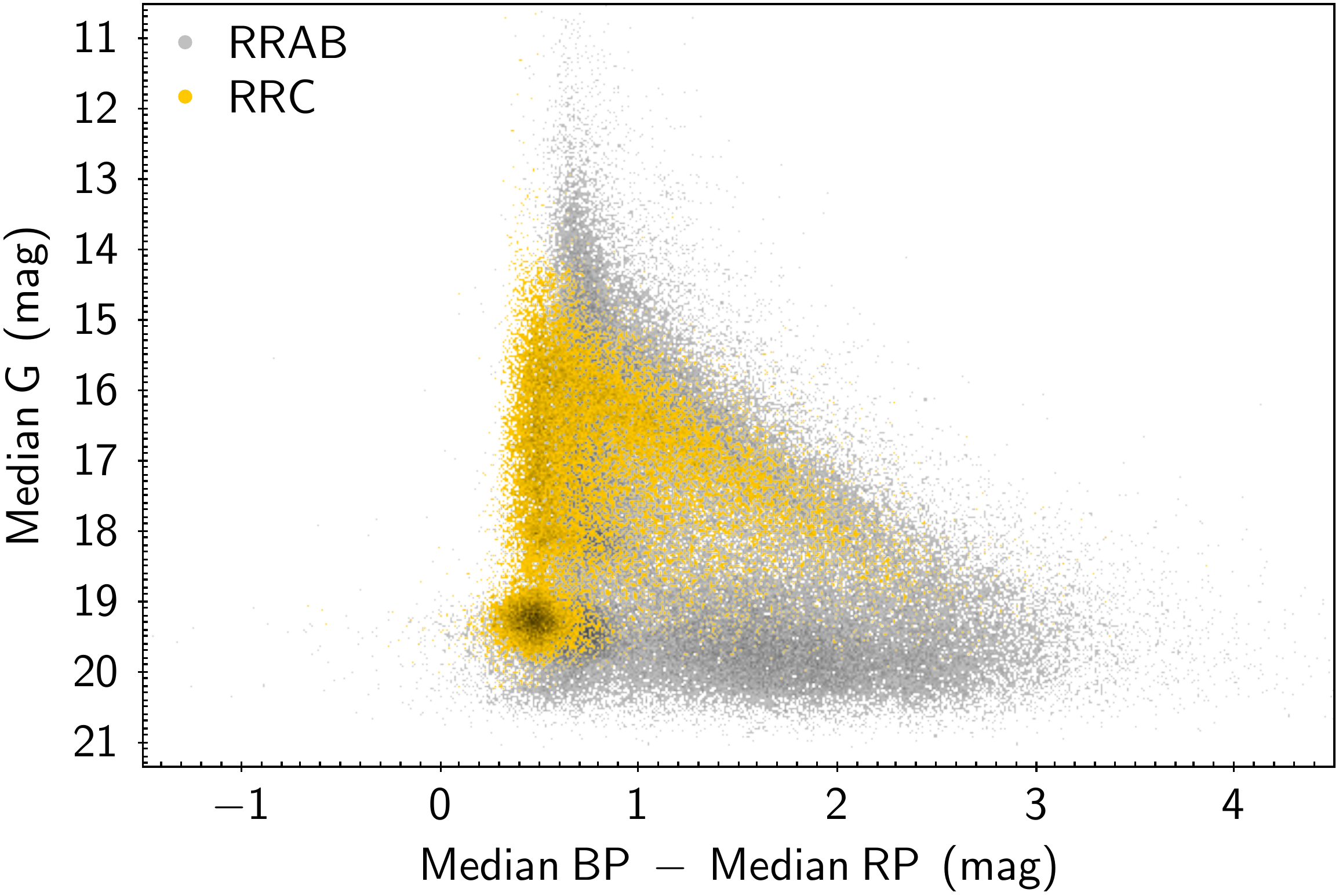}\\(b)\\ \vspace*{0.2cm}
  \includegraphics[width=\hsize]{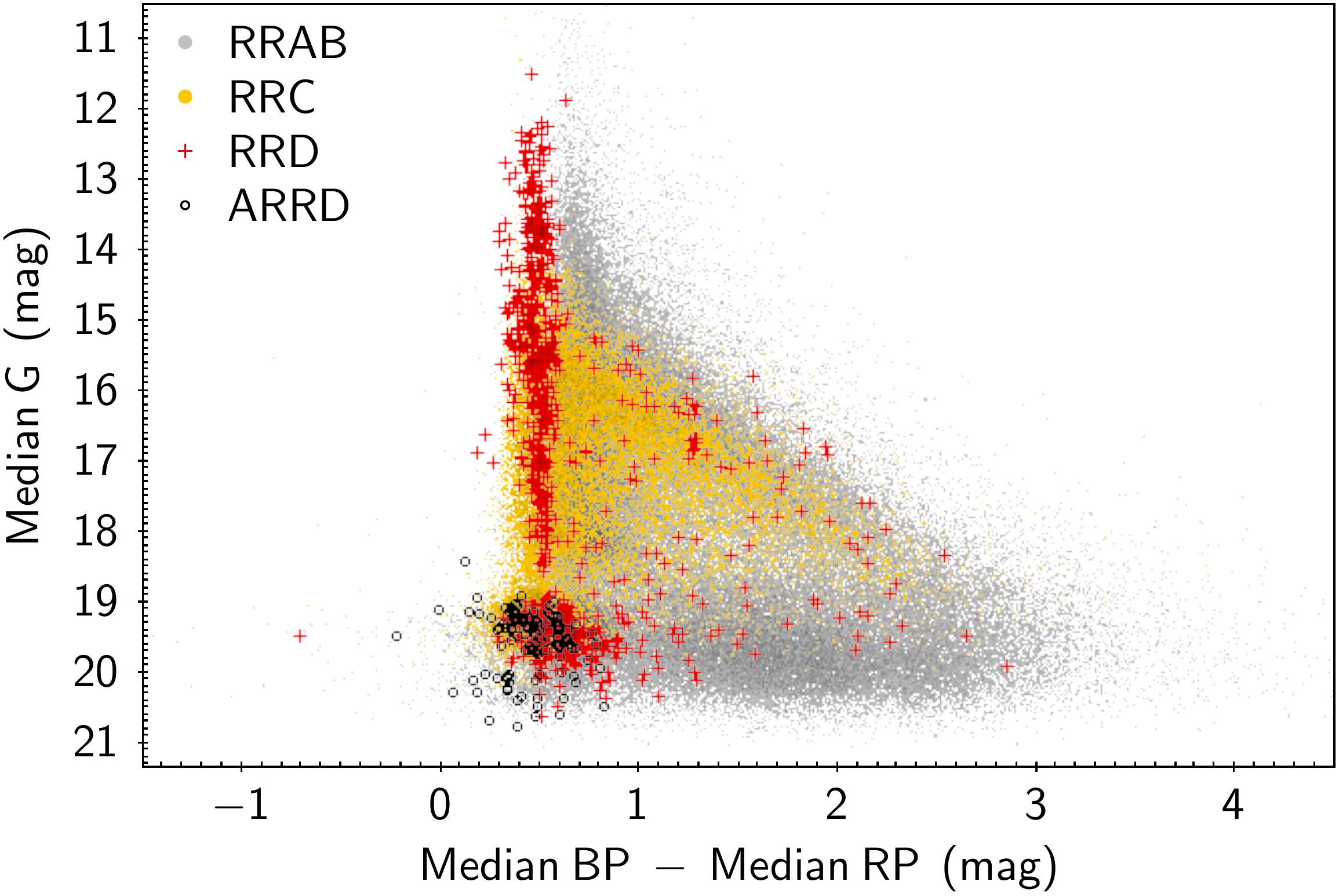}\\(c)
  \caption{Colour-magnitude diagram (as median \gmag-band magnitude vs.\ median~\gbp$-$~median~\grp) of the RR\,Lyrae classifications. The three panels show the fundamental mode (labelled RRAB, grey dots), first-overtone (labelled RRC, orange dots), and (anomalous) double-mode (labelled ARRD/RRD, black circles/red crosses) RR\,Lyrae subtypes, cumulatively, in panels (a), (b), and (c), respectively. See text for the explanation of the main features.}
  \label{fig:RR_color_mag}
\end{figure}

\subsection{RR Lyrae stars \label{sec:res-rr}}

\begin{figure}
\centering
  \includegraphics[width=\hsize]{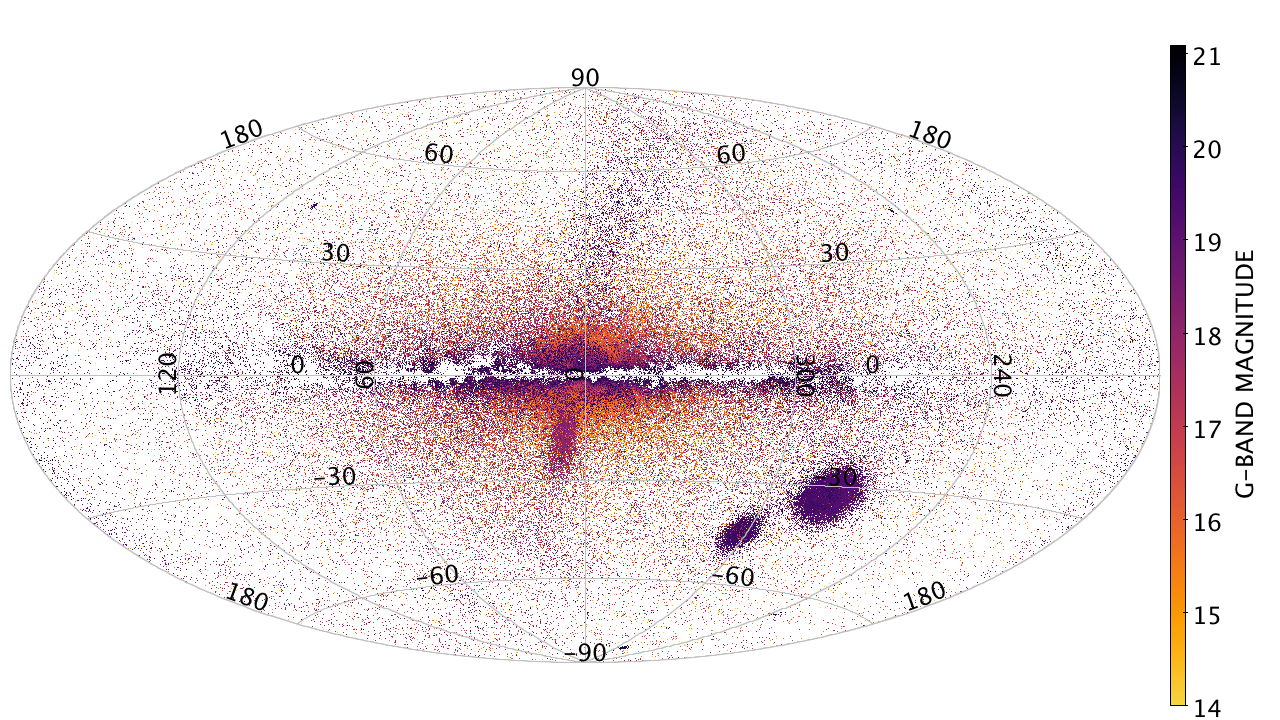}
  \caption{Distribution of the RR\,Lyrae classifications (all subtypes) in the sky (Galactic coordinates in degrees) colour-coded by the median \gmag-band magnitude as indicated in the legend on the right-hand side (values beyond the legend range share the colour of the closest value in the legend).}
  \label{fig:RR_sky_mag}
\end{figure}

The RR\,Lyrae classifications include 195\,780 candidates, further subclassified into 162\,469~fundamental mode (RRAB), 32\,370~first-overtone (RRC), 834~double-mode (RRD), and 107~anomalous double-mode (ARRD) subtypes. Their distribution in a colour-magnitude diagram is shown in Fig.~\ref{fig:RR_color_mag} in terms of median \gmag magnitude versus \bpminrp colour.
The unreddened colour of RRAB candidates is centred between \bpminrp of 0.6 and 0.7~mag, as inferred from the sources in the Galactic halo that form the vertical structure observed in Fig.~\ref{fig:RR_color_mag}(a). 
The slightly reddened clump around \gmag$\approx 18$--18.4~mag and \bpminrp$\approx 0.8$~mag is associated with about 3~thousand candidates in the Sagittarius dwarf spheroidal galaxy, which is visible just below the Galactic bulge in Fig.~\ref{fig:RR_sky_mag}.
The faintest unreddened clump at about \gmag$\approx 18.8$--20~mag is primarily due to almost 20~thousand RRAB candidates in the Magellanic Clouds and to classifications that belong to the Sagittarius stream above the Galactic bulge.
Some of the fainter objects are still related to the Magellanic Clouds and others to dwarf spheroidal galaxies \citep{2018arXiv180502079C}.
The diagonal branch in Fig.~\ref{fig:RR_color_mag}(a) is due to candidates that are reddened and extinguished mostly by the Galactic dust in the disc and bulge regions, with redder and fainter candidates located closer to the Galactic equator.
Objects between the vertical and diagonal features, including the horizontal overdensity at the faint end with \bpminrp$>$1~mag, are dominated by 20--25\% of likely misclassified objects in the Galactic disc or bulge.

The RRC candidates show similar structures to those of RRAB candidates in Fig.~\ref{fig:RR_color_mag}(b), with an offset of almost 0.2~mag bluer on average than RRAB candidates, a reduced contamination at the faint end, and about 5 and 0.5~thousand candidates in the regions of the Magellanic Clouds and Sagittarius dwarf spheroidal galaxy, respectively. 
About 20\%\ of the RRD candidates shown in Fig.~\ref{fig:RR_color_mag}(c) are in the Magellanic Clouds, and the others are distributed mostly in the halo.
More than half (about 60\%) of the ARRD candidates are in the Magellanic Clouds because the training set included only OGLE-IV samples from this region; other ARRD candidates are scattered across the sky.

The processing of the RR\,Lyrae star prototype was particularly unfortunate in \gaia~DR2: the published values of its mean photometry and parallax were inaccurate \citep{2018arXiv180409375A,2018arXiv180409365G} and it was missed also from the all-sky classification of variables because the sampling of the signal made the time-series statistics (without time-series modelling) similar to eclipsing binary variations. Only 10 of the 26 valid measurements in the \gmag band were in the faint half of the magnitude range, leading to a positive skewness of the magnitude distribution (typical of eclipsing binaries), and the random forest probability of this object to belong to the eclipsing binary class became slightly higher than the one of an RRAB type (0.5 versus 0.4, respectively).
The SOS module for RR\,Lyrae stars received this object correctly classified by the independent classification run limited to sources with at least 20~FoV transits in the \gmag band (which could take advantage of the periodicity information), but eventually it discarded this candidate because a key Fourier parameter was not sufficiently accurate \citep{2018arXiv180502079C}.

\begin{figure}
\centering
  \includegraphics[width=\hsize]{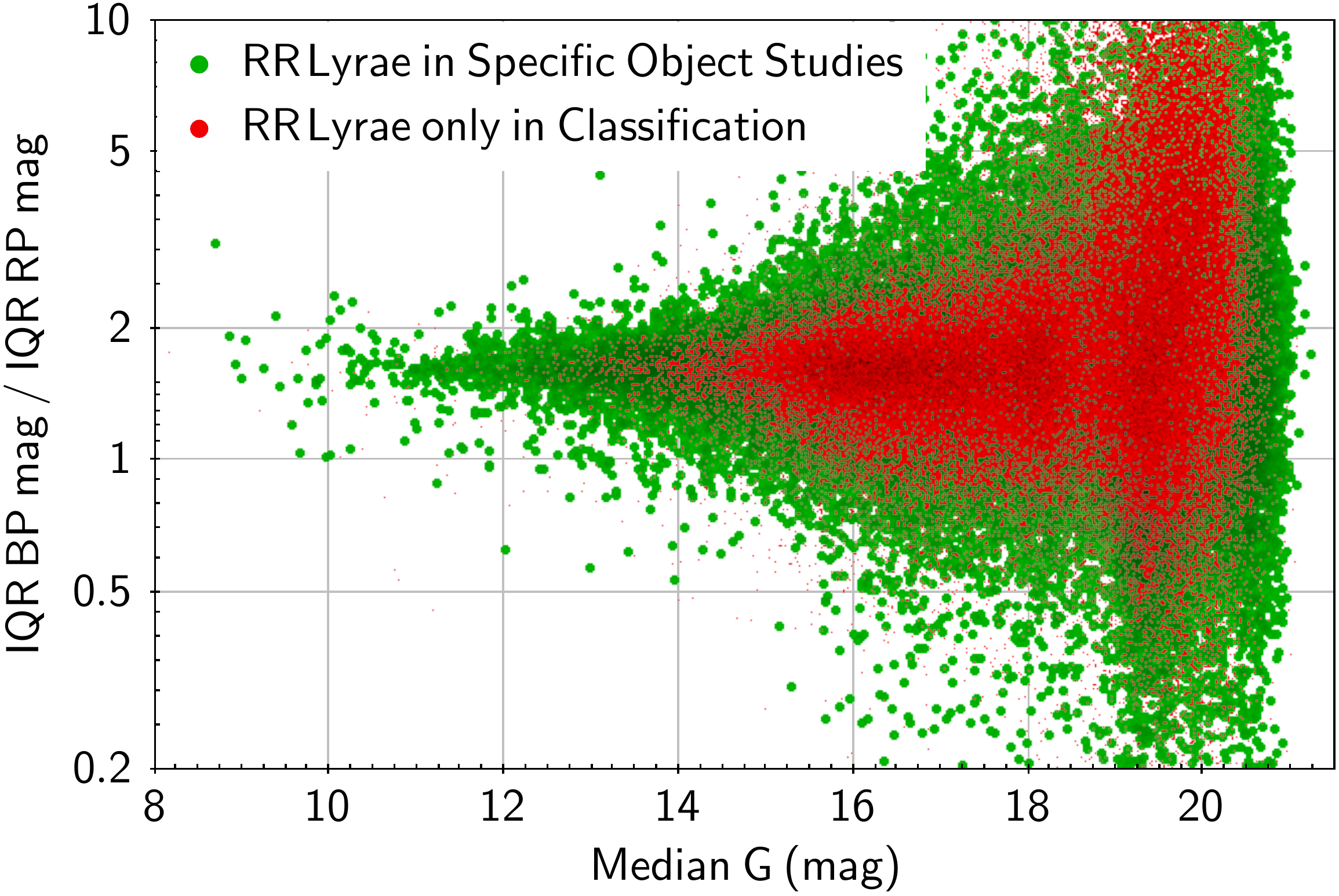}\\(a)\\ \vspace*{0.2cm}
  \includegraphics[width=\hsize]{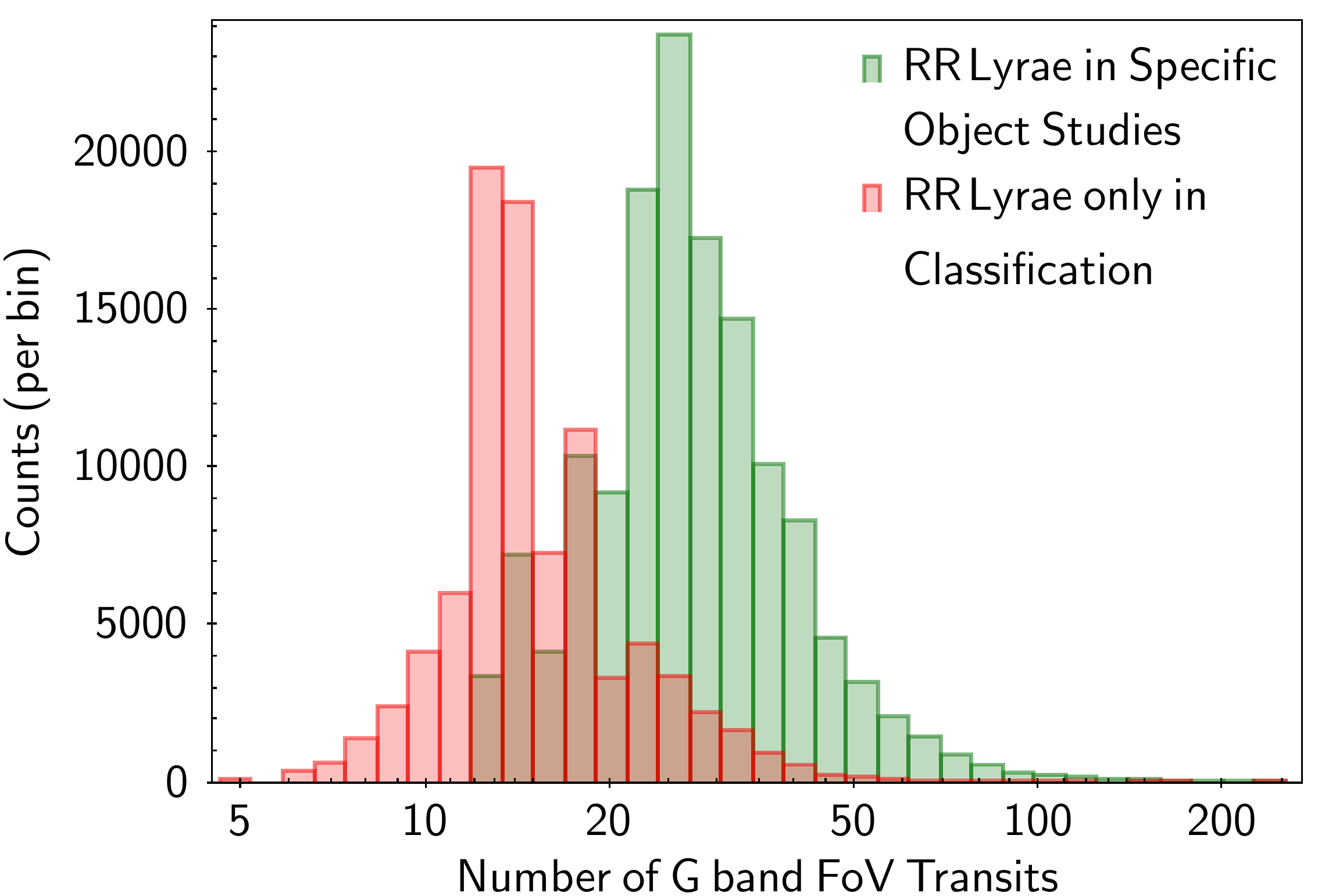}\\(b)
  \caption{Top panel (a): Distribution of the ratios of the interquartile range (IQR) in the \gbp versus \grp bands vs.\ the median \gmag-band magnitude for the RR\,Lyrae candidates in the SOS \citep[green dots;][]{2018arXiv180502079C} vs.\ those present only in the classification results (red dots). Bottom panel (b): Distribution of the number of \gmag-band FoV transits for each source for the RR\,Lyrae candidates in the SOS (green bars) vs.\ those present only in the classification results (red bars).}
  \label{fig:SOS_CLASS_rr}
\end{figure}

\begin{figure}
\centering
  \includegraphics[width=\hsize]{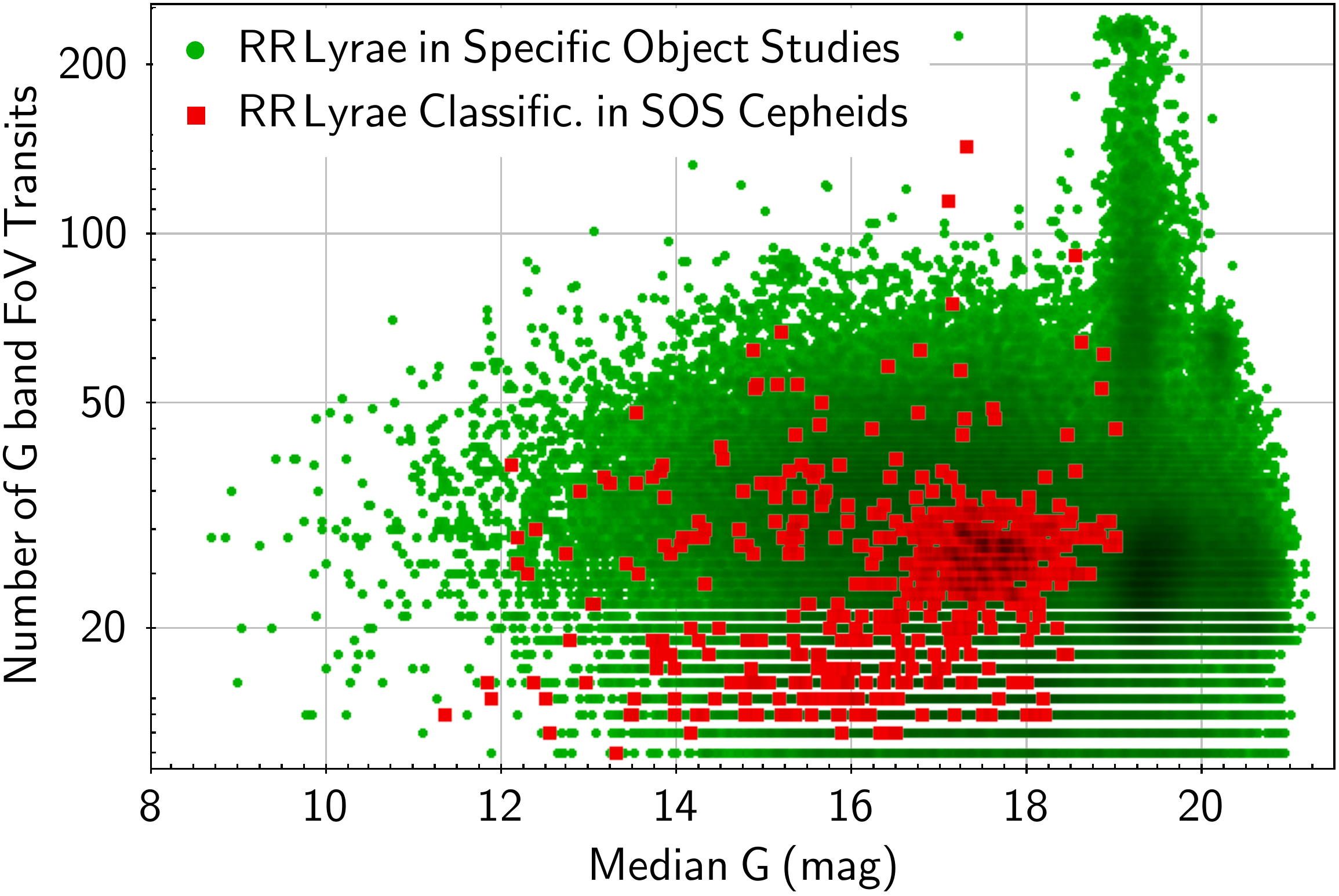}
  \caption{RR\,Lyrae classifications (red squares) that were reclassified as Cepheids in the SOS \citep[green dots;][]{2018arXiv180502079C} as a function of median \gmag-band magnitude and number of \gmag-band FoV transits per source.}
  \label{fig:SOS_CLASS_rr_cep}
\end{figure}

The RR\,Lyrae candidates with at least 12~FoV transits in the \gmag  band were considered for validation by the dedicated SOS module with Fourier modelling, as described in \citet{2018arXiv180502079C}, although this lower limit was not always sufficient to guarantee the confirmation of the related classifications.
In total, 88\,120~RR\,Lyrae classifications were not confirmed in SOS. 
Because stellar pulsations are expected to exhibit larger variations in the \gbp than in the \grp band, the distribution of the ratios of the interquartile range (IQR) in the \gbp versus \grp band is shown as a function of median \gmag-band magnitude in Fig.~\ref{fig:SOS_CLASS_rr}(a) for the RR\,Lyrae candidates in SOS versus those present only in the classification results. Although the IQR is increasingly influenced by the photometric noise towards fainter magnitudes, 78\,194 of the RR\,Lyrae classifications (89\%\ of the candidates not confirmed in SOS) have IQR(\gbp)/IQR(\grp)$>$1.
The distributions of the number of \gmag-band FoV transits for each source for the SOS-confirmed versus unconfirmed RR\,Lyrae candidates are presented in Fig.~\ref{fig:SOS_CLASS_rr}(b) and are found to exhibit two distinct peaks, highlighting the importance of the number of observations in the SOS modelling of these objects and thus their confirmation process.
A total of 618~RR\,Lyrae classifications are reclassified as Cepheids in SOS, as shown in Fig.~\ref{fig:SOS_CLASS_rr_cep} as a function of median \gmag-band magnitude and number of \gmag-band FoV transits for each source: 593 and 436 (96 and 71\%) are labelled as RRAB and have classification scores lower than 0.5, respectively.
Of the RR\,Lyrae stars that are misclassified as Cepheids, those that form a clump with median \gmag of approximately 16.5--18.5~mag and with a similar number of observations amount to slightly more than half of the sample and are located in the region of the Magellanic Clouds. The other misclassified candidates are scattered across most of the sky, with a higher occurrence in the region of the Galactic bulge.

\begin{figure}
\centering
  \includegraphics[width=\hsize]{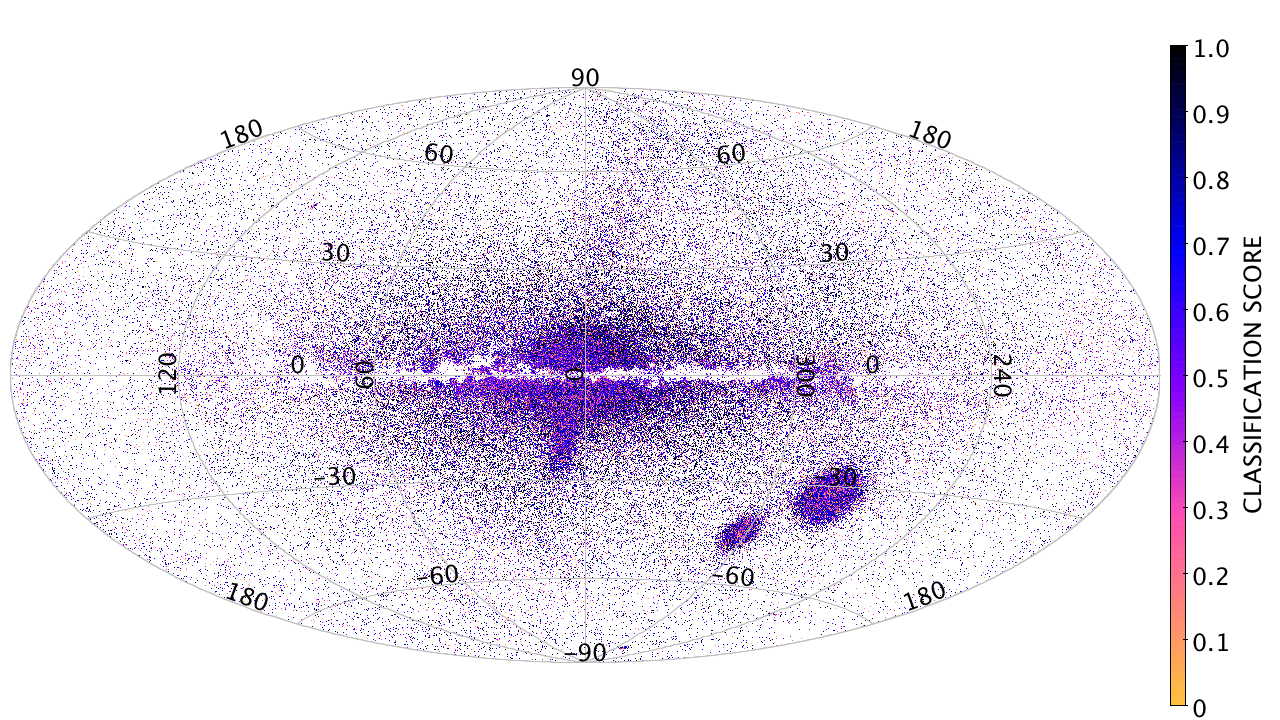}
  \caption{Distribution of the RR\,Lyrae classifications (all subtypes) in the sky (Galactic coordinates in degrees) colour-coded by the classification score (see Section~\ref{sec:class_score}) as indicated in the legend on the right-hand side.}
  \label{fig:RR_sky_score}
\end{figure}

\begin{figure}
\centering
  \includegraphics[width=\hsize]{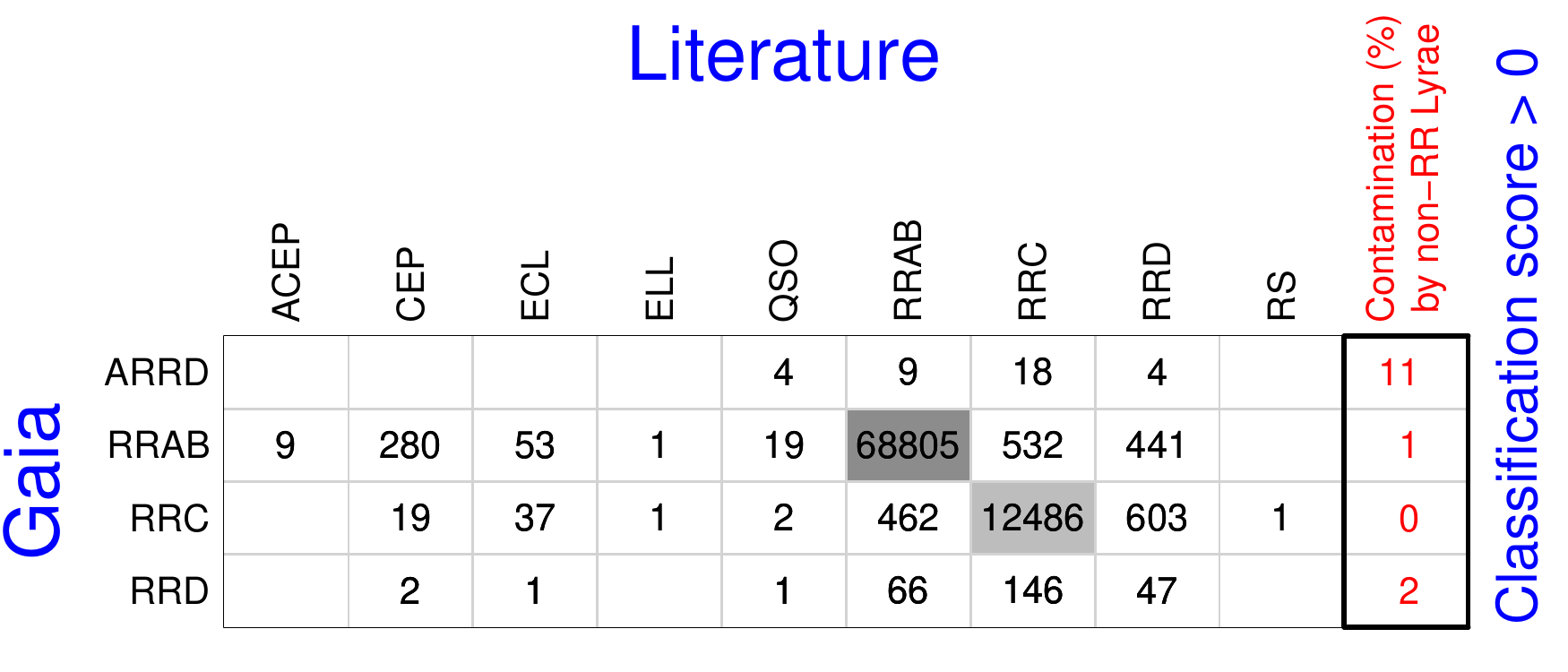}
  \caption{Comparison between the number RR\,Lyrae classifications of any score per subtype (in rows) and the classes found for these objects in the literature (in columns), among the 494~thousand crossmatched sources with \gmag-band range greater than 0.1~mag. Contamination rates (rounded per-cent values) with objects that do not belong to the RR\,Lyrae superclass are listed in the last column in red. Darker shaded squares indicate higher occurrences.}  
  \label{fig:RR_XM_CM_counts}
\end{figure}

\begin{figure}
\centering
  \includegraphics[width=\hsize]{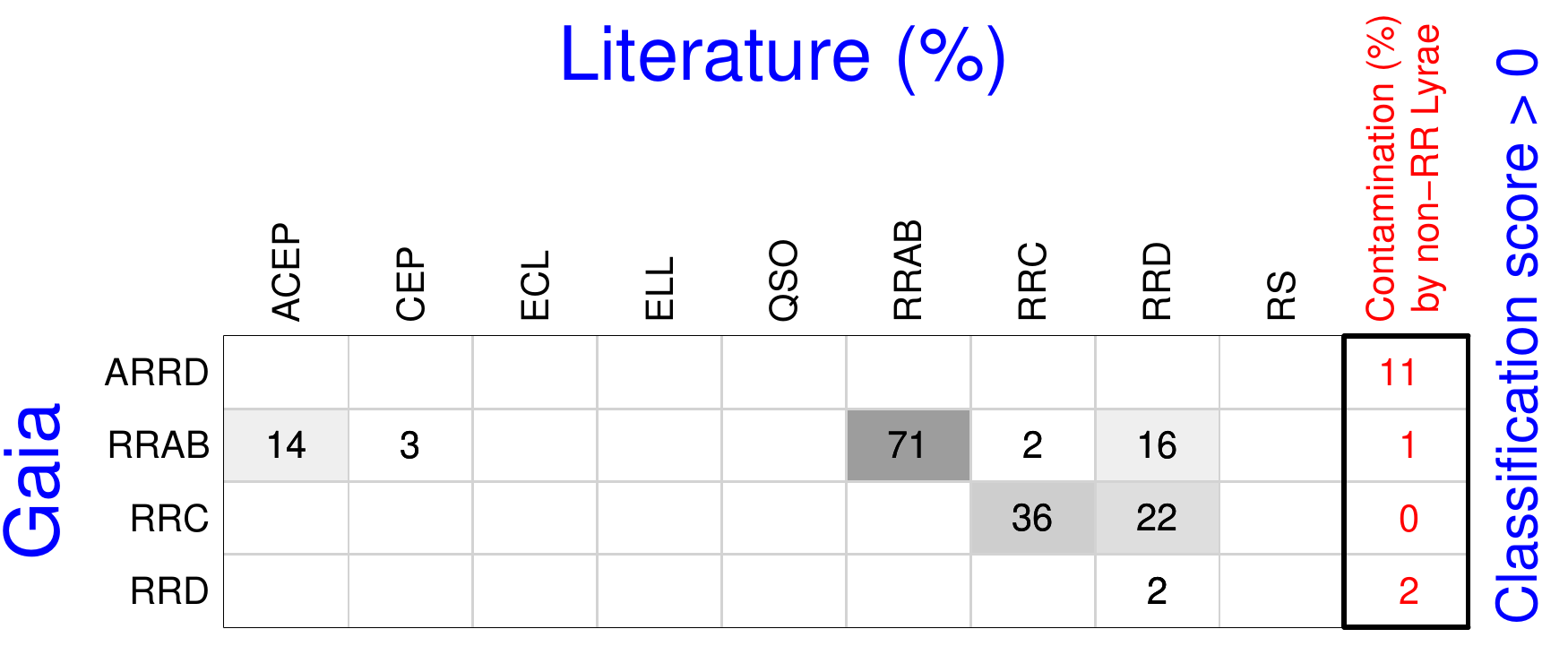}
  \caption{Same as Fig.~\ref{fig:RR_XM_CM_counts}, but in terms of (rounded) per-cent values with respect to the total number of objects for each class in the literature. Columns do not sum to 100\%\ because only rows relevant to RR\,Lyrae classifications are shown (and completeness is limited).}
  \label{fig:RR_XM_CM_percent}
\end{figure}

The distribution in the sky of the classification scores of RR\,Lyrae types is presented in Fig.~\ref{fig:RR_sky_score}, without particularly noticeable variations as a function of sky region. 
A comparison of the apparent completeness and contamination rates of RR\,Lyrae candidates of any score, employing sources that were cross matched with the literature, is shown in Figs.~\ref{fig:RR_XM_CM_counts} (in counts) and~\ref{fig:RR_XM_CM_percent} (in percentage), after excluding all training-set objects.
Most of the confusion seems to be among the RR\,Lyrae subtypes, and RRAB and RRD represent the most and the least complete, respectively.
Contamination rates are clearly underestimated, but they should not be over-interpreted because of the reasons mentioned in the beginning of Section~\ref{sec:results}. Non-RR\,Lyrae contaminating classes include, in order of relevance, Cepheids (CEP), eclipsing binaries (ECL), and quasars (QSO). 

One source of unexpected contamination (not included in Figs.~\ref{fig:RR_XM_CM_counts} and~\ref{fig:RR_XM_CM_percent}) includes galaxies, especially as the \gmag-band line spread function fitting of extended objects might return different flux levels as the \gaia spacecraft scan angle rotates (S. Cheng \& S. Koposov, NYC \gaia Sprint 2018\footnote{\url{http://gaia.lol/2018NYC.html}}). 
The source identifiers of 982 likely galaxies in the 140\,784~RR\,Lyrae classifications confirmed by the dedicated SOS module are listed in \citet{2018arXiv180502079C}.

\begin{figure}
\centering
  \includegraphics[width=\hsize]{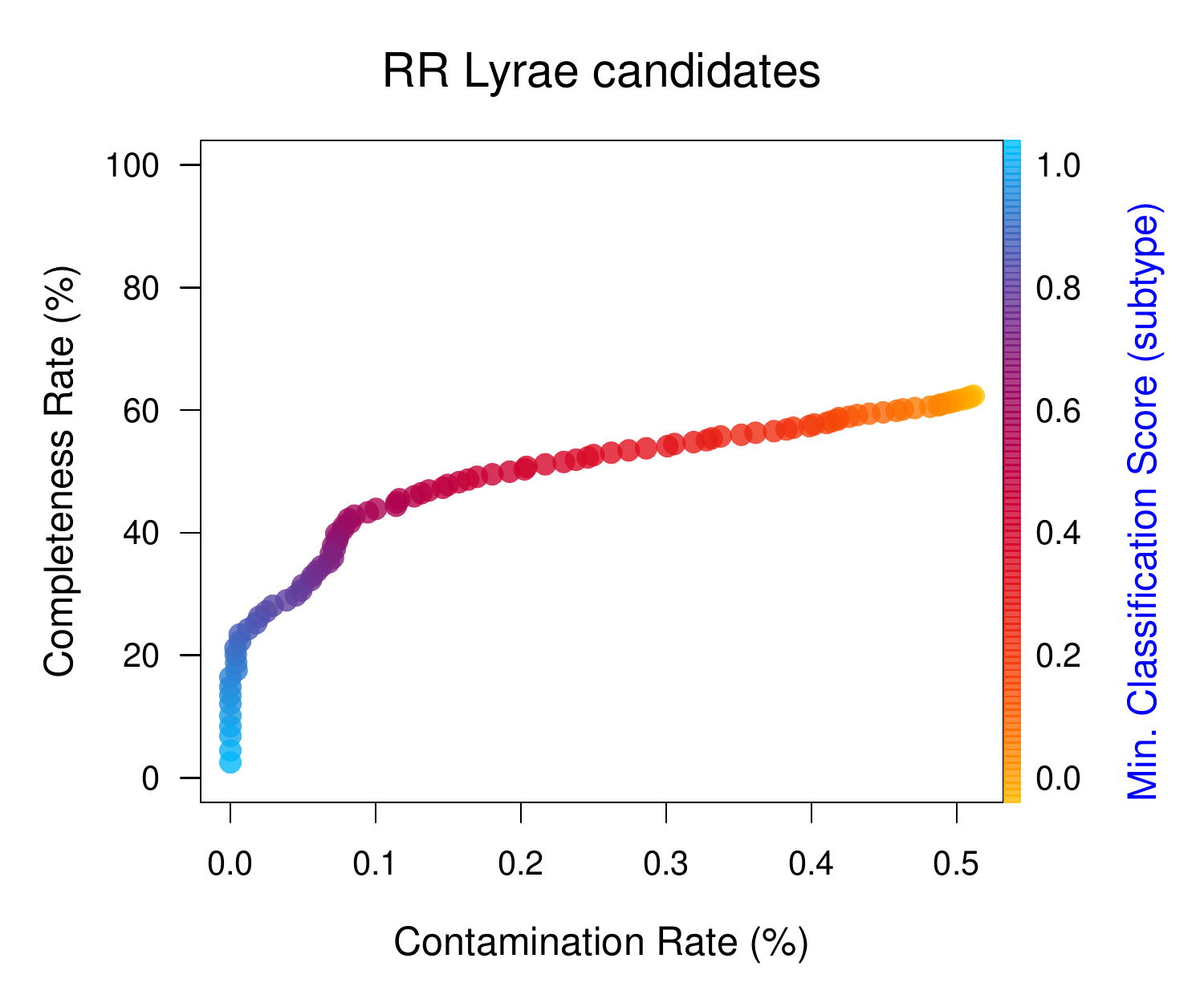}
  \caption{Completeness vs.\ contamination rates of RR\,Lyrae classifications (all subtypes) as a function of minimum classification score (defined in Section~\ref{sec:class_score} and colour-coded as indicated in the legend on the right-hand side).}
  \label{fig:RR_XM_CC}
\end{figure}

Samples with higher completeness and lower contamination can often be selected by applying thresholds to classification scores and other quantities, such as brightness, variation amplitude, number of observations, and sky region.
The dependence of the apparent completeness and contamination rates (derived from crossmatched sources) on minimum classification scores for the RR\,Lyrae subtypes is shown in Fig.~\ref{fig:RR_XM_CC}, generally confirming the expected trend that higher score thresholds increase the completeness to contamination ratio more efficiently than lower score limits.
In addition to the SOS module results described in \citet{2018arXiv180502079C} as part of the \gaia variability pipeline, an independent validation of the RR\,Lyrae classifications was performed with stars observed in selected \textit{K2} fields of the \textit{Kepler} space telescope \citep{2018arXiv180511395M}.

\subsection{Cepheids \label{sec:res-cep}}

\begin{figure}
\centering
  \includegraphics[width=\hsize]{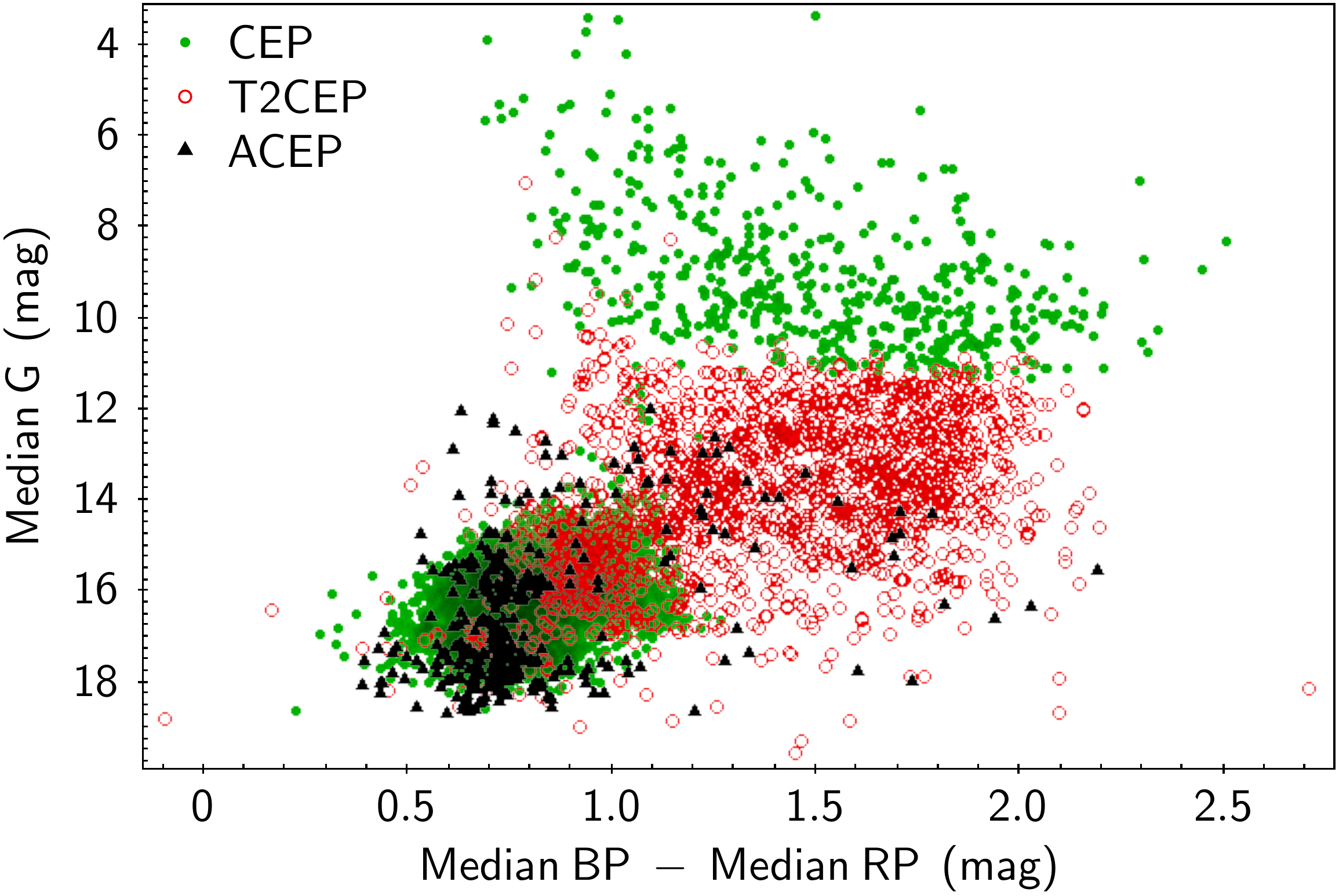}
  \caption{Colour-magnitude diagram (as median \gmag-band magnitude vs.\ median~\gbp$-$~median~\grp) of the Cepheid classifications, including the classical (labelled CEP, green dots), type-II (labelled T2CEP, red circles), and anomalous (labelled ACEP, black triangles) Cepheids.}
  \label{fig:CEP_color_mag}
\end{figure}

\begin{figure*}
\centering
  \includegraphics[width=\hsize]{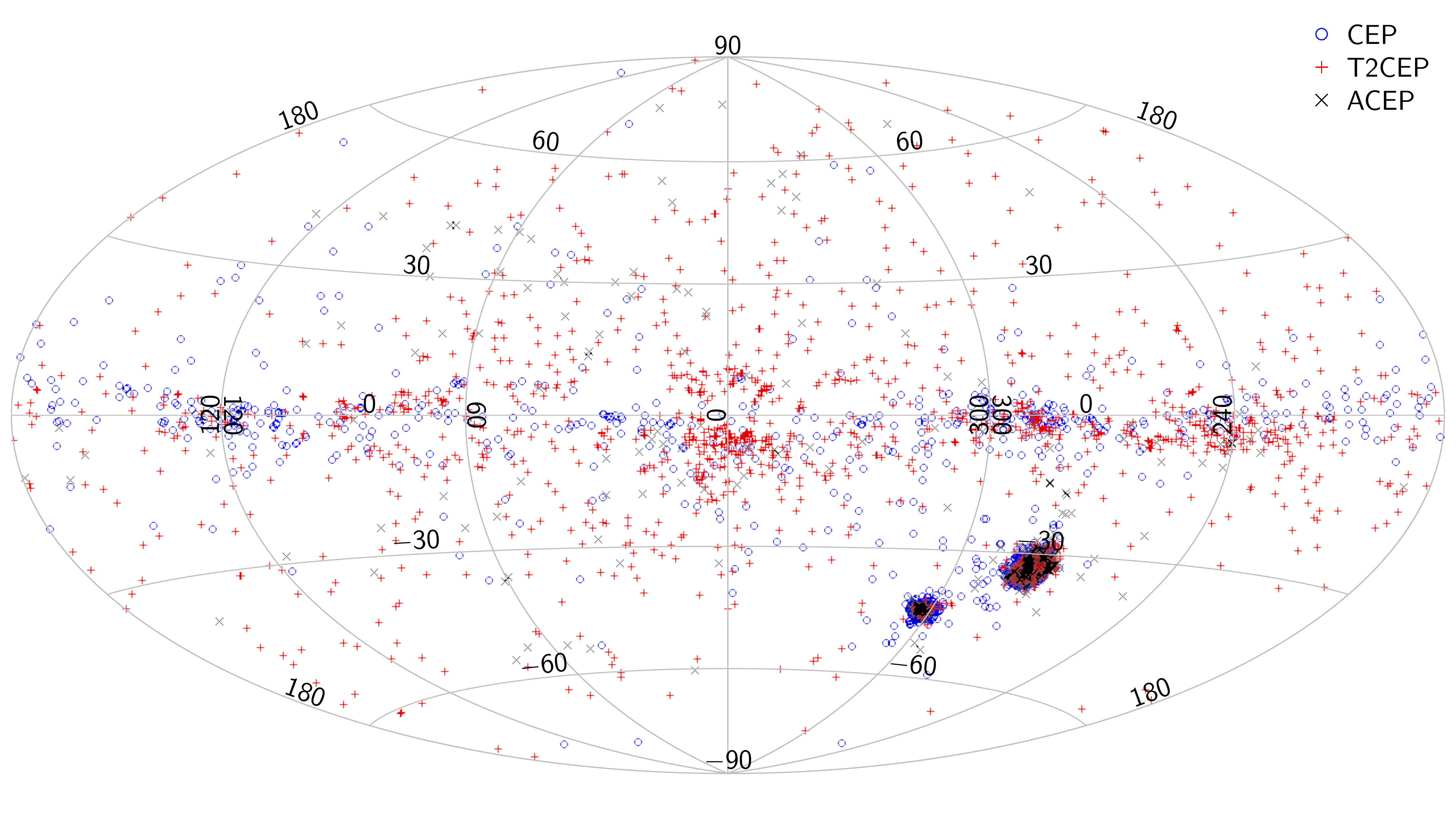}
  \caption{Distribution of the Cepheid subtype classifications in the sky (Galactic coordinates in degrees), including the classical (CEP, blue circles), type-II (T2CEP, red~`$+$' marks), and anomalous (ACEP, black~`$\times$' marks) Cepheids.}
  \label{fig:CEP_sky_type}
\end{figure*}

The Cepheid classifications include 8550 candidates, further subclassified into 6493~classical (or $\delta$) Cepheids (CEP), 1743~type-II Cepheids (T2CEP), and 314~anomalous Cepheids (ACEP). Their distribution in a colour-magnitude diagram is shown in Fig.~\ref{fig:CEP_color_mag} in terms of median \gmag magnitude versus \bpminrp colour for each subtype.
Most of the classical Cepheids fainter than \gmag$ \approx 13$~mag (about 90\%\ of all CEP classifications) are located in the region of the Magellanic Clouds, while the remaining (almost 10\%) brighter candidates are found primarily in correspondence of the Galactic disc, as shown also from the sky distribution in Fig.~\ref{fig:CEP_sky_type}.
The small number of classical Cepheids classified in the range of \gmag$\approx 11$--13~mag is due to a training-set bias because about half of the classical Cepheid sample was derived from \textit{Hipparcos} classifications and the other half from the region of the Magellanic Clouds. Thus, the CEP classifications represent the magnitude ranges of the Magellanic Clouds and of the \textit{Hipparcos} survey (with magnitude limit at \gmag$\approx 11$~mag). 
Except for a handful of candidates, type-II Cepheid classifications are fainter than \gmag$ \approx 10$~mag, and almost one-third of them is associated with the Magellanic Clouds (typically bluer than \bpminrp$ \approx 1.5$~mag), while the rest is scattered in the Galactic halo, with slight overdensities below the Galactic bulge and in parts of the disc.
The anomalous Cepheid classifications are fainter than \gmag$ \approx 12$~mag, and slightly more than half of them are found in the Magellanic Clouds (similar to the training-set composition and thus likely biased by it), while the other candidates are scattered across the whole sky.

\begin{figure}
\centering
  \includegraphics[width=\hsize]{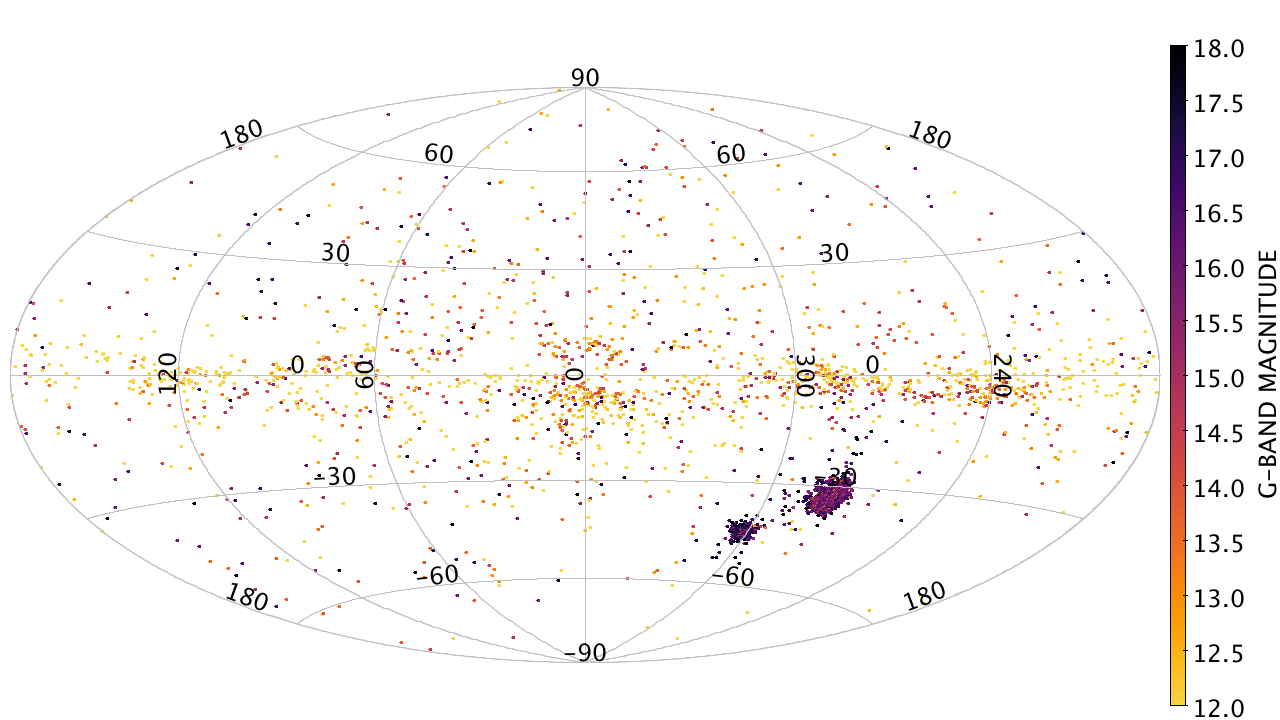}
  \caption{Same as Fig.~\ref{fig:RR_sky_mag}, but for Cepheid classifications (all subtypes).}
  \label{fig:CEP_sky_mag}
\end{figure}

\begin{figure}
\centering
  \includegraphics[width=\hsize]{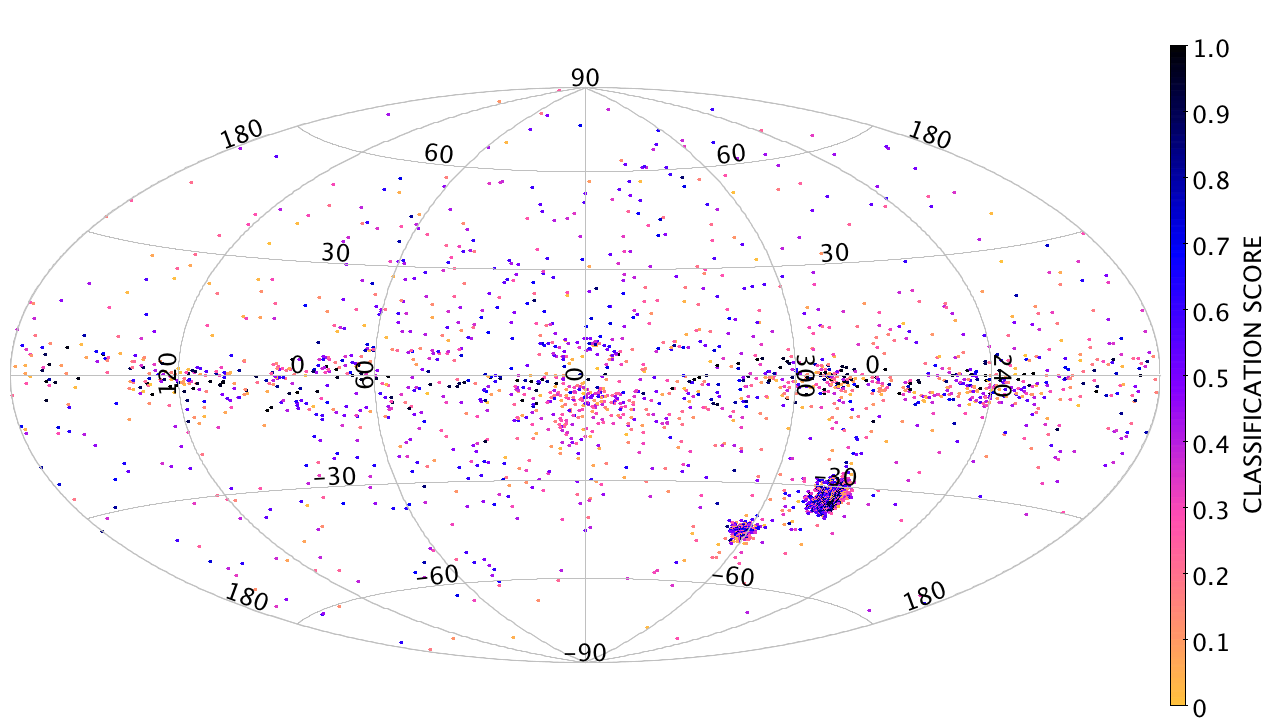}
  \caption{Same as Fig.~\ref{fig:RR_sky_score}, but for Cepheid classifications (all subtypes).}
  \label{fig:CEP_sky_score}
\end{figure}

The distribution in the sky of the median \gmag-band magnitudes of the classified Cepheids, shown in Fig.~\ref{fig:CEP_sky_mag}, confirms the findings from Figs.~\ref{fig:CEP_color_mag} and~\ref{fig:CEP_sky_type} that except for the Magellanic Clouds, the Galactic disc includes mostly bright classical Cepheid classifications and that fainter candidates (near the disc or in the halo) are classified as type-II or anomalous Cepheids.
The distribution of the classification scores of the Cepheid candidates in the sky is presented in Fig.~\ref{fig:CEP_sky_score} and suggests less certain identifications in the region of the Sagittarius dwarf spheroidal galaxy and in a some patches of the Galactic disc that are correlated with the overdensities of type-II Cepheid classifications.

\begin{figure}
\centering
  \includegraphics[width=\hsize]{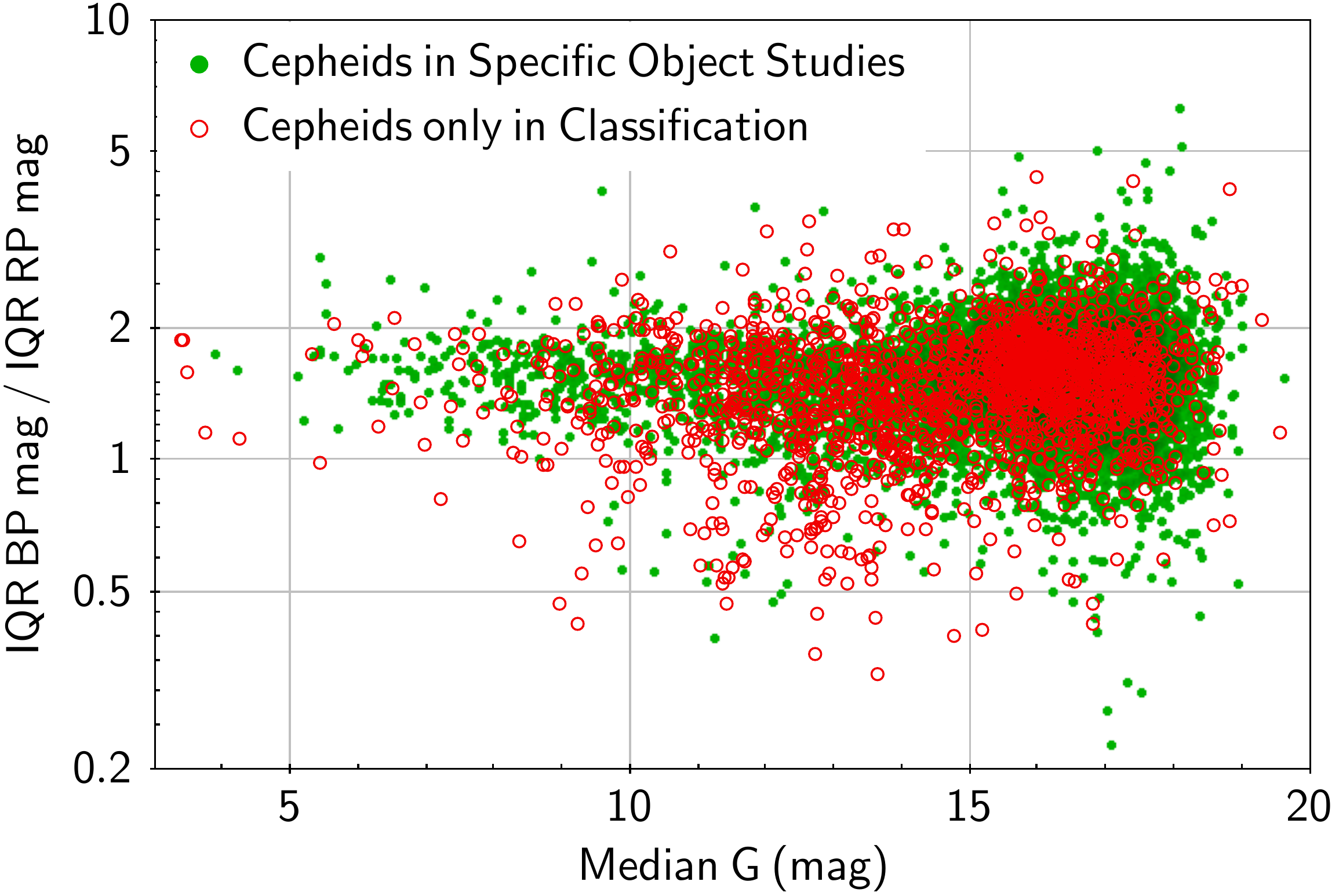}
  \caption{Same as Fig.~\ref{fig:SOS_CLASS_rr}(a), but for the Cepheid candidates in the SOS \citep[green dots;][]{2018arXiv180502079C} vs.\ those present only in the classification results (red circles).}
  \label{fig:SOS_CLASS_cep}
\end{figure}

\begin{figure}
\centering
  \includegraphics[width=\hsize]{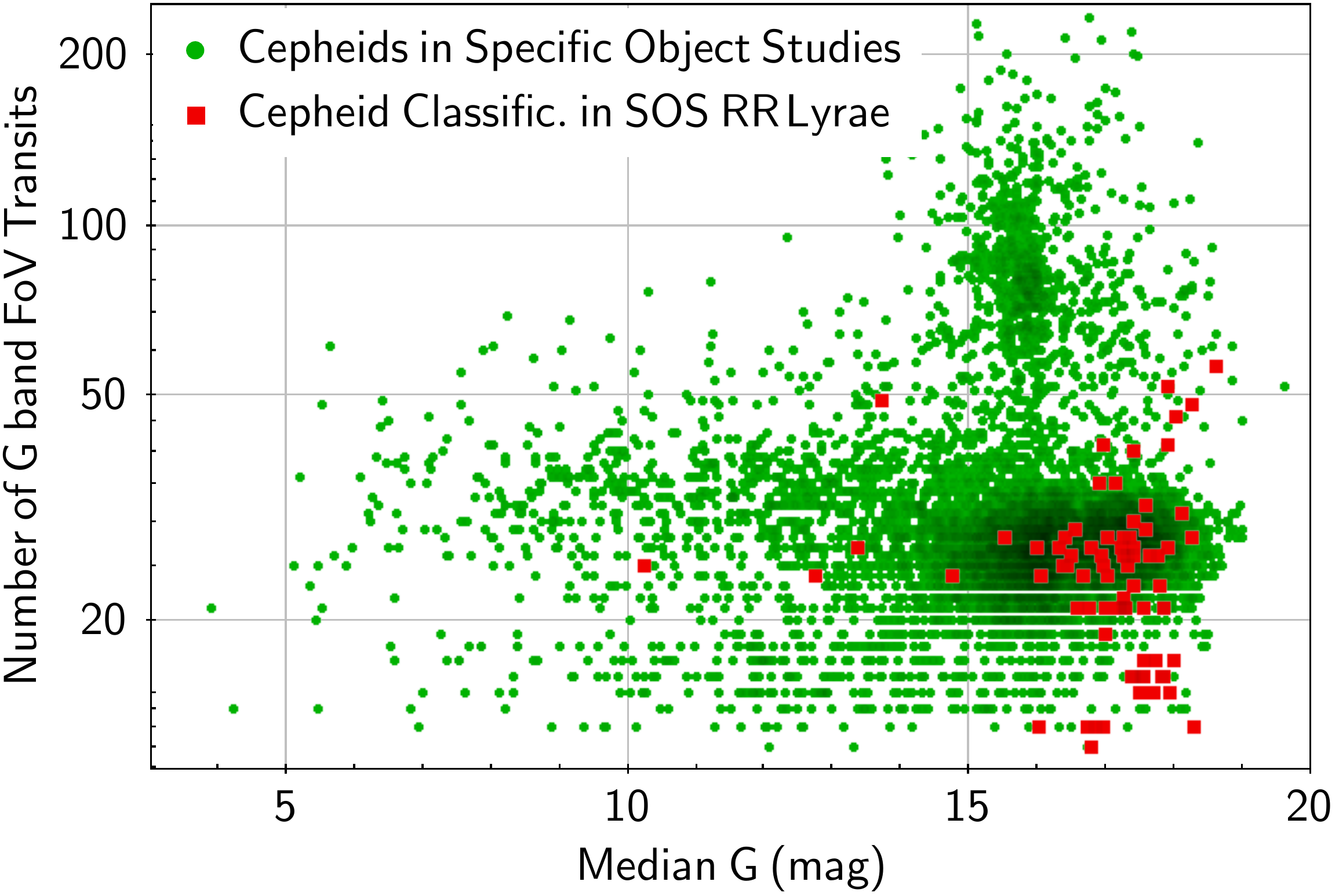}
  \caption{Same as Fig.~\ref{fig:SOS_CLASS_rr_cep}, but for the Cepheid classifications (red squares) that were reclassified as RR\,Lyrae stars in the SOS \citep[green dots;][]{2018arXiv180502079C}.}
  \label{fig:SOS_CLASS_cep_rr}
\end{figure}

Similar to the RR\,Lyrae classifications (Section~\ref{sec:res-rr}), the Cepheid candidates with at least 12~FoV transits in the \gmag  band were considered for validation by the dedicated SOS module \citep{2018arXiv180502079C} and a total of 1863~Cepheid classifications were not confirmed in SOS.
The distribution of the IQR in the \gbp versus \grp band is shown as a function of median \gmag-band magnitude in Fig.~\ref{fig:SOS_CLASS_cep} for the Cepheid candidates in SOS versus those present only in the classification results, and 1654 of the latter (89\%\ of the candidates not confirmed in SOS) have IQR(\gbp)/IQR(\grp)$>$1. 
A total of 77~Cepheid classifications are reclassified as RR\,Lyrae stars in SOS, as shown in Fig.~\ref{fig:SOS_CLASS_cep_rr} as a function of median \gmag-band magnitude and number of \gmag-band FoV transits per source: 40 and 72 of them (52 and 94\%) are labelled as ACEP and have classification scores lower than 0.5 (typically at the faint end), respectively.
In the special case of the $\delta$\,Cephei prototype, the latter is correctly identified as a classical Cepheid in the classification results, but it is absent from the SOS results because the source colours, the recovered period, and the related Fourier parameters were affected negatively by the limited sampling of this object.

\begin{figure}
\centering
  \includegraphics[width=\hsize]{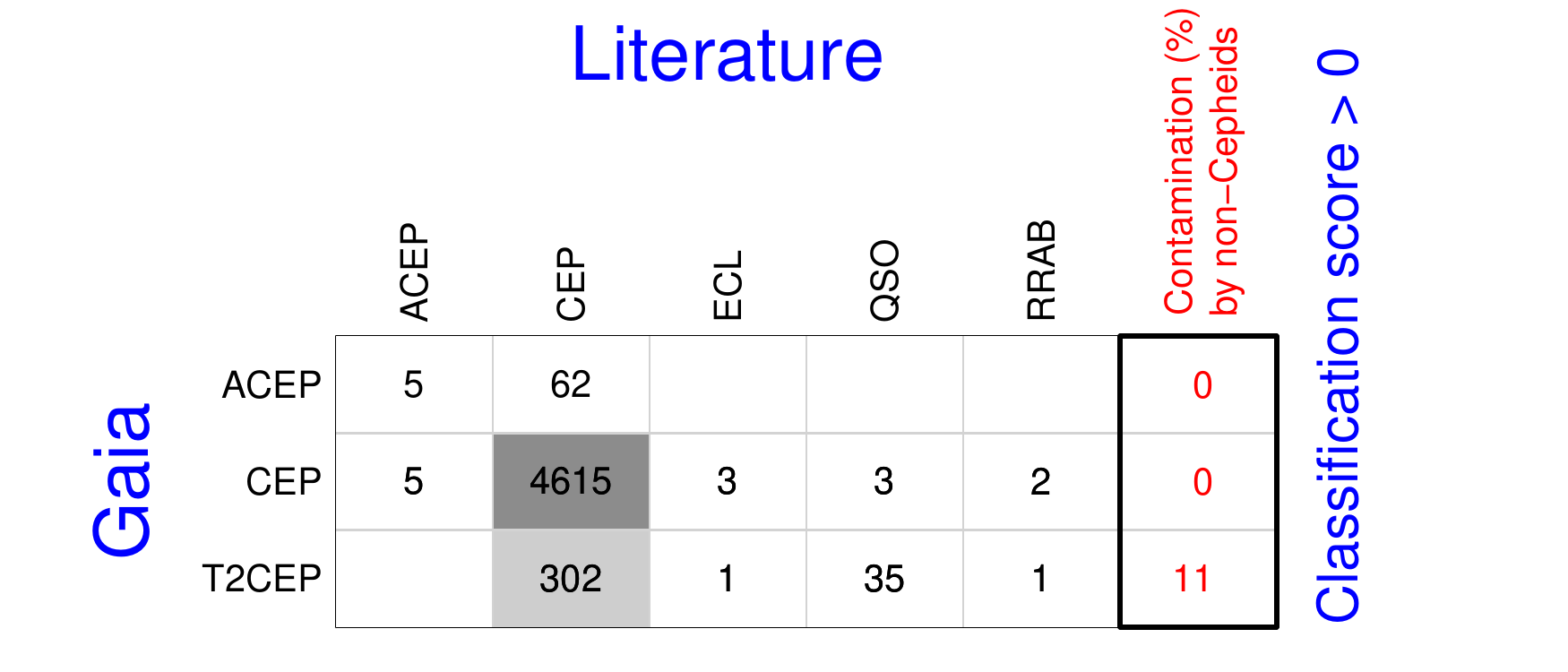}
  \caption{Same as Fig.~\ref{fig:RR_XM_CM_counts}, but for Cepheid classifications.}
  \label{fig:CEP_XM_CM_counts}
\end{figure}

\begin{figure}
\centering
  \includegraphics[width=\hsize]{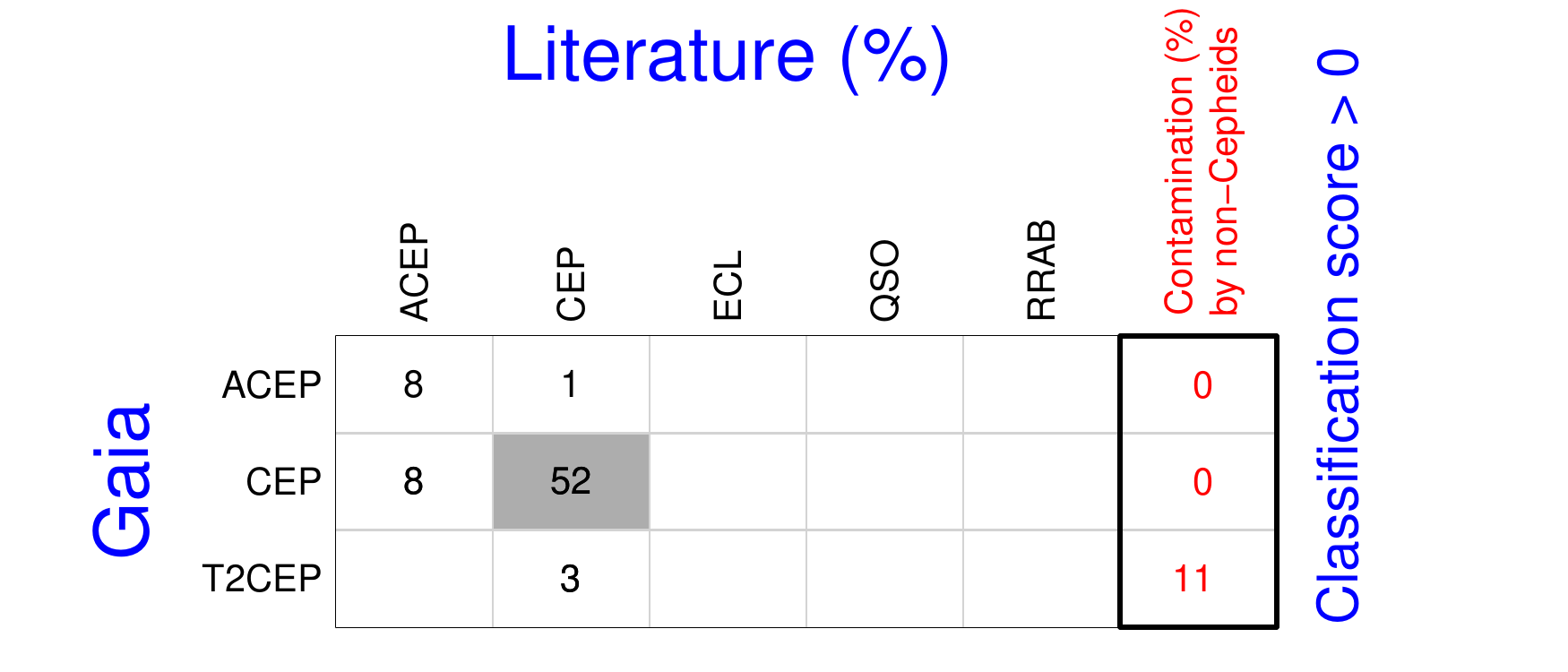}
  \caption{Same as Fig.~\ref{fig:RR_XM_CM_percent}, but for Cepheid classifications.}
  \label{fig:CEP_XM_CM_percent}
\end{figure}

A comparison of the apparent completeness and contamination rates of Cepheid candidates of any classification score, employing sources that were cross matched with the literature, is shown in Figs.~\ref{fig:CEP_XM_CM_counts} (in counts) and~\ref{fig:CEP_XM_CM_percent} (in percentage), after excluding all training-set objects (and all of the T2CEP crossmatched sources, as all of them were used in the training set).
The completeness of the classical Cepheids, much better represented in the training set, is superior to the anomalous Cepheid classifications, as expected.
Contamination rates are clearly underestimated, but they should not be over-interpreted because of the reasons mentioned in the beginning of Section~\ref{sec:results}. A small set of quasars (QSO) seems to be the primary source of contamination, especially for T2CEP classifications. 

\begin{figure}
\centering
  \includegraphics[width=\hsize]{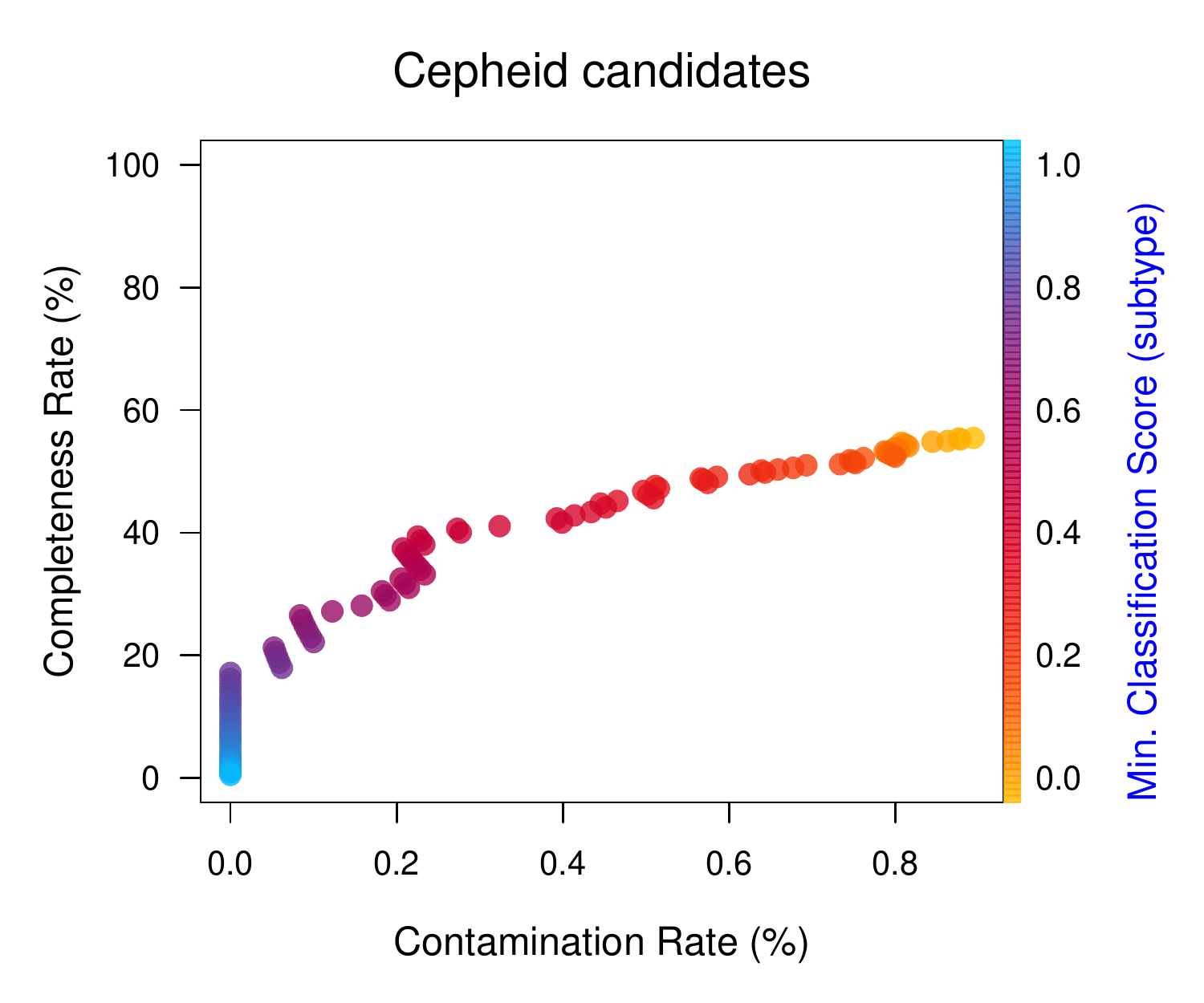}
  \caption{Same as Fig.~\ref{fig:RR_XM_CC}, but for Cepheid classifications.}
  \label{fig:CEP_XM_CC}
\end{figure}

The dependence of the apparent completeness and contamination rates (derived from crossmatched sources) on minimum classification scores for the Cepheid subtypes is shown in Fig.~\ref{fig:CEP_XM_CC}. It generally confirms the expected trend that higher score thresholds increase the completeness-to-contamination ratio more efficiently than lower score limits.

In addition to the SOS module results described in \citet{2018arXiv180502079C} as part of the \gaia variability pipeline, an independent validation of a small sample of Cepheid classifications was performed with stars observed in selected \textit{K2} fields of the \textit{Kepler} space telescope \citep{2018arXiv180511395M}.
The set of \gaia~DR2 Cepheids in the SOS results has recently been reclassified \citep{2018arXiv181010486R}, providing a sample with negligible contamination and more accurate Cepheid subclassifications.

\subsection{$\delta$\,Scuti and SX\,Phoenicis stars}

\begin{figure}
\centering
  \includegraphics[width=\hsize]{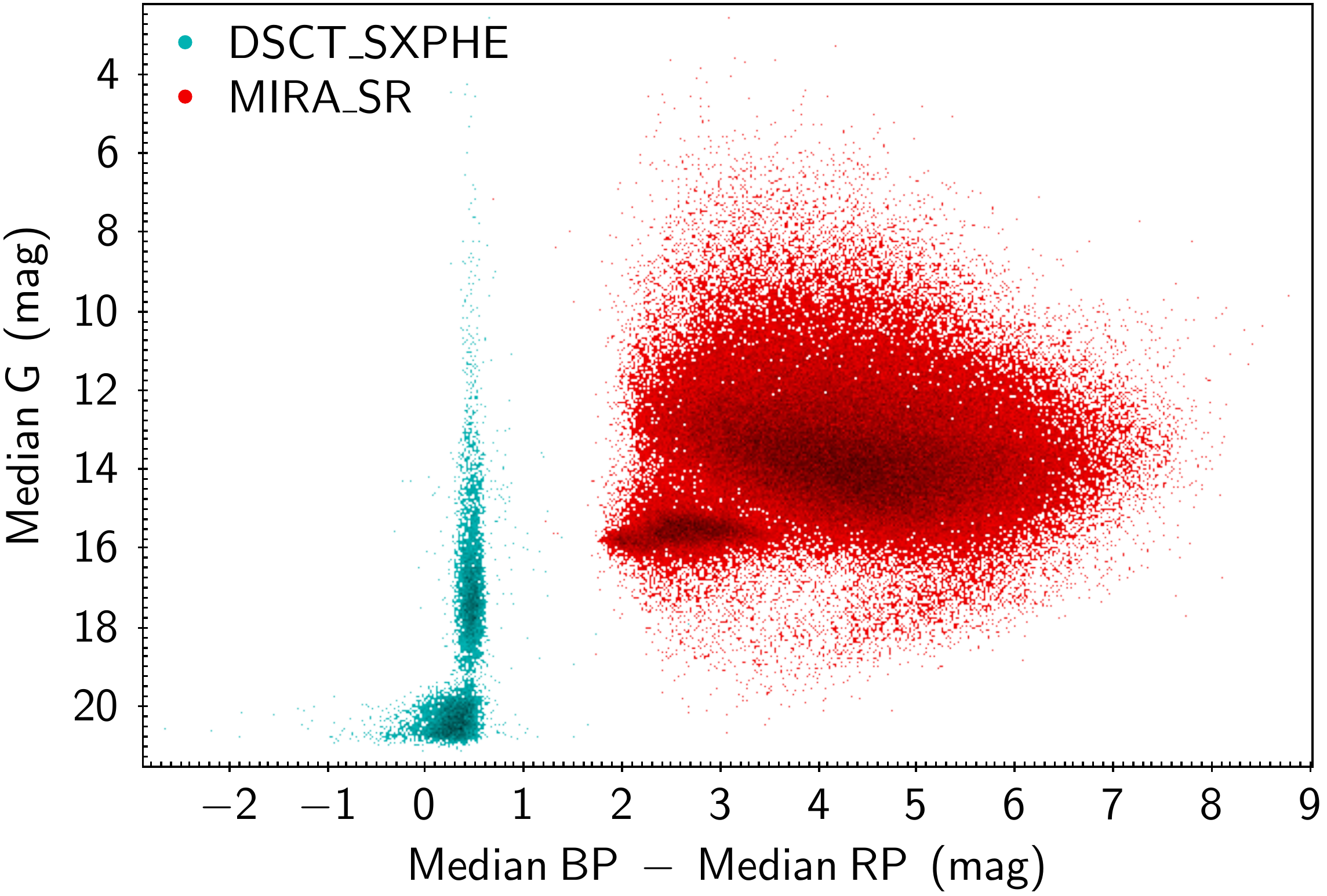}
  \caption{Colour-magnitude diagram (as median \gmag-band magnitude vs.\ median~\gbp$-$~median~\grp) of the $\delta$\,Scuti/SX\,Phoenicis (labelled DSCT\_SXPHE, in cyan) and Mira/semiregular (labelled MIRA\_SR, in red) classifications.}
  \label{fig:DSCT_LPV_color_mag}
\end{figure}

\begin{figure}
\centering
  \includegraphics[width=\hsize]{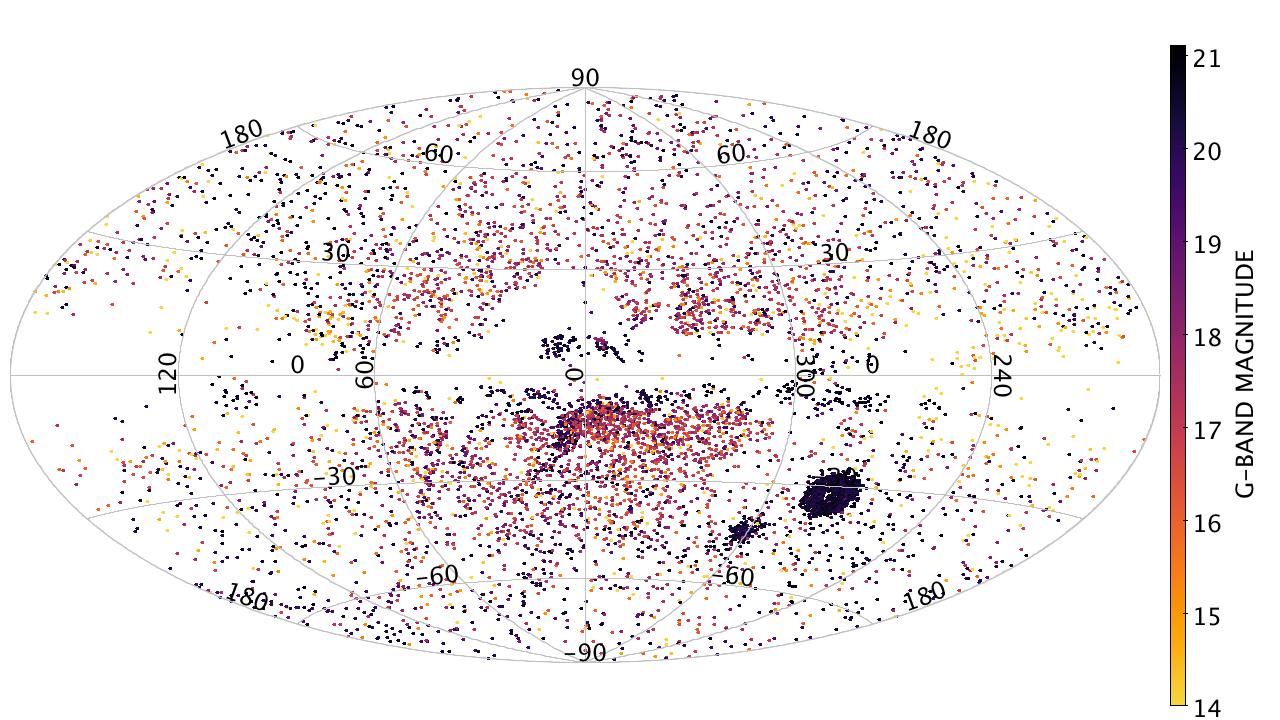}
  \caption{Same as Fig.~\ref{fig:RR_sky_mag}, but for $\delta$\,Scuti/SX\,Phoenicis classifications.}
  \label{fig:DSCT_sky_mag}
\end{figure}

\begin{figure}
\centering
  \includegraphics[width=\hsize]{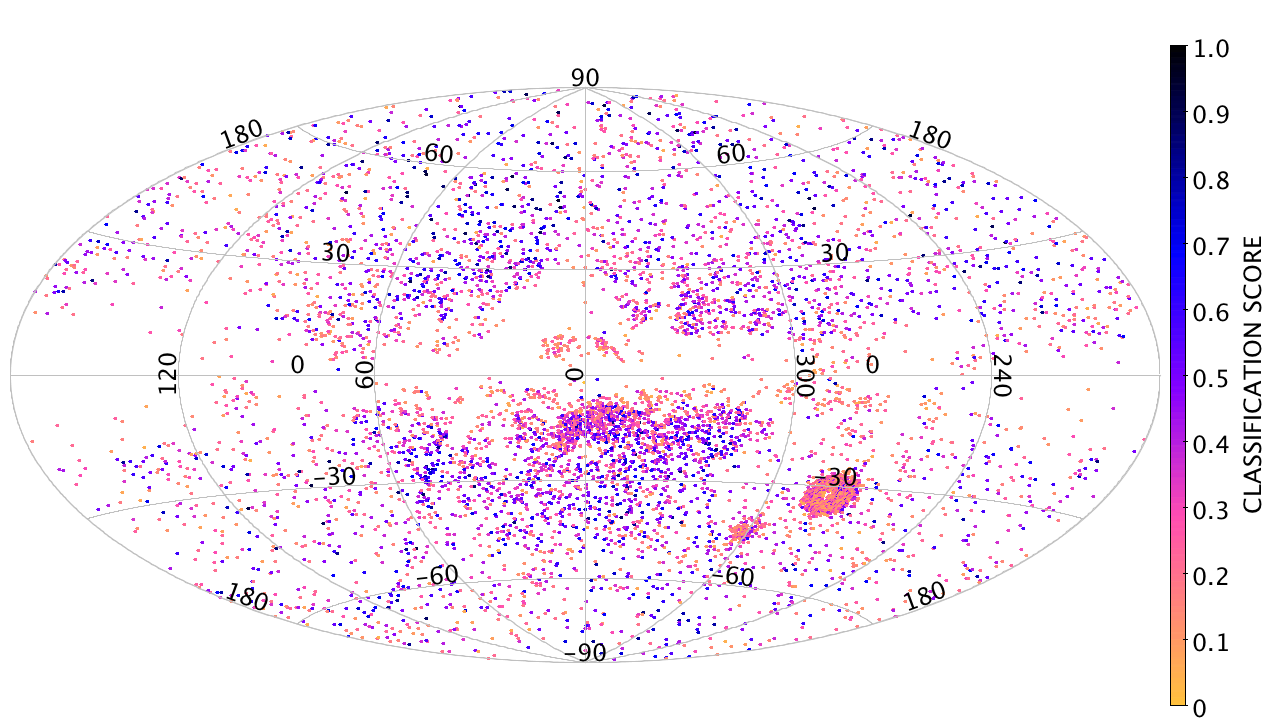}
  \caption{Same as Fig.~\ref{fig:RR_sky_score}, but for $\delta$\,Scuti/SX\,Phoenicis classifications.}
  \label{fig:DSCT_sky_score}
\end{figure}

The $\delta$\,Scuti and SX\,Phoenicis classifications include 8882 candidates and are merged together (labelled as DSCT\_SXPHE) because their \gaia light curves can appear very similar (without metallicity information). Their distribution in a colour-magnitude diagram is shown in Fig.~\ref{fig:DSCT_LPV_color_mag} in terms of median \gmag magnitude versus \bpminrp colour.
In general, most of the $\delta$\,Scuti/SX\,Phoenicis candidates are identified in the Magellanic Clouds, in the Galactic halo (as expected in particular for the metal-poor SX\,Phoenicis candidates), and in the Sagittarius dwarf spheroidal galaxy (see Fig.~\ref{fig:DSCT_sky_mag}).
The vast majority of candidates is not affected by Galactic extinction, likely because reddened $\delta$\,Scuti/SX\,Phoenicis samples are insufficiently represented in the training set.
Approximately half of the candidates are in the clump between median \gmag of 19.5 and 20.7~mag (about the \gaia magnitude limit): half of these are located in the Magellanic Clouds, while the others are partly in the Sagittarius dwarf spheroidal galaxy and partly scattered in the Galactic halo.
Brighter candidates are distributed mostly in the Galactic halo, with a larger number density in the region below the Galactic bulge, as shown in Fig.~\ref{fig:DSCT_sky_mag}.

The distribution in the sky of the classification scores of the $\delta$\,Scuti and SX\,Phoenicis candidates is presented in Fig.~\ref{fig:DSCT_sky_score}, and a comparison with Fig.~\ref{fig:DSCT_sky_mag} indicates a definite reduction of score levels in the identifications of candidates fainter than a median \gmag of about 19.5~mag, especially in the Magellanic Clouds and around the Galactic bulge. For sources with median \gmag between 17 and 19.5~mag, the correlation between brighter candidates and higher classification scores is much weaker than at the faint end, and it becomes negligible for $\delta$\,Scuti/SX\,Phoenicis candidates brighter than median \gmag$\approx 17$~mag.

\begin{figure}
\centering
  \includegraphics[width=\hsize]{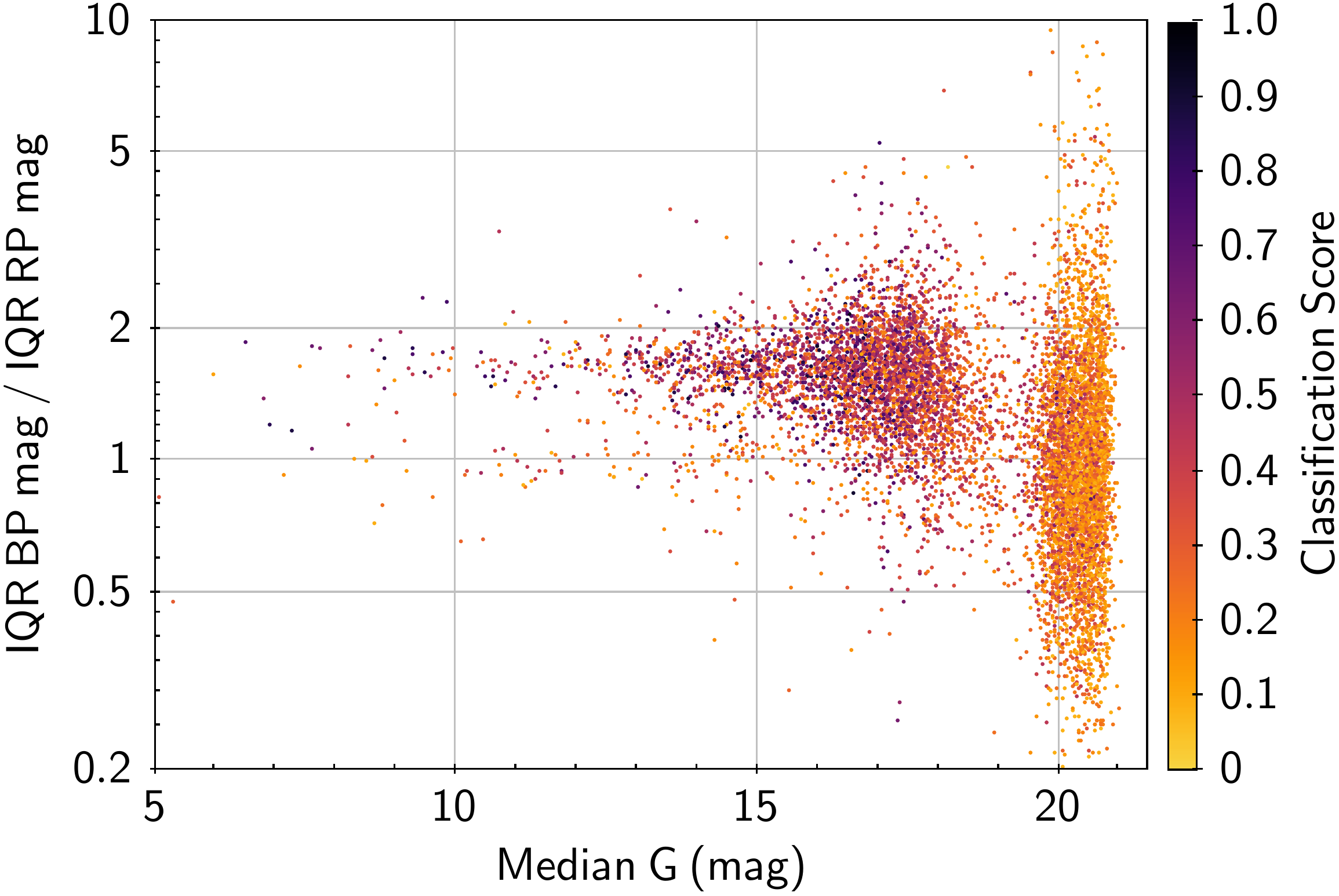}
  \caption{Distribution of the ratios of the interquartile range (IQR) in the \gbp vs.\ \grp bands as a function of the median \gmag-band magnitude for the $\delta$\,Scuti/SX\,Phoenicis candidates. The classification score is colour-coded as indicated in the legend on the right-hand side.}
  \label{fig:DSCT_iqr_bp_rp_ratio}
\end{figure}

The $\delta$\,Scuti/SX\,Phoenicis classifications are not verified by a dedicated SOS module.
Although a number of candidates could be confirmed by their period and a comparison of the Fourier amplitudes and phases of the first two harmonics, a large number of insufficiently sampled candidates would remain unconfirmed (but not necessarily rejected).  We therefore opted for a simple check of the light variations of the $\delta$\,Scuti/SX\,Phoenicis candidates in different bands, and let the community study these candidates in further detail (possibly with additional data).
As illustrated for Cepheids and RR\,Lyrae stars (in Sections~\ref{sec:res-cep} and \ref{sec:res-rr}), $\delta$\,Scuti/SX\,Phoenicis pulsators are also expected to exhibit larger variations in the \gbp  than in the \grp band, although these intrinsically fainter stars have lower amplitudes on average (for a given apparent brightness), and their IQR estimates are therefore more strongly affected by noise.  The distribution of the ratios of the IQR in the \gbp versus \grp band is shown as a function of median \gmag-band magnitude in Fig.~\ref{fig:DSCT_iqr_bp_rp_ratio}. As the IQR is more strongly influenced by the photometric noise especially towards fainter magnitudes, the following fractions of $\delta$\,Scuti/SX\,Phoenicis candidates with IQR(\gbp)/IQR(\grp)$>$1 inferred from this distribution represent likely lower limits\footnote{If the number of measurements in \grp is sufficiently smaller than in \gbp, the former could artificially reduce the IQR(\grp) and boost the IQR(\gbp)/IQR(\grp) ratio. However, the distribution of the difference of the number of measurements in \grp and \gbp is rather symmetric, relieving concerns of such biases for the fractions of sources with IQR(\gbp)/IQR(\grp)$>$1.}: 0.87 ($=$576/664), 0.94 ($=$1942/2077), 0.91 ($=$3795/4189), and 0.67 ($=$5981/8882), for source subsets with median \gmag brighter than 15, 17, 19, and any magnitude, respectively. 
Higher classification scores further support the same conclusion, as the overall fraction of 0.67 increases to 0.75 ($=$4449/5944), 0.80 ($=$3251/4057), 0.86 ($=$2213/2578), and 0.90 ($=$1366/1513) in subsets of candidates associated with scores greater than 0.2, 0.3, 0.4, and 0.5, respectively.

\begin{figure}
\centering
  \includegraphics[width=\hsize]{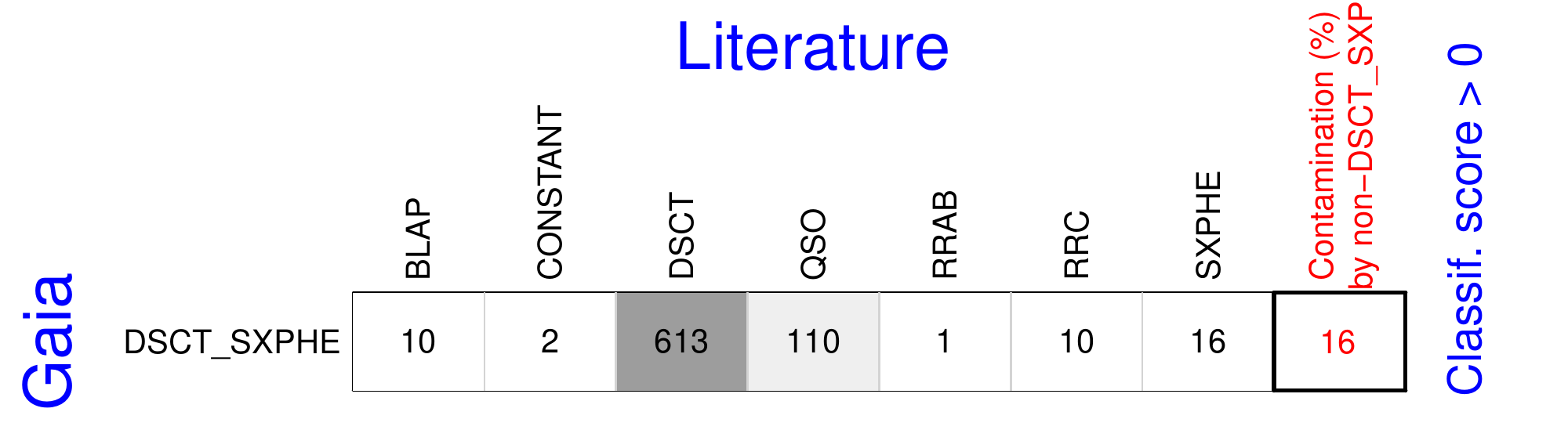}
  \caption{Same as Fig.~\ref{fig:RR_XM_CM_counts}, but for $\delta$\,Scuti/SX\,Phoenicis classifications. In this case, the training-set objects were included in the computation of the completeness and contamination rates.}
  \label{fig:DS_XM_CM_counts}
\end{figure}

\begin{figure}
\centering
  \includegraphics[width=\hsize]{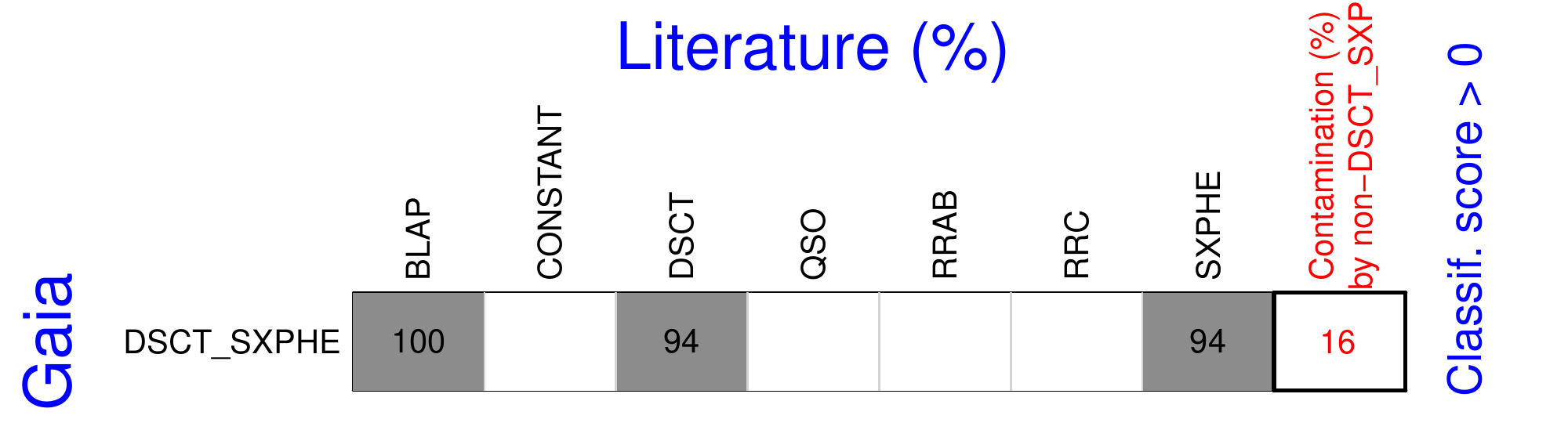}
  \caption{Same as Fig.~\ref{fig:RR_XM_CM_percent}, but for $\delta$\,Scuti/SX\,Phoenicis classifications.  In this case, the training-set objects were included in the computation of the completeness and contamination rates.}
  \label{fig:DS_XM_CM_percent}
\end{figure}

\begin{figure}
\centering
  \includegraphics[width=\hsize]{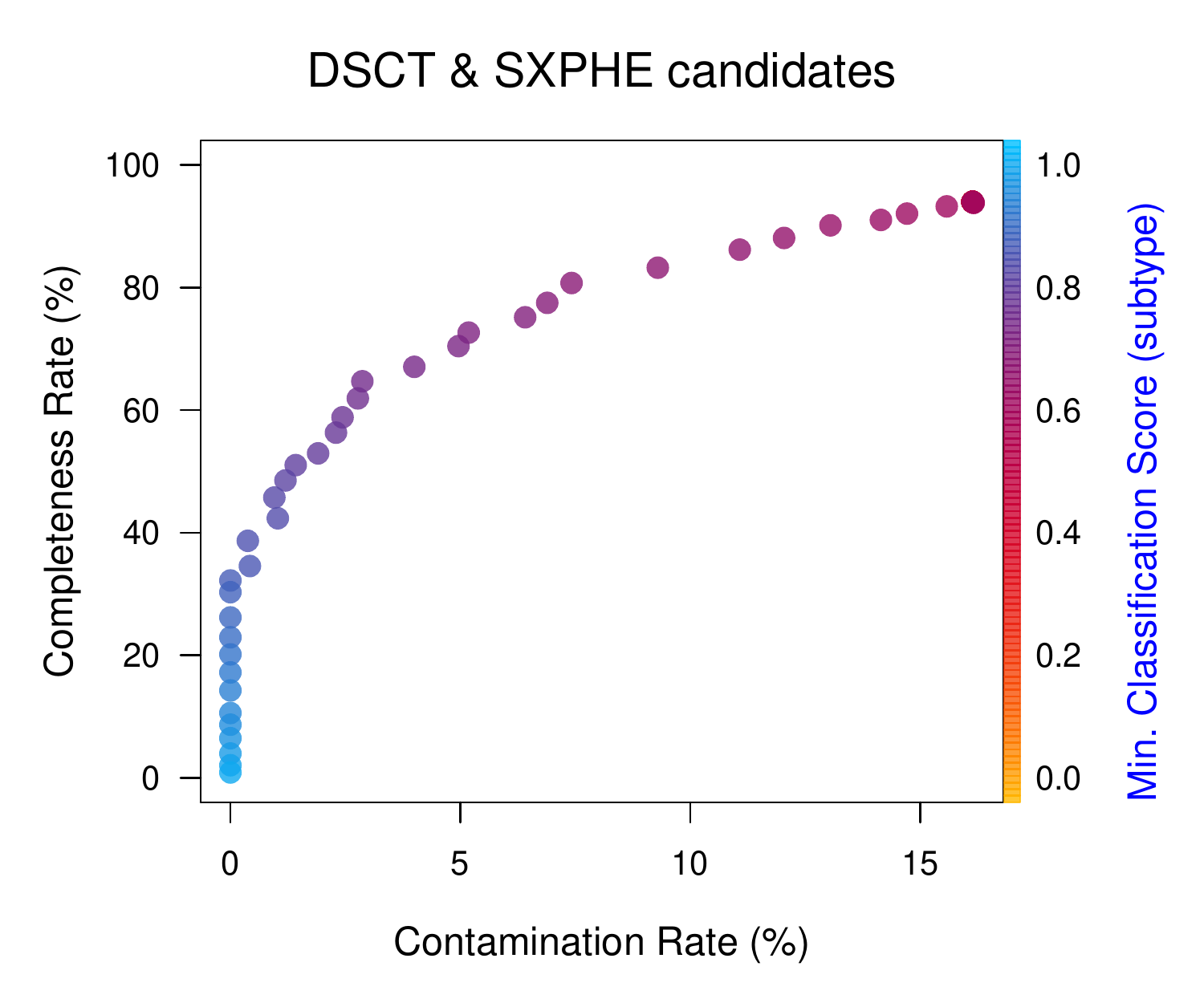}
  \caption{Same as Fig.~\ref{fig:RR_XM_CC}, but for $\delta$\,Scuti/SX\,Phoenicis classifications.  In this case, the training-set objects were included in the computation of the completeness and contamination rates.}
  \label{fig:DS_XM_CC}
\end{figure}

As mentioned in Section~\ref{sec:validation}, all of the cross-matched $\delta$\,Scuti/SX\,Phoenicis stars were included in the training set, so that the comparison with the literature employed for Figs.~\ref{fig:DS_XM_CM_counts}--\ref{fig:DS_XM_CC} is expected to be overly optimistic, but it is still included to show the pitfalls even in the best-case scenario.
The apparent completeness and contamination rates of $\delta$\,Scuti/SX\,Phoenicis candidates of any score are shown in Figs.~\ref{fig:DS_XM_CM_counts} (in counts) and~\ref{fig:DS_XM_CM_percent} (in percentage), suggesting that  QSOs constitute the main source of contamination.  The BLAP classifications were merged with the DSCT\_SXPHE class (as explained in Section~\ref{sec:impl}), therefore the trained BLAP sources were included among the $\delta$\,Scuti/SX\,Phoenicis candidates.

The dependence of the apparent completeness and contamination rates on minimum classification scores for the $\delta$\,Scuti/SX\,Phoenicis candidates is shown in Fig.~\ref{fig:DS_XM_CC}. This generally confirms the expected trend that higher score thresholds increase the completeness-to-contamination ratio more efficiently than lower score limits.

\subsection{Mira and semiregular variables}

\begin{figure}
\centering
  \includegraphics[width=\hsize]{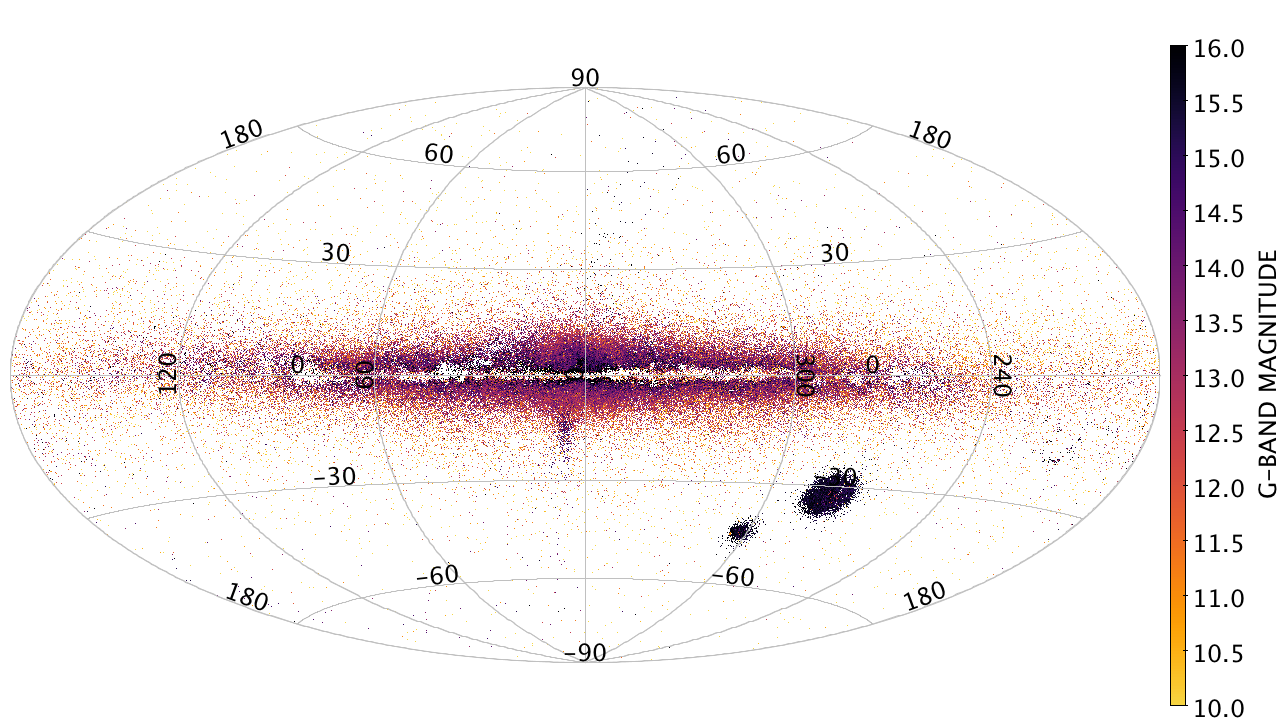}
  \caption{Same as Fig.~\ref{fig:RR_sky_mag}, but for Mira/semiregular classifications.}
  \label{fig:LPV_sky_mag}
\end{figure}

The Mira and semiregular classifications include 150\,757 candidates of long-period variables (labelled MIRA\_SR). 
Their distribution in a colour-magnitude diagram is shown in Fig.~\ref{fig:DSCT_LPV_color_mag} in terms of median \gmag magnitude versus \bpminrp colour.
As apparent in Fig.~\ref{fig:LPV_sky_mag}, most of the long-period candidates are associated with the Galactic disc and bulge, which form the largest clump visible in Fig.~\ref{fig:DSCT_LPV_color_mag}, and approximately 8\%\ of the candidates are in the region of the Magellanic Clouds, which give rise to the secondary clump centred at median \gmag$ \approx 15$ to 16~mag and bluer than \bpminrp$\approx 4$~mag.
The bright end includes candidates that are distributed more uniformly in the sky, while the faint end is represented primarily by extinguished objects, mostly close to the Galactic equator and some of them in the Magellanic Clouds.

\begin{figure}
\centering
  \includegraphics[width=\hsize]{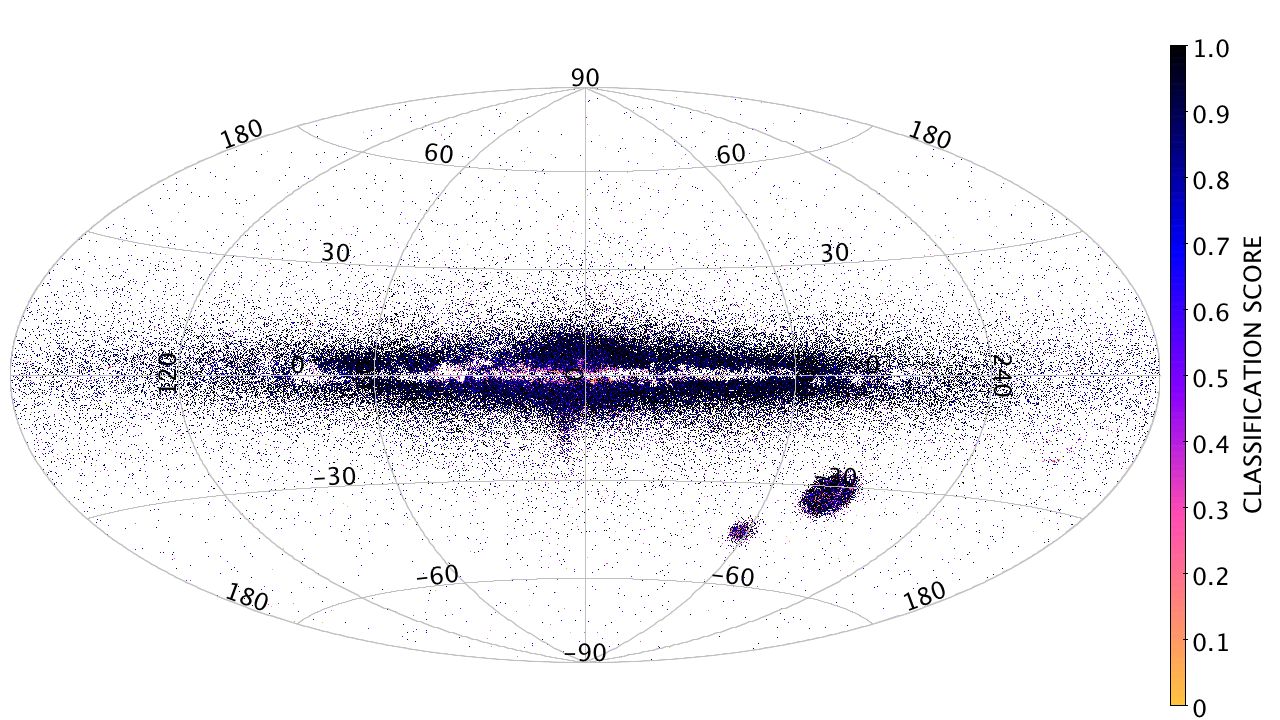}
  \caption{Same as Fig.~\ref{fig:RR_sky_score}, but for Mira/semiregular classifications.}
  \label{fig:LPV_sky_score}
\end{figure}

The distribution in the sky of the classification scores of  Mira and semiregular variables is presented in Fig.~\ref{fig:LPV_sky_score} and seems rather uninformative because it is dominated by high score values throughout the sky (about 80\% of the candidates have a score greater than 0.8), except for stars close to the Galactic equator (in particular in the region of the bulge), which are classified with reduced confidence.

As mentioned in Section~\ref{sec:validation}, the selection of reliable Mira and semiregular classifications was particularly challenging because the training set was not fully representative. Instead of filtering contaminating objects out by means of a validation classifier, 
the Mira and semiregular candidates were passed directly to the SOS module dedicated to long-period variables, and the published subset of these candidates followed from the selections described in \citet{2018arXiv180502035M}.
We note that among the contaminants of the Mira and semiregular candidates published in \gaia~DR2 is a small set (smaller than 1\%) of young stellar objects \citep[see details in][]{2018arXiv180502035M} and 99 sources that are part of the AllWISEAGN catalogue \citep{2015ApJS..221...12S}.

\section{Conclusions \label{sec:conclusions} }

The first all-sky classifications of four main classes of high-amplitude pulsating stars (RR\,Lyrae, Cepheid, $\delta$\,Scuti/SX\,Phoenicis, and Mira/semiregular) with intermediate \gaia data include already large numbers of candidates that can be used in many applications, especially those requiring a combination of a certain degree of homogeneity on the large scales and sufficient statistics of specific variability types.
Training-set biases are common in automated supervised classification methods, and their severity was reduced by source sampling (through the large number of sources that could be cross matched with the literature) and semi-supervised techniques. Some of the remaining biases were described for the interpretation of the results of specific (sub)classes, when applicable.
The classified subtypes of RR\,Lyrae and Cepheids can be inaccurate, especially if much less common than others of the same family, without making use of periods, Fourier parameters, or enough observations, but are provided in the attempt to increase the purity of the most common subtype. Parent classes are more reliable, and their use is suggested whenever possible.
Validation classifiers helped alleviate the presence of contaminants in the classification results at the cost of some reduction in completeness, and selections of even more reliable candidates were achieved by dedicated SOS modules \citep{2018arXiv180502079C, 2018arXiv180502035M}.

The next \gaia data release will include many more variable stars and variability classes, with improved classification accuracy through the increased number of measurements after about three years of observations (making models possible for the majority of sources) and the use of new data types (some of which were already available in \gaia~DR2, but not early enough to be used in the variability pipeline), 
in addition to improvements in the astrometric and photometric data. More details on the classification results (such as classification attributes and probability arrays) are also planned to be published in the future.

\begin{acknowledgements}
This work has made use of data from the European Space Agency (ESA) mission \gaia (\url{https://www.cosmos.esa.int/gaia}), processed by the \gaia Data Processing and Analysis Consortium (DPAC, \url{https://www.cosmos.esa.int/web/gaia/dpac/consortium}). 
Funding for the DPAC has been provided by national institutions, some of which participate in the \gaia Multilateral Agreement, 
which include, 
for Switzerland, the Swiss State Secretariat for Education, Research and Innovation through the ESA Prodex program, the `Mesures d'accompagnement', the `Activit\'{e}s Nationales Compl\'{e}mentaires', the Swiss National Science Foundation, and the Early Postdoc.Mobility fellowship;
for Belgium, the BELgian federal Science Policy Office (BELSPO) through PRODEX grants;
for Italy, Istituto Nazionale di Astrofisica (INAF) and the Agenzia Spaziale Italiana (ASI) through grants I/037/08/0,  I/058/10/0,  2014-025-R.0, and 2014-025-R.1.2015 to INAF (PI M.G. Lattanzi);
for Hungary, the Lend\"ulet grants LP2014-17 and LP2018-7 from the Hungarian Academy of Sciences, and the NKFIH grants K-115709, PD-116175, PD-121203 from the Hungarian National Research, Development, and Innovation Office. L.M.\ and E.P.\ have been supported by the J\'anos Bolyai Research Scholarship of the Hungarian Academy of Sciences.
This work made use of software from Postgres-XL (\url{https://www.postgres-xl.org}), Java (\url{https://www.oracle.com/java/}), Weka~\citep{Weka-citation}, R~\citep{R-citation}, and TOPCAT/STILTS \citep{2005ASPC..347...29T}.
\end{acknowledgements}

\bibliographystyle{aa}
\raggedbottom
\bibliography{34616corr_updated}

\begin{appendix}

\section{Training classes \label{app:classes}}
      The definitions of training-set class labels are listed below, highlighting the class labels published in \gaia~DR2 with a bold font. 
      \begin{enumerate}
         \item \textbf{ACEP}: Anomalous Cepheids.
         \item ACV: $\alpha^2$\,Canum Venaticorum-type stars.
         \item ACYG: $\alpha$\,Cygni-type stars.
         \item \textbf{ARRD}: Anomalous double-mode RR\,Lyrae stars.
         \item BCEP: $\beta$\,Cephei-type stars.
         \item BLAP: Blue large amplitude pulsators.
         \item \textbf{CEP}: Classical ($\delta$) Cepheids.
         \item CONSTANT: Objects whose variations (or absence thereof) are consistent with
            those of constant sources.
         \item CV: Cataclysmic variables of unspecified type.
         \item \textbf{DSCT}: $\delta$\,Scuti-type stars.
         \item ECL: Eclipsing binary stars.
         \item ELL: Rotating ellipsoidal variable stars (in close binary systems).
         \item FLARES: Magnetically active stars displaying flares.
         \item GCAS: $\gamma$\,Cassiopeiae-type stars.
         \item GDOR: $\gamma$\,Doradus-type stars.
         \item \textbf{MIRA}: Long-period variable stars of the $o$ (omicron) Ceti type (Mira).
         \item OSARG: OGLE small-amplitude red giant variable stars. 
         \item QSO: Optically variable quasi-stellar extragalactic sources.
         \item ROT: Rotation modulation in solar-like stars due to magnetic activity (spots).
         \item \textbf{RRAB}: Fundamental-mode RR\,Lyrae stars.
         \item \textbf{RRC}: First-overtone RR\,Lyrae stars.
         \item \textbf{RRD}: Double-mode RR\,Lyrae stars.
         \item RS: RS\,Canum Venaticorum-type stars.
         \item SOLARLIKE: Stars with solar-like variability induced by magnetic activity
            (flares, spots, and rotational modulation).
         \item SPB: Slowly pulsating B-type stars.
         \item SXARI: SX\,Arietis-type stars.
         \item \textbf{SXPHE}: SX\,Phoenicis-type stars.
         \item \textbf{SR}: Long-period variable stars of the semiregular type.
         \item \textbf{T2CEP}: Type-II Cepheids.
      \end{enumerate}

\section{Sample ADQL queries \label{app:queries}}
Documentation, examples, and support to access the \gaia data are available in the \gaia archive help web-page\footnote{\url{http://gea.esac.esa.int/archive-help/index.html}}.
Sample queries related to the classified variable stars, in some cases after SOS processing, are presented in the appendices of \citet{2018arXiv180502079C}, \citet{2018arXiv180409373H}, and \citet{2018arXiv180502035M}. 
We here add a few additional examples for the extraction of classification information combined with astrometry, photometry, or SOS results available in other archive tables.

In order to identify the 618 RR\,Lyrae classifications that were reclassified as Cepheids in SOS and retrieve information related to the classifier results, the number of \gmag-band FoV transits and their median magnitude, the \bpminrp colour from the medians of each band,  the SOS reclassification labels, and the Galactic latitude and longitude in degrees, the ADQL query is as follows.

{\small
\begin{verbatim}
SELECT c.source_id, best_class_name, best_class_score, 
       num_selected_g_fov, median_mag_g_fov, 
       median_mag_bp-median_mag_rp AS med_bp_rp,
       type_best_classification, l, b
FROM gaiadr2.vari_classifier_result AS c
INNER JOIN gaiadr2.vari_time_series_statistics AS stat 
        ON c.source_id = stat.source_id
INNER JOIN gaiadr2.vari_cepheid AS cep
        ON c.source_id = cep.source_id
INNER JOIN gaiadr2.gaia_source AS s
        ON c.source_id = s.source_id 
WHERE best_class_name = 'RRAB' OR 
      best_class_name = 'RRC' OR best_class_name='RRD' 
      OR best_class_name = 'ARRD'
\end{verbatim}
}

To select $\delta$\,Scuti/SX\,Phoenicis candidates with parallax and epoch photometry that satisfy the conditions of IQR(\gbp)/IQR(\grp)$>$1.5, median \gmag brighter than 17~mag, classification score greater than 0.5, and relative parallax precision better than 20\%, sorted by decreasing IQR ratio, the corresponding ADQL query is as follows.

{\small
\begin{verbatim}
SELECT s.source_id, median_mag_g_fov, best_class_score,
       iqr_mag_bp/iqr_mag_rp AS bp_rp_iqr_ratio,
       parallax, parallax_error, epoch_photometry_url
FROM gaiadr2.vari_classifier_result AS c
INNER JOIN gaiadr2.vari_time_series_statistics AS stat
        ON c.source_id = stat.source_id
INNER JOIN gaiadr2.gaia_source AS s
        ON c.source_id = s.source_id 
WHERE best_class_name = 'DSCT_SXPHE' AND
      iqr_mag_bp/iqr_mag_rp > 1.5 AND
      median_mag_g_fov < 17 AND
      best_class_score > 0.5 AND
      parallax_over_error > 5
ORDER BY bp_rp_iqr_ratio DESC
\end{verbatim}
}

\end{appendix}

\end{document}